\documentclass[prd,onecolumn,showpacs,showkeys,floatfix,superscriptaddress,11pt]{revtex4}

\usepackage[utf8]{inputenc} 
\usepackage[T1]{fontenc}    
\usepackage{hyperref}       
\usepackage{url}            
\usepackage{booktabs}       
\usepackage{amsfonts}       
\usepackage{nicefrac}       
\usepackage{microtype}      
\usepackage{lipsum}		
\usepackage{graphicx}
\usepackage{placeins}
\usepackage{natbib}
\usepackage{doi}
\usepackage{graphicx}
\usepackage{subfigure}
\usepackage{epsf}
\usepackage{bm}
\usepackage{amsmath}
\usepackage{amsfonts}
\usepackage{amssymb}
\usepackage{graphicx}
\usepackage{tabularx}
\usepackage{color}%
\providecommand{\U}[1]{\protect\rule{.1in}{.1in}}

\newcommand{\be}{\begin{equation}}
\newcommand{\ee}{\end{equation}}

\newcommand{\mincir}{\raise
-3.truept\hbox{\rlap{\hbox{$\sim$}}\raise4.truept\hbox{$<$}\ }}
\newcommand{\magcir}{\raise
-3.truept\hbox{\rlap{\hbox{$\sim$}}\raise4.truept\hbox{$>$}\ }}

\providecommand{\U}[1]{\protect\rule{.1in}{.1in}}

\usepackage{tikz,xcolor,hyperref}

\definecolor{lime}{HTML}{A6CE39}
\DeclareRobustCommand{\orcidicon}{%
	\begin{tikzpicture}
	\draw[lime, fill=lime] (0,0) 
	circle [radius=0.16] 
	node[white] {{\fontfamily{qag}\selectfont \tiny ID}};
	\draw[white, fill=white] (-0.0625,0.095) 
	circle [radius=0.007];
	\end{tikzpicture}
	\hspace{-2mm}
}

\foreach \x in {A, ..., Z}{%
	\expandafter\xdef\csname orcid\x\endcsname{\noexpand\href{https://orcid.org/\csname orcidauthor\x\endcsname}{\noexpand\orcidicon}}
}

\begin{document}

\title{Global dynamics in Einstein-Gauss-Bonnet scalar field cosmology with matter}

\author{Alfredo D. Millano\orcidC{}}
\email{alfredo.millano@alumnos.ucn.cl}
\affiliation{Departamento de Matem\'{a}ticas, Universidad Cat\'{o}lica del Norte, Avda.
Angamos 0610, Casilla 1280 Antofagasta, Chile}

\author{Genly Leon\orcidA{}}
\email{genly.leon@ucn.cl}
\affiliation{Departamento de Matem\'{a}ticas, Universidad Cat\'{o}lica del Norte, Avda.
Angamos 0610, Casilla 1280 Antofagasta, Chile}
\affiliation{Institute of Systems Science, Durban University of Technology, PO Box 1334,
Durban 4000, South Africa}

\author{Andronikos Paliathanasis\orcidB{}}
\email{anpaliat@phys.uoa.gr}
\affiliation{Institute of Systems Science, Durban University of Technology, PO Box 1334,
Durban 4000, South Africa}
\affiliation{Departamento de Matem\'{a}ticas, Universidad Cat\'{o}lica del Norte, Avda.
Angamos 0610, Casilla 1280 Antofagasta, Chile}

\begin{abstract}
We study the dynamics of the field equations in a four-dimensional isotropic and homogeneous spatially flat Friedmann--Lema\^{\i}tre--Robertson--Walker geometry in the context of Einstein-Gauss-Bonnet theory with a matter source and a scalar field coupled to the Gauss-Bonnet scalar. In this theory, the Gauss-Bonnet term contributes to the field equations. The mass of the scalar field depends on the potential function and the Gauss-Bonnet term. For the scalar field potential, we consider the exponential function and the coupling function between the scalar field and the Gauss-Bonnet scalar is considered to be the linear function. Moreover, the scalar field can have a phantom behaviour. We consider a set of dimensionless variables and write the field equations into a system or algebraic-differential equations. For the latter, we investigate the equilibrium points and their stability properties. In order to perform a global analysis of the asymptotic dynamics, we use compactified variables. This gravitational theory can explain the Universe's recent and past acceleration phases. Therefore, it can be used as a toy model for studying inflation or as a dark energy candidate. 
\end{abstract}
\date{\today}
\maketitle

\newpage

\section{Introduction}

In the theory of General Relativity, the physical space is described by a
four-dimensional Riemannian manifold \cite{ae1}, and Ricci's scalar of the Levi-Civita connection expresses the Lagrangian of the
field equations.
In \cite{lv1}, it has been shown that the Einstein-Hilbert Action Integral of
General Relativity generated by Ricci's scalar with or without the
cosmological constant term is the unique Action which gives second-order
field equations in a four-dimensional manifold. That is not true in
higher-order theories, where in \cite{lv2}, the most generic Action Integral
was presented, providing second-order differential equations in an
arbitrary dimensional spacetime. The so-called Lovelock gravity is the natural
extension of General Relativity.

General Relativity is a well-tested theory for the description of
astrophysical phenomena \cite{grt3} and compact objects \cite{grt1,grt2};
nevertheless, General Relativity fails to explain the observational phenomena
in cosmological scales. The cosmological observations indicate that the Universe at present is under an acceleration phase known as late-time
acceleration \cite{rr1,Teg}. However, it was proposed that the Universe
had been under a previous acceleration phase in its very early stages. The
inflationary mechanism can solve various observational phenomena such as the
horizon problem, the flatness problem, the homogeneity of the Universe and
other observations \cite{guth, Aref1}.

For the description of inflation, a scalar field is introduced in gravitational
theory; in the slow-roll limit, the scalar field potential dominates the
cosmological fluid and drives the dynamics for acceleration to occur \cite{newinf}. Furthermore, scalar fields have been introduced as
dark energy candidates for the description of the late-time acceleration, see
for instance \cite{q4,q15,q17,q21,q22,q24,q25,q27} and references therein.
Besides, scalar fields can attribute the degrees of freedom provided in the
field equations from the introduction of geometric invariants during the
modification of the Einstein-Hilbert Action Integral \cite{q12}. There is
a\ taxonomy of modified theories of gravity proposed in the
literature, which is divided into DE models linked to a fluid with the capability
of accelerating the Universe and models in which the Einstein field equations of the General Theory of Relativity are
modified; see the review articles \cite{md1,md2,md3}.

Gauss-Bonnet gravity belongs to the family of Lovelock's theory, where the
Gauss-Bonnet scalar is introduced in the Action Integral \cite{pan1}. However, the Gauss-Bonnet scalar is a topological
invariant in a four-dimensional manifold, meaning it does not provide any terms in the field
equations. In \cite{dd1}, to overpass this problem, the authors
introduced a re-scale on the Gauss-Bonnet coupling constant such that a
singular limit is introduced in Lovelock's gravity in the limit of the
four dimensions. With the latter, the Gauss-Bonnet term introduces non-trivial
terms to gravitational dynamics, and the field equations remain free from
Ostrogradsky instabilities.\ The introduction of a nonlinear function of the
Gauss-Bonnet scalar is another attempt to introduce non-trivial dynamical terms
in the field equations in four-dimensional gravity \cite{gbm01,gbm03,bb1}.

We are interested in the Einstein-Gauss-Bonnet scalar field
gravity in this work, where a scalar field coupled to the Gauss-Bonnet term is introduced in
the gravitational Action Integral. The coupling function ensures that the
Gauss-Bonnet term survives during the variation and affects the gravitational
dynamics. In this theory, the mass of the scalar field depends on the
Gauss-Bonnet component. The theory has been studied before in cosmological
scales \cite{in1,in2} and in astrophysical objects \cite{in4,bb2}. In the
limit of a spatially flat Friedmann--Lema\^{\i}tre--Robertson--Walker\ (FLRW)
the phase-space analysis for the field equations performed in \cite{dn11,dn12,dn13}.
It was found that the only equilibrium point where the Gauss-Bonnet term
contributes to the cosmological fluid is that of the de Sitter universe.
Nevertheless, in \cite{dn3}, a systematic analysis of the phase-space presented
where it was found that the new scaling solutions are supported in the
Einstein-Gauss-Bonnet scalar theory, where the Gauss-Bonnet term contributes to
the cosmological fluid. In the following, we extend the analysis presented in
\cite{dn3}, where we introduce an ideal gas in the field equations. The latter
is necessary to investigate if the Einstein-Gauss-Bonnet scalar
theory can reproduce the cosmological history and to infer the theory's viability.

The dynamical analysis of the gravitational field equations is a powerful
method for the analysis of the asymptotic dynamics of the theory
\cite{dn1,dn2}. Gravity is a nonlinear theory, and even in cosmological studies
where the field equations are ordinary differential equations, exact and
analytic solutions are challenging to be found. Moreover, we can study asymptotic solutions' existence conditions and stability
properties by analyzing the dynamics. Thus, we can solve the initial value
problem and reconstruct the cosmological evolution and history \cite{dn3,dn4}.
The method has been widely applied in various gravitational models in
cosmological studies \cite{dn5,dn6,dn7,dn8} and for analyzing
compact objects \cite{dn9,dn10}. The structure of the paper is as follows.

In section \ref{II}, the gravitational theory of our consideration, which is that of the four-dimensional
Einstein-Gauss-Bonnet theory with a scalar field coupled to the Gauss-Bonnet term is presented. We consider a quintessence and a phantom scalar field. In section \ref{III}, we perform a detailed analysis of the phase-space for
the exponential scalar field potential~$V\left(  \phi\right)  =V_{0} e^{\lambda\phi}$ and the linear coupling function $f(\phi)=f_{0} \phi$.  In section \ref{IV}, we consider the case where the scalar field is massless.  Section \ref{V} is devoted to study the case where the model has no scalar field potential. Finally, in section \ref{con}, we summarize our results.

\section{Einstein-Gauss-Bonnet scalar field 4D Cosmology with matter}
\label{II}

The gravitational theory of our consideration is that of the four-dimensional
Einstein-Gauss-Bonnet theory with a scalar field coupled to the Gauss-Bonnet
term. Hence, the gravitational Action Integral reads \cite{bb1}
\begin{equation}
S=\int d^{4}x\sqrt{-g}\left(  \frac{R}{2}-\frac{\varepsilon}{2}g_{\mu\nu}%
\phi^{;\mu}\phi^{;\nu}-V\left(  \phi\right)  -f\left(  \phi\right)
G+L_{matter}\right)  ,\label{ai.01}%
\end{equation}
where $R$ is the Ricci scalar of the metric tensor $g_{\mu\nu}$, $\phi$ is the scalar field, which inherits the symmetries of the background space,~parameter
$\varepsilon$ takes the values $\varepsilon=\pm1$ indicates if the scalar
field $\phi$ is quintessence $\left(  \varepsilon=+1\right)  $ or phantom
$\left(  \varepsilon=-1\right)  ,$ $V\left(  \phi\right)  ~$is the scalar
field potential, $G$ is the Gauss-Bonnet term, $f\left(  \phi\right)  $ is the
coupling function, which is considered to be a non-constant and $L_{matter}$ is
the Lagrangian for the matter source. For an ideal gas with energy
density $\rho_{m}$, the latter Lagrangian reads $L_{matter}=\rho_{m}$.

For a spatially flat Friedmann--Lema\^{\i}tre--Robertson--Walker (FLRW)
geometry with scale factor $a\left(  t\right)  $ and line element%
\begin{equation}
ds^{2}=-dt^{2}+a^{2}\left(  t\right)  \left(  dr^{2}+r^{2}\left(  d\theta
^{2}+\sin^{2}\theta d\varphi^{2}\right)  \right), \label{ww.16}%
\end{equation}
the Ricci scalar and the Gauss-Bonnet scalars are%
\begin{equation}
R=6\left(  2H^{2}+\dot{H}\right),
\end{equation}%
and 
\begin{equation}
G=24H^{2}\left(  \dot{H}+H^{2}\right).
\end{equation}
in which $H=\frac{\dot{a}}{a}$ is the Hubble function, where a dot means
derivative with respect to the independent variable $t$, that is $\dot
{a}=\frac{da}{dt}$.

Thus, from the Action Integral (\ref{ai.01}) we can write the point-like
Lagrangian for the field equations%
\begin{equation}
L\left(  a,\dot{a},\phi,\dot{\phi}\right)  =-3a\dot{a}^{2}+\frac{\varepsilon
}{2}a^{3}\dot{\phi}^{2}+8\dot{a}^{3}f_{,\phi}\dot{\phi}-a^{3}V\left(
\phi\right)  - a^{3}\rho_{m},\label{ww.02}%
\end{equation}
where for the matter source it holds
\begin{equation}
\dot{\rho}_{m}+3H\left(  \rho_{m}+p_{m}\right)  =0,
\label{cons-matter}
\end{equation}
in which $p_{m}$ is the pressure for the matter source. Hence, for a
constant equation of state parameter, i.e. $p_m=w_m \rho_m$, it follows
\begin{equation}
\rho_{m}=\rho_{0}a^{-3(1+w_{m})},\label{rhom}%
\end{equation}
from where it follows that (\ref{ww.02}) reads \cite{bb1}%
\begin{equation}
L\left(  a,\dot{a},\phi,\dot{\phi}\right)  =-3a\dot{a}^{2}+\frac{\varepsilon
}{2}a^{3}\dot{\phi}^{2}+8\dot{a}^{3}f_{,\phi}\dot{\phi}-a^{3}V\left(
\phi\right)  -\rho_{0}a^{-3 w_{m}}.
\end{equation}
For $f_{,\phi}=0$, the latter Lagrangian function describes
the scalar field theory without the Gauss-Bonnet term. Indeed in a
four-dimensional spacetime, the Gauss-Bonnet term is a total derivative, and its
contribution to the Euler-Lagrange equation is eliminated.

The gravitational field equations follow from the variation of the latter
Lagrangian with respect to the dynamical variables $\left\{  a,\phi\right\}
$, while the constraint equation is the Hamiltonian function.

Indeed, the field equations are %

\begin{align}
&  -48H^{3}\dot{\phi}{f^{\prime}(\phi)}^2-2\rho_{m}-2V(\phi)-\epsilon\dot{\phi}%
^{2}=0,\label{gen-syst-1}\\
&  -16H\dot{H}\dot{\phi}{f^{\prime}(\phi)}^3\dot{\phi}f^{\prime}(\phi)+\frac{1}%
{2}\epsilon\dot{\phi}^{2}-V(\phi)+w_{m}\rho_{m}  +H^{2}\left(  -8\dot{\phi}^{2}f^{\prime\prime}(\phi)-8\ddot{\phi}f^{\prime
}(\phi)+3\right)  +2\dot{H}=0,\label{gen-syst-2}\\
&  3H\left(  -8H\left(  \dot{H}+H^{2}\right)  f^{\prime}(\phi)-\epsilon
\dot{\phi}\right)  -V^{\prime}(\phi)-\epsilon\ddot{\phi}=0.\label{gen-syst-3}%
\end{align}

The effective density and pressure of the scalar field are given by
\begin{align}
\rho_{\phi} &  =\frac{1}{2}\dot{\phi}\left(  48H^{3}f^{\prime}(\phi
)+\epsilon\dot{\phi}\right)  +V(\phi),\\
p_{\phi} &  =\frac{8H^{2}f^{\prime}(\phi)V^{\prime}(\phi)}{-8\epsilon
H\dot{\phi}{f^{\prime}(\phi)}^4{f^{\prime}(\phi)}^2+\epsilon}-\frac{\epsilon V(\phi)}{-8\epsilon
H\dot{\phi}{f^{\prime}(\phi)}^4{f^{\prime}(\phi)}^2+\epsilon}\nonumber\\
&  +\frac{192H^{6}{f^{\prime}(\phi)}^2+\epsilon\dot{\phi}\left(  16H^{2}\left(
\dot{\phi}f^{\prime\prime}(\phi)-4Hf^{\prime}(\phi)\right)  -\epsilon\dot
{\phi}\right)  }{16\epsilon H\dot{\phi}f^{\prime}(\phi)-2\left(
96H^{4}{f^{\prime}(\phi)}^2+\epsilon\right)  },
\end{align}
where we can define the effective equation of state (EoS) $\omega_{\phi}%
=\frac{p_{\phi}}{\rho_{\phi}}.$

In the following, we shall perform a detailed analysis of the phase-space for
the exponential scalar field potential~$V\left(  \phi\right)  =V_{0}%
e^{\lambda\phi}$ and the linear coupling function $f(\phi)=f_{0} \phi$.

\section{Linear coupling}
\label{III}

The field equations \eqref{gen-syst-1}, \eqref{gen-syst-2} and \eqref{gen-syst-3} become%
\begin{align}
& -48 f_0 H^3 \dot{\phi}+6 H^2-2
   \rho_m-2 V(\phi )-\epsilon 
   \dot{\phi}^2=0, \label{main-syst-1}\\
&-16 f_0 H \dot{H} \dot{\phi}+H^2 \left(3-8 f_0 \ddot{\phi}\right)-16 f_0 H^3 \dot{\phi}    +2  \dot{H}-V(\phi )+w_m \rho_m+\frac{1}{2} \epsilon  \dot{\phi}^2=0, \label{main-syst-2}\\
& -3H \left(8 f_0 H    \left(\dot{H}+H^2\right)+\epsilon  \dot{\phi}\right)-V'(\phi )-\epsilon  \ddot{\phi}=0.\label{main-syst-3}
\end{align}
together with the equation \eqref{cons-matter}. 

\subsection{Dynamical system in dimensionless variables}
\label{III-A}

In order to study the phase space, we introduce the following normalized dimensionless variables, 
\begin{equation}\label{newvars}
    {x}=\frac{\phi '}{\sqrt{6} \sqrt{H^2+1}}, \; {y}=  \frac{\sqrt{V(\phi )}}{\sqrt{3}
   \sqrt{H^2+1}}, \;z=\frac{\rho_m}{3
   \left(H^2+1\right)},\; \eta=\frac{H}{\sqrt{1+H^2}}.
\end{equation}

With these definitions, the first modified Friedmann equation is written in the algebraic form as
\begin{equation}\label{Friedmann-new-var}
  6 \left(\eta ^2-1\right) \left(-\eta
   ^2+\epsilon  x^2+y^2+z\right)-48
   \sqrt{6} f_0 \eta ^3 x=0.
\end{equation}

Using eq. \eqref{Friedmann-new-var} we can find the following definition for $z$

\begin{equation}
    \label{z-definition}
    z=\frac{8 \sqrt{6} f_0 \eta ^3 x}{\eta
   ^2-1}+\eta ^2-\epsilon  x^2-y^2.
\end{equation}
 Observe that when $x=y=0$, we acquire  $z= \eta^2$, which means $\Omega_{m}= \rho_m/(3H^2)=z/\eta^2=1$, and we have matter-dominated solutions.

By combining \eqref{newvars} and \eqref{z-definition} we can write system \eqref{main-syst-2}-\eqref{main-syst-3} as follows

\begin{align}
    &\frac{dx}{d\tau}=\frac{1}{K}\Bigg[\eta  \left(192 f_0^2 \eta ^4 x
   \left(\eta ^2-3 w_m\right)+4
   \sqrt{6} f_0 \left(\eta ^2-1\right)
   \eta  \left(\eta ^2 \left(2 (3
   w_m-1) \epsilon  x^2-3
   w_m-1\right)+3 (w_m-5) \epsilon 
   x^2\right)\right)\nonumber \\ &-\left(\eta
   ^2-1\right) y^2 \left(\sqrt{6}
   \left(\eta ^2 (\lambda -12 f_0
   (w_m+1))-\lambda \right)+3 \eta 
   x \left(16 f_0 \lambda +\eta ^2
   (8 f_0 \lambda +w_m \epsilon
   +\epsilon )-(w_m+1) \epsilon
   \right)\right)\nonumber \\ &+\eta\left(3 \epsilon  \left(\eta
   ^2-1\right)^2 x \left((w_m+1)
   \eta ^2+x^2 (\epsilon -w_m
   \epsilon )-2\right)\right)\Bigg], \label{newsyst-1}\\
    &\frac{dy}{d\tau}=\frac{y}{4K}\Bigg[384 f_0^2 \eta ^7+6 \epsilon 
   \left(\eta ^2-1\right)^2 \eta 
   \left(\frac{8 \sqrt{6} f_0 w_m
   \eta ^3 x}{\eta ^2-1}+x^2
   (\epsilon -w_m \epsilon )-(w_m+1)
   \left(y^2-\eta ^2\right)\right)\nonumber \\ &-16
   f_0 \left(\eta ^2-1\right) \eta
   ^3 \left(\sqrt{6} \epsilon  \eta 
   x+3 \lambda  y^2\right)+2 \sqrt{6}
   \lambda  x\Bigg], \label{newsyst-2}\\
    &\frac{d\eta}{d\tau}=\frac{1}{K}\Bigg[\left(\eta ^2-1\right) \left(192 f_0^2
   \eta ^6+3 \epsilon  \left(\eta
   ^2-1\right)^2 \left(\frac{8 \sqrt{6}
   f_0 w_m \eta ^3 x}{\eta
   ^2-1}+x^2 (\epsilon -w_m \epsilon
   )-(w_m+1) \left(y^2-\eta
   ^2\right)\right)\right)\nonumber \\  &+\left(-8 f_0 \left(\eta
   ^2-1\right)^2 \eta ^2 \left(\sqrt{6}
   \epsilon  \eta  x+3 \lambda 8
   y^2\right)\right)\Bigg], \label{newsyst-3}
\end{align}
where we defined $K:=K(x,y,\eta,\epsilon,f_0)=192 f_0^2 \eta ^4+2 \epsilon 
   \left(\eta ^2-1\right) \left(8 \sqrt{6}
   f_0 \eta  x+\eta ^2-1\right)$ and introduce the time derivative 
$d f/d\tau =1/\sqrt{1+H^2}df/dt.$ We will also consider $-1\leq \eta \leq 1$ and $0\leq w_m\leq 1.$

\subsection{General case for $\epsilon=1$}
\label{III-B}
\begin{figure}[h]
    \centering
    \includegraphics[scale=0.5]{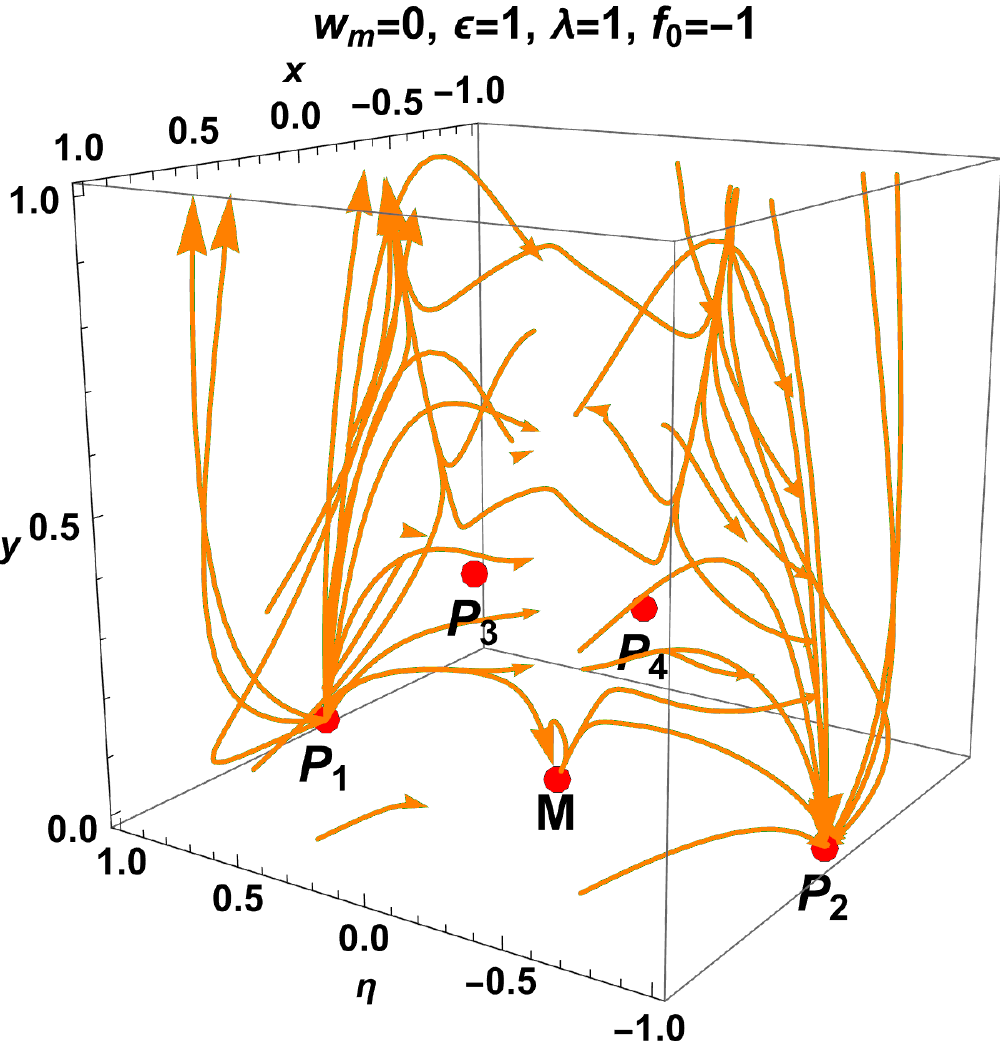}
    \includegraphics[scale=0.5]{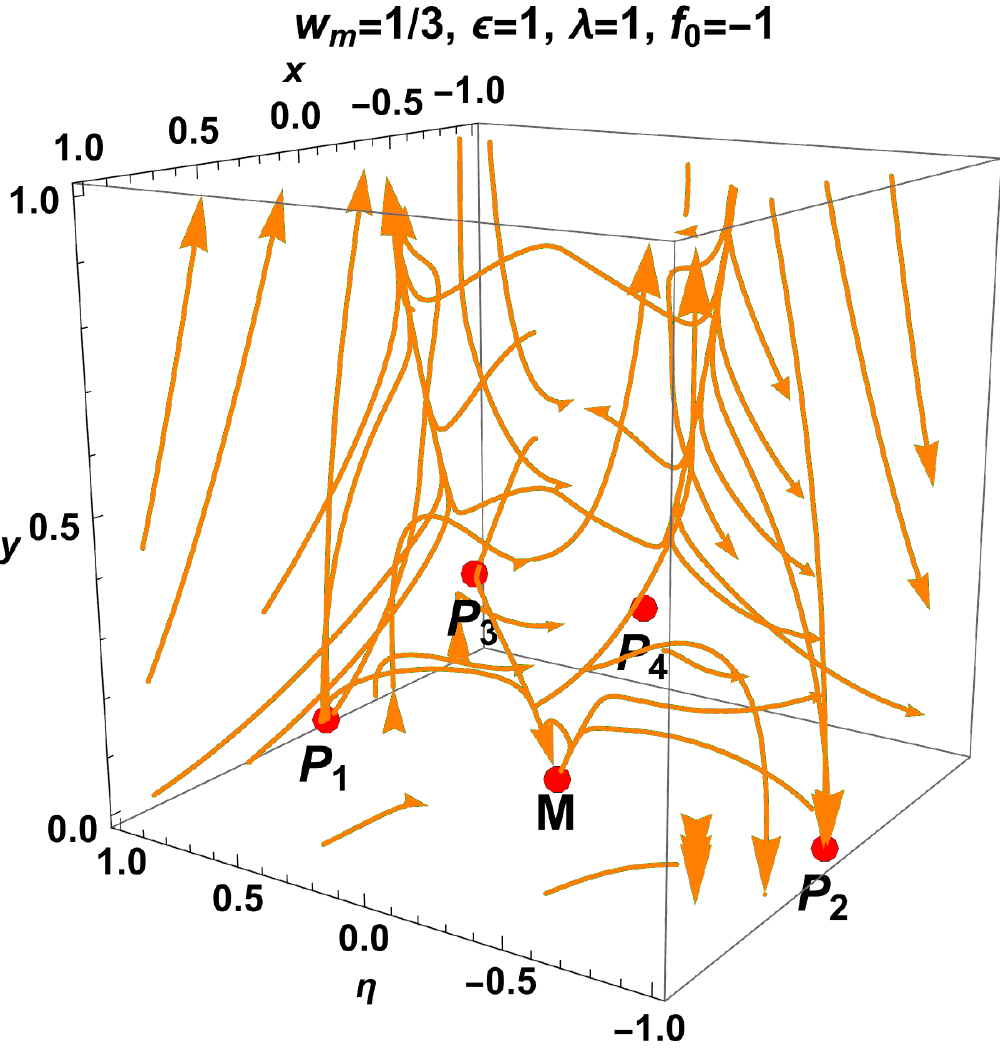}
    \includegraphics[scale=0.5]{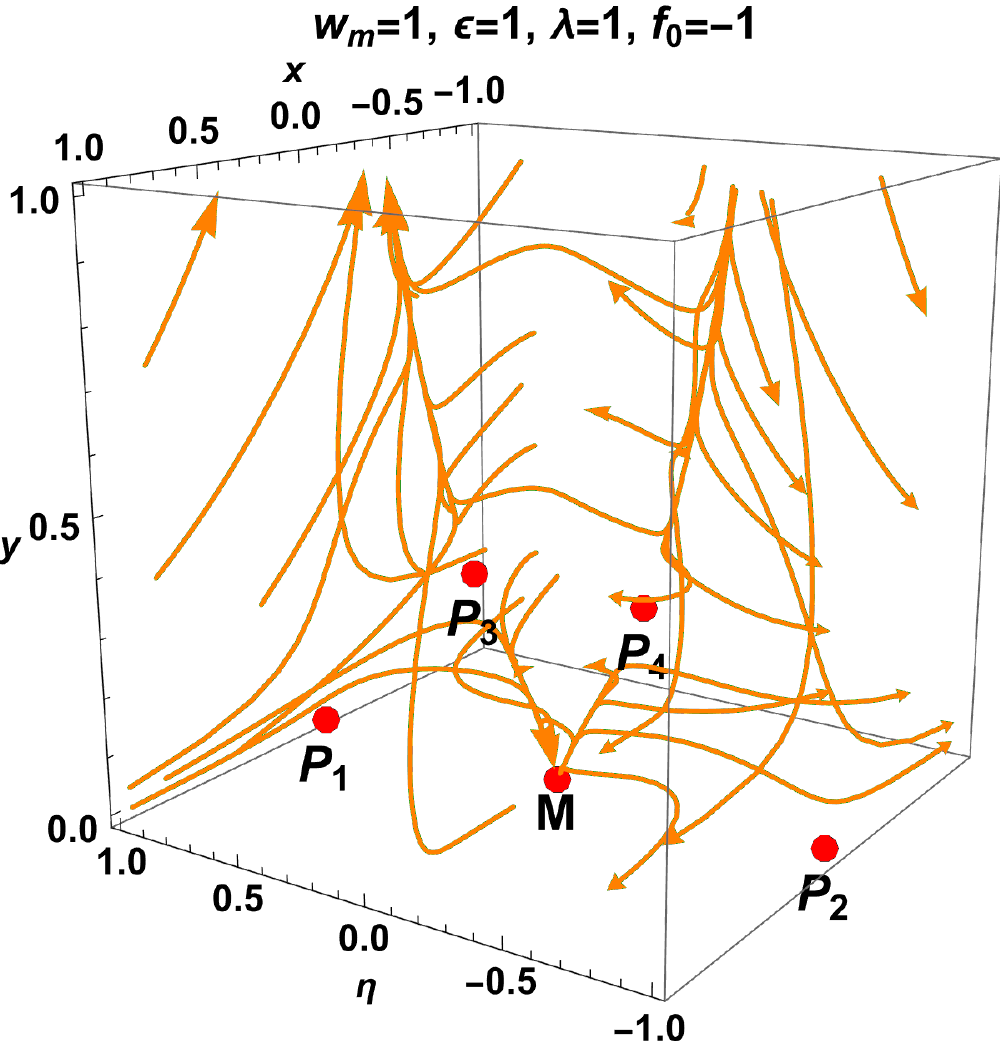}
    \caption{Phase-space analysis for system \eqref{newsyst-1}-\eqref{newsyst-3} for $\epsilon=1$ and different values of the parameters $\lambda, f_0.$ Here we consider $Y>0$ and the three cases $w_m=0, \frac{1}{3}, 1.$}
    \label{fig:5}
\end{figure}

The equilibrium points for system \eqref{newsyst-1}-\eqref{newsyst-3} for $\epsilon=1$ in the coordinates $(x,y,\eta)$ are the following:
\begin{enumerate}
    \item $M=(0,0,0),$ with eigenvalues $\{0,0,0\}.$ The asymptotic solution is that of the Minkowski spacetime.
    \item $P_{1,2}=(0,0,\pm 1),$ with eigenvalues $\{\pm 1,\pm 2,\pm (1-3 w_m)\}.$ These points describe a universe dominated by the Gauss-Bonnet term, and they verify that $\omega_{\phi}=-\frac{1}{3}$ and $q=0.$ These points are 
    \begin{enumerate}
        \item $P_1$ is a source ($P_2$ is a sink) for $0\leq w_m< \frac{1}{3},$
        \item saddles for $\frac{1}{3}<w_m\leq 1$,
        \item non-hyperbolic for $w_m=\frac{1}{3}.$
    \end{enumerate}
    \item $P_3=(0,\sqrt{\frac{\lambda }{\lambda -8 f_0}},\sqrt{\frac{\lambda }{\lambda -8 f_0}}).$ This point exists for $f_0=0$ and $\lambda \neq 0$ or $f_0<0$ and $\lambda\geq 0$ or  $f_0>0$ and $\lambda \leq 0.$ The eigenvalues are $\left\{-\frac{3 \sqrt{\lambda } (w_m+1)}{\sqrt{\lambda -8 f_0}},-\frac{\sqrt{\lambda } \left(3 \sqrt{3
   \lambda ^2+2}+\sqrt{51 \lambda ^2+18}\right)}{2 \sqrt{3 \lambda ^2+2} \sqrt{\lambda -8
   f_0}},\frac{\sqrt{\lambda } \left(\sqrt{51 \lambda ^2+18}-3 \sqrt{3 \lambda ^2+2}\right)}{2 \sqrt{3 \lambda
   ^2+2} \sqrt{\lambda -8 f_0}}\right\}.$ This point describes a de Sitter universe, and we verify that $\omega_{\phi}=-1$ and $q=-1.$ We also verify that the point is a saddle.
    \item $P_4=(0,\sqrt{\frac{\lambda }{\lambda -8 f_0}},-\sqrt{\frac{\lambda }{\lambda -8 f_0}}),$ with eigenvalues \newline $\left\{\frac{3 \sqrt{\lambda } (w_m+1)}{\sqrt{\lambda -8 f_0}},\frac{\sqrt{\lambda } \left(3 \sqrt{3
   \lambda ^2+2}-\sqrt{51 \lambda ^2+18}\right)}{2 \sqrt{3 \lambda ^2+2} \sqrt{\lambda -8
   f_0}},\frac{\sqrt{\lambda } \left(3 \sqrt{3 \lambda ^2+2}+\sqrt{51 \lambda ^2+18}\right)}{2 \sqrt{3 \lambda
   ^2+2} \sqrt{\lambda -8 f_0}}\right\}.$ This point existence conditions, values for $\omega_{\phi}, q$, physical interpretation and stability are the same as $P_3.$
\end{enumerate}
In Figure \ref{fig:5} we present the stability analysis for system \eqref{newsyst-1}-\eqref{newsyst-3} with $\epsilon=1$ and different values of the parameters $\lambda$ and $f_0.$ We consider $y>0$; however, the system is unbounded, suggesting nontrivial dynamics at infinity.  We also considered the three cases $w_m=0$ (dust), $\frac{1}{3}$ (radiation) and $1$ (stiff matter). A summary of the results of this section is presented in Table \ref{tab:4}. 

\begin{table}[ht!]
    \caption{Equilibrium points of system \eqref{newsyst-1}-\eqref{newsyst-3} for $\epsilon=1$ with their stability conditions. Also includes the value of $\omega_{\phi}$ and $q.$}
    \label{tab:4}
\newcolumntype{C}{>{\centering\arraybackslash}X}
\centering
 \setlength{\tabcolsep}{3mm}
\begin{tabularx}{\textwidth}{ccccccc}
\toprule 
  \text{Label}  & \; $x$& $y$& $\eta$ & \text{Stability}& $\omega_{\phi}$&$q$\\
  \midrule  
          $M$ & $0$ & $0$& $0$  & non-hyperbolic & indeterminate & indeterminate\\  \midrule 
          $P_1$ & $0$ & $0$ & $1$ & source for $0\leq w_m<1/3$ & &  \\
          &&&& saddle for $1/3<w_m\leq 1$ & & \\
          &&&& non-hyperbolic for $w_m=1/3$  & $-\frac{1}{3}$ & $0$\\  \midrule
          $P_2$ & $0$ & $0$ & $-1$ & sink for $0\leq w_m<1/3$ && \\
          &&&& saddle for $1/3<w_m\leq 1$ &&\\
          &&&& non-hyperbolic for $w_m=1/3$  & $-\frac{1}{3}$ & $0$\\  \midrule
          $P_3$ & $0$ & $\sqrt{\frac{\lambda }{\lambda -8 f_0}}$ & $\sqrt{\frac{\lambda }{\lambda -8 f_0}}$ & saddle & $-1$ &$-1$\\  \midrule          
          $P_4$ & $0$ & $\sqrt{\frac{\lambda }{\lambda -8 f_0}}$ & $-\sqrt{\frac{\lambda }{\lambda -8 f_0}}$ & saddle  & $-1$ & $-1$\\
        \bottomrule
    \end{tabularx}
\end{table}

\FloatBarrier

\begin{figure}[h]
    \centering
    \includegraphics[scale=0.45]{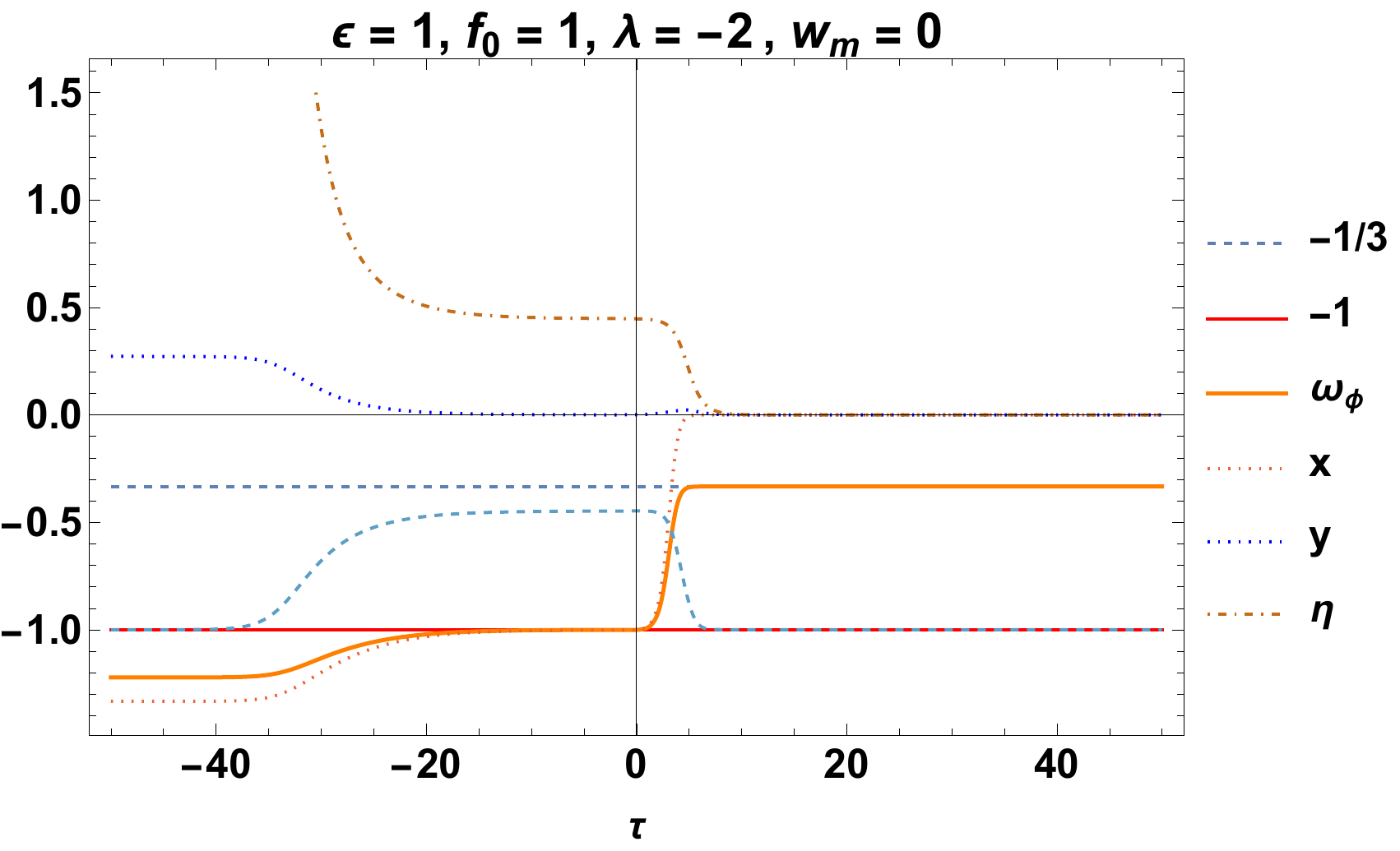}
    \includegraphics[scale=0.45]{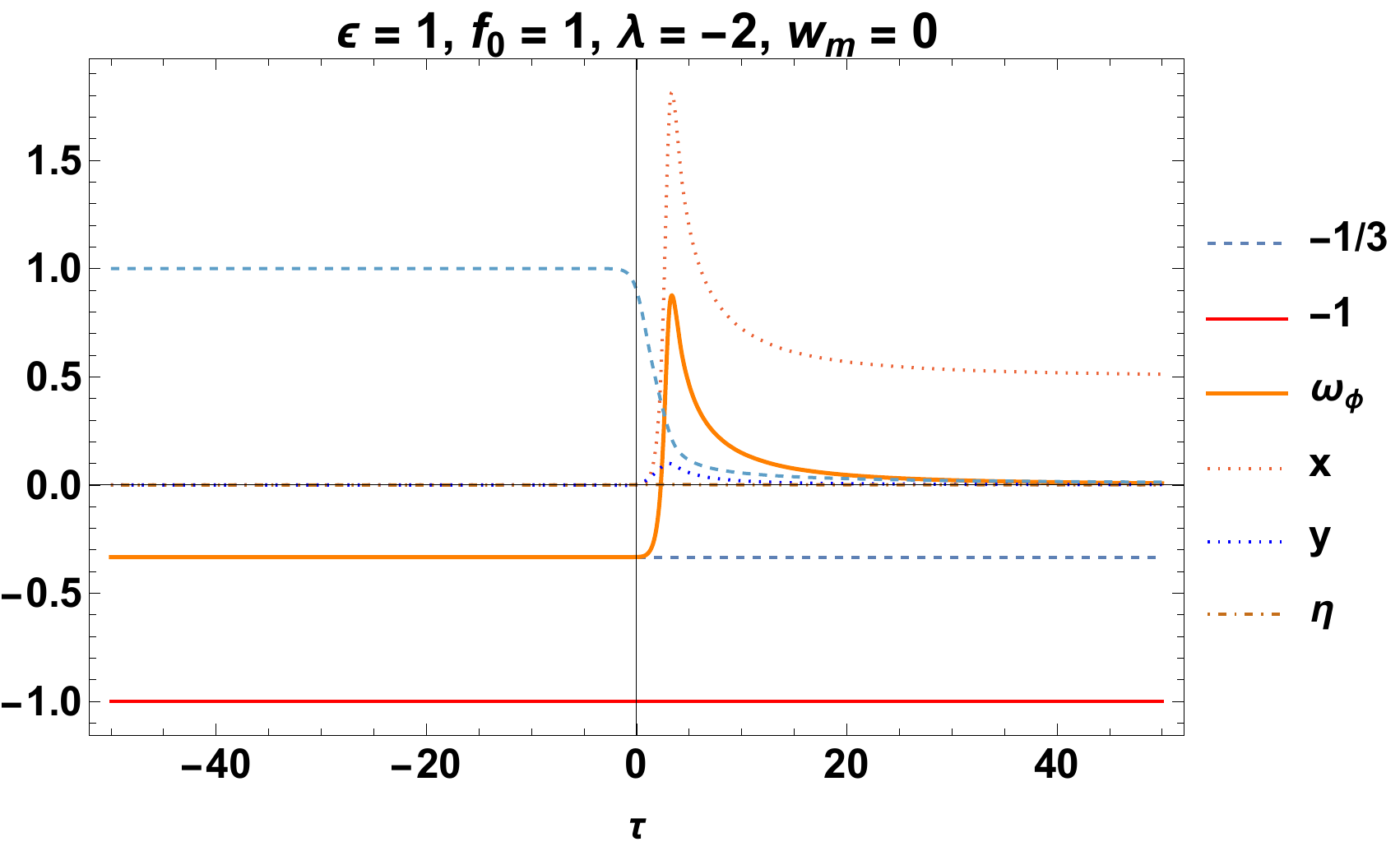}
    \caption{$\omega_\phi(\tau)$, $x(\tau)$, $y(\tau)$, and $\eta(\tau)$ evaluated at a solution of system \eqref{newsyst-1}- \eqref{newsyst-3} for $\epsilon=1.$ The initial conditions for the left plot are $x(0)=0.001, \quad y(0)=\sqrt{\frac{\lambda }{\lambda -8 f_0}},  \quad \eta (0)=-\sqrt{\frac{\lambda }{\lambda -8 f_0}}$ (i.e., near~the saddle point $P_3$). The~solution is past asymptotic to $\omega_{\phi} =-1$ ($q=-1$), then remains near the de Sitter point $P_3$, then tending asymptotically to $\omega_{\phi} =-\frac{1}{3}$ (the Gauss-Bonnet point $P_2$) from~below. The initial conditions for the plot on the right are $x(0)=0.001,\quad y(0)=0.001,  \quad \eta (0)=0.9$ (i.e., near~the source point $P_1$). The solution is past asymptotic to $\omega_{\phi} =-\frac{1}{3}, q=0$ (zero acceleration), then it grows to $\omega_{\phi},q>0$, finally, it tends asymptotically to $\omega_{\phi}=0, \quad q=\frac{1}{2}$ describing a matter-dominated solution. }
    \label{fig:weff1}
\end{figure}

 Figure \ref{fig:weff1} displays the expressions $\omega_\phi(\tau)$, $x(\tau)$, $y(\tau)$, and $\eta(\tau)$ evaluated at a solution of system \eqref{newsyst-1}- \eqref{newsyst-3} for $\epsilon=1$ for the initial conditions for the left plot are $x(0)=0.001, \quad y(0)=\sqrt{\frac{\lambda }{\lambda -8 f_0}},  \quad \eta (0)=-\sqrt{\frac{\lambda }{\lambda -8 f_0}}$ (i.e., near~the saddle point $P_3$). The~solution is past asymptotic to $\omega_{\phi} =-1$ ($q=-1$), then remains near the de Sitter point $P_3$, then tending asymptotically to $\omega_{\phi} =-\frac{1}{3}$ (the Gauss-Bonnet point $P_2$) from~below. The initial conditions for the plot on the right are $x(0)=0.001,\quad y(0)=0.001,  \quad \eta (0)=0.9$ (i.e., near~the source point $P_1$). The solution is past asymptotic to $\omega_{\phi} =-\frac{1}{3}, q=0$ (zero acceleration), then it grows to $\omega_{\phi},q>0$, finally, it tends asymptotically to $\omega_{\phi}=0, \quad q=\frac{1}{2}$ describing a matter-dominated solution.
  
\FloatBarrier
\subsection{General case for $\epsilon=-1$}
\label{III-C}
\begin{figure}[h]
    \centering
    \includegraphics[scale=0.5]{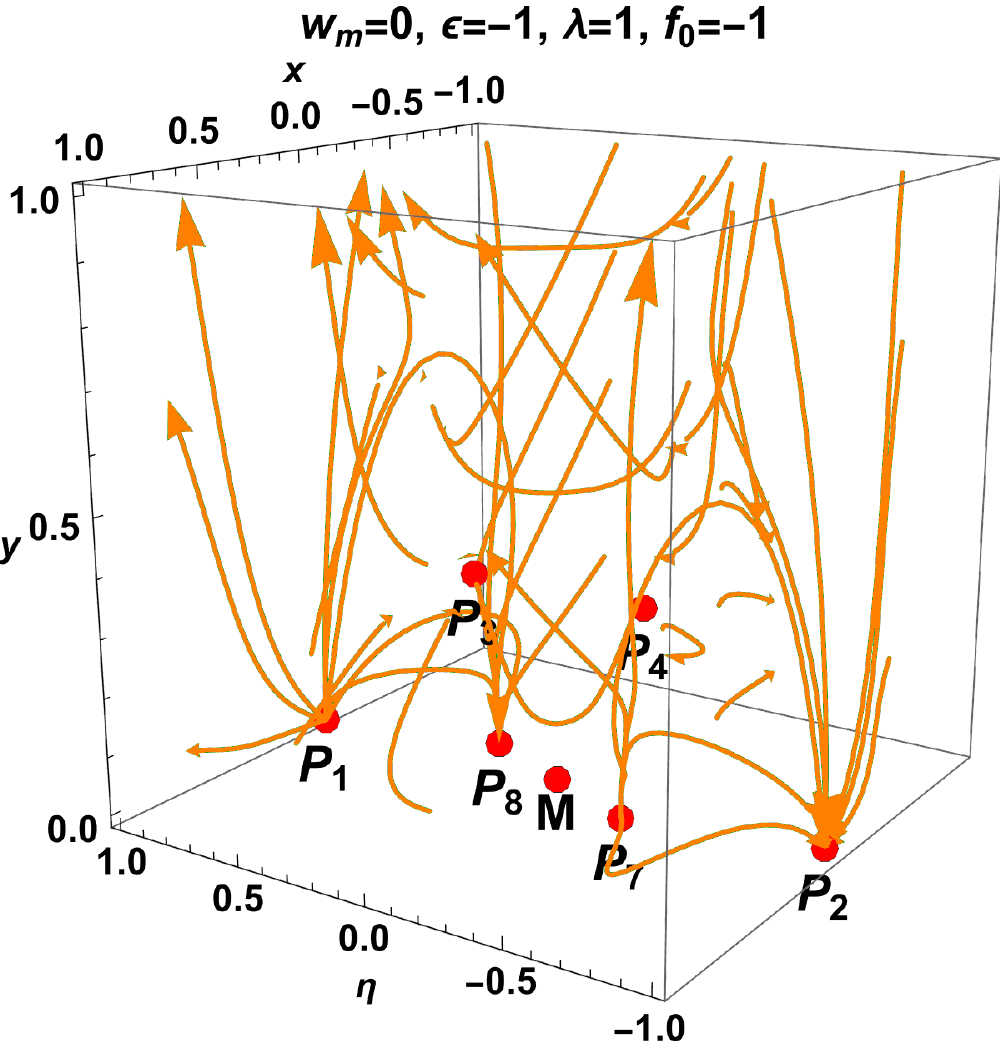}
    \includegraphics[scale=0.5]{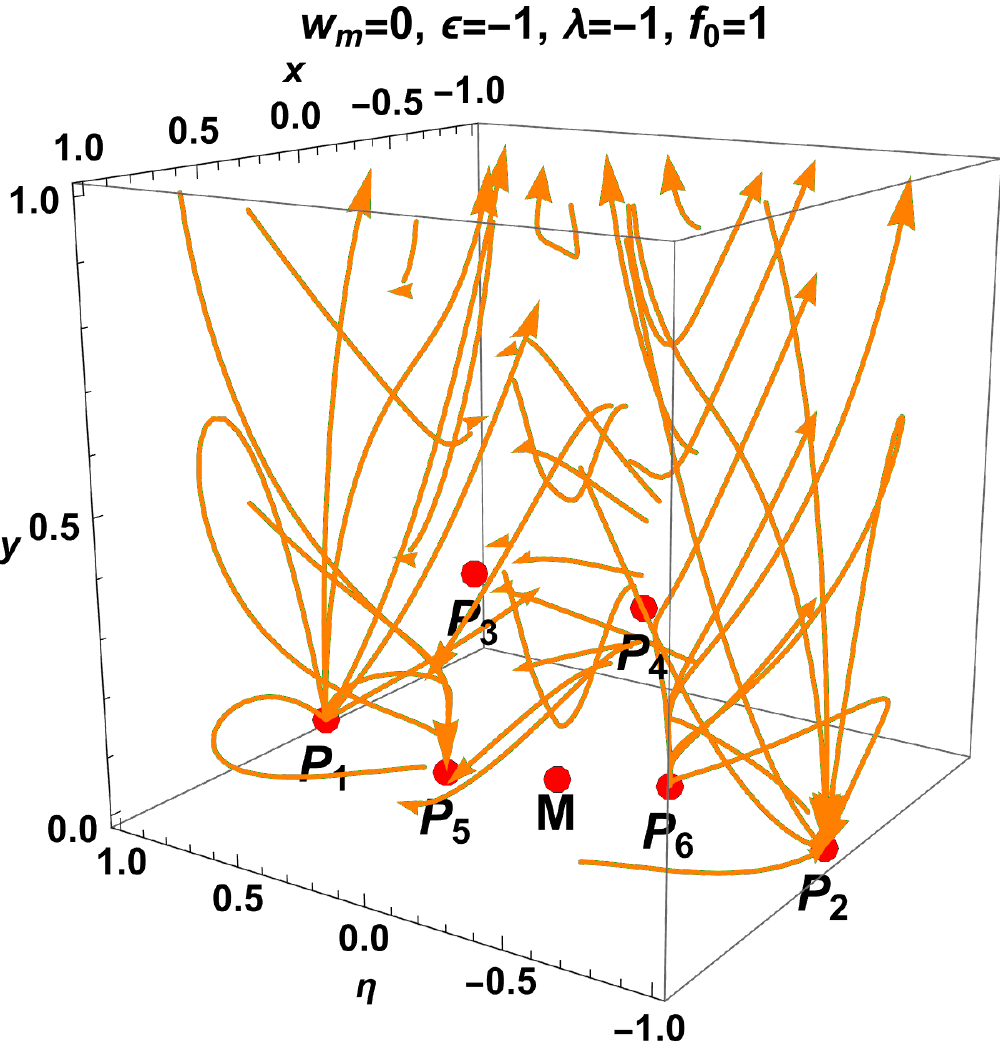}
    \includegraphics[scale=0.5]{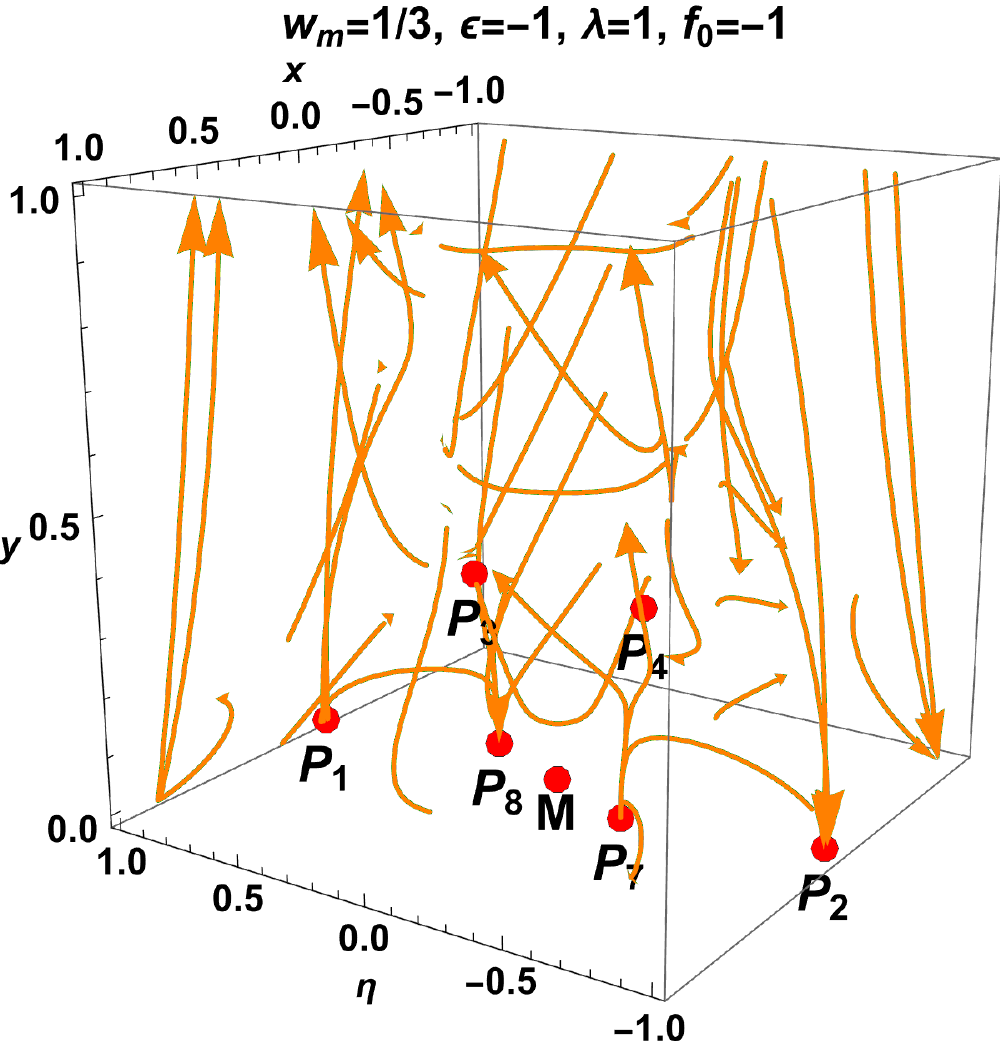}
    \includegraphics[scale=0.5]{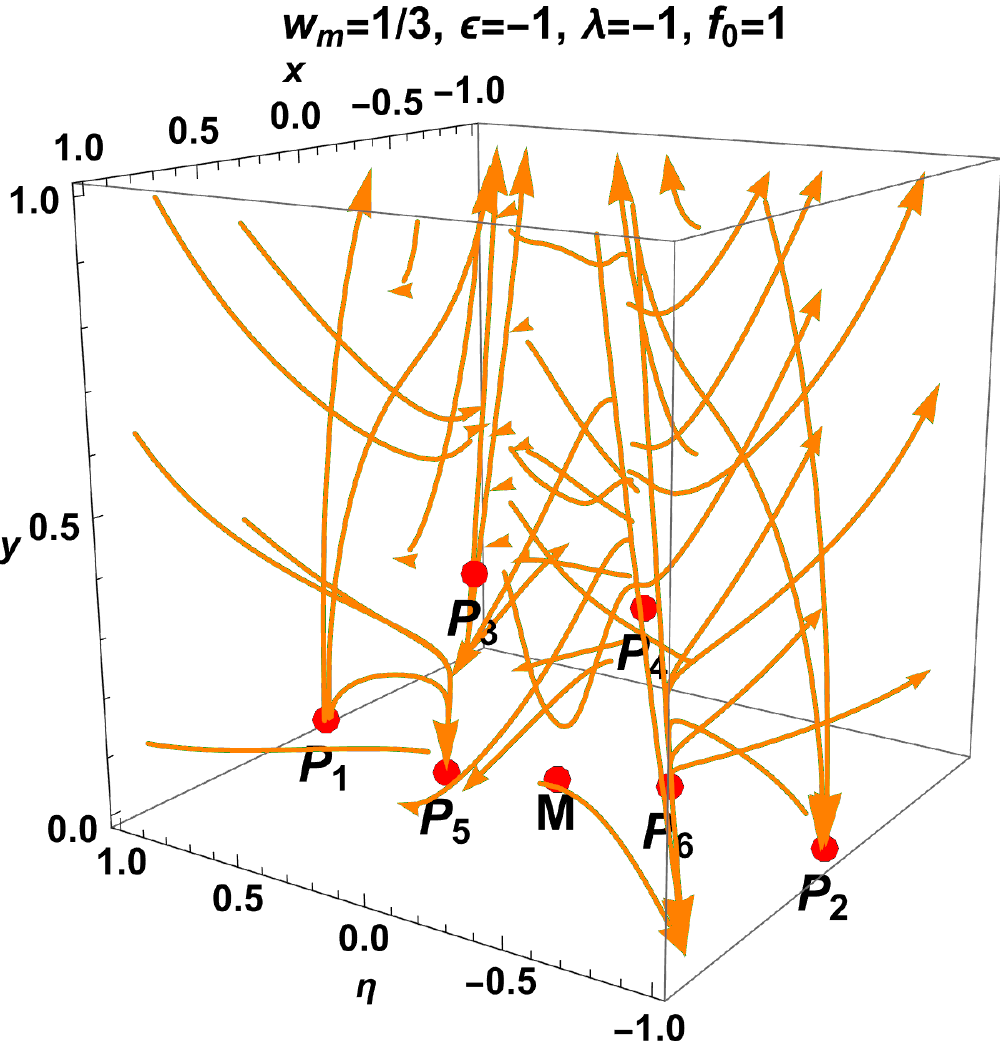}
    \includegraphics[scale=0.5]{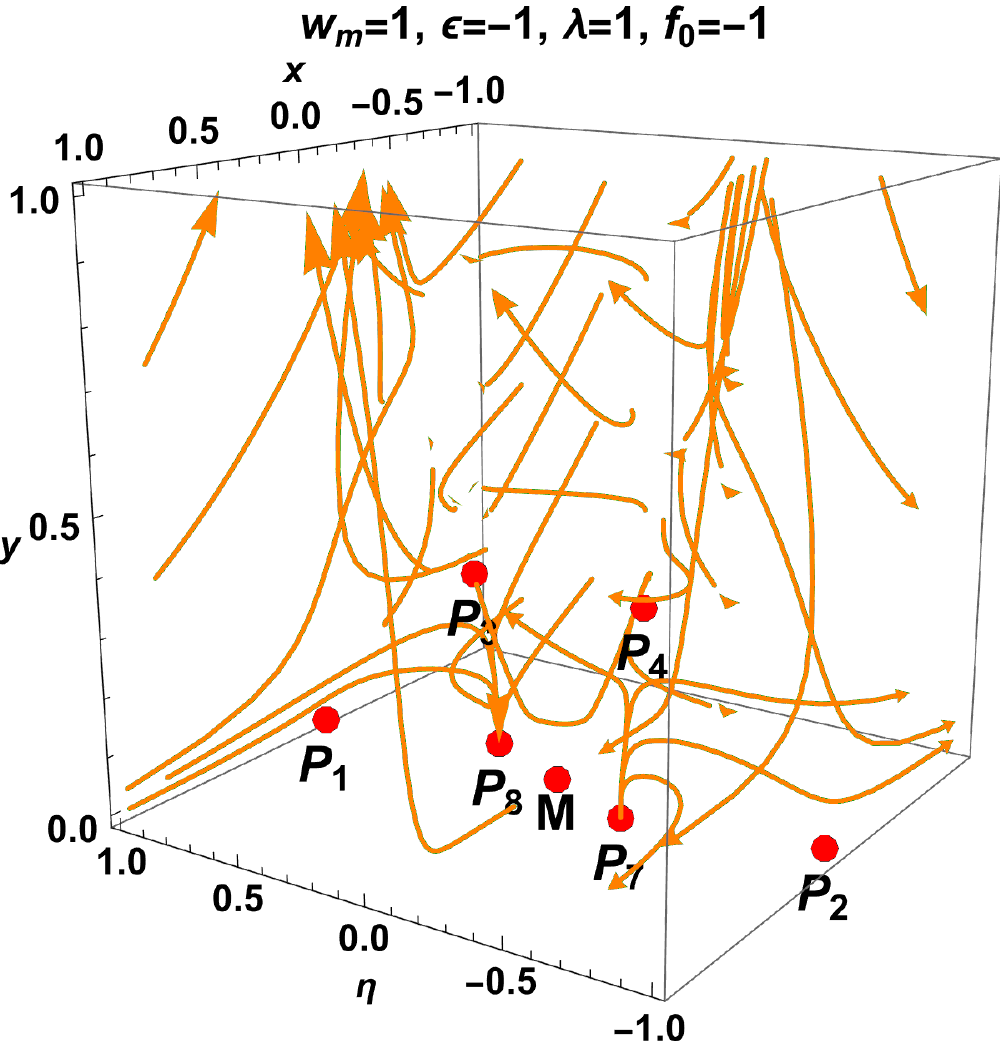}
    \includegraphics[scale=0.5]{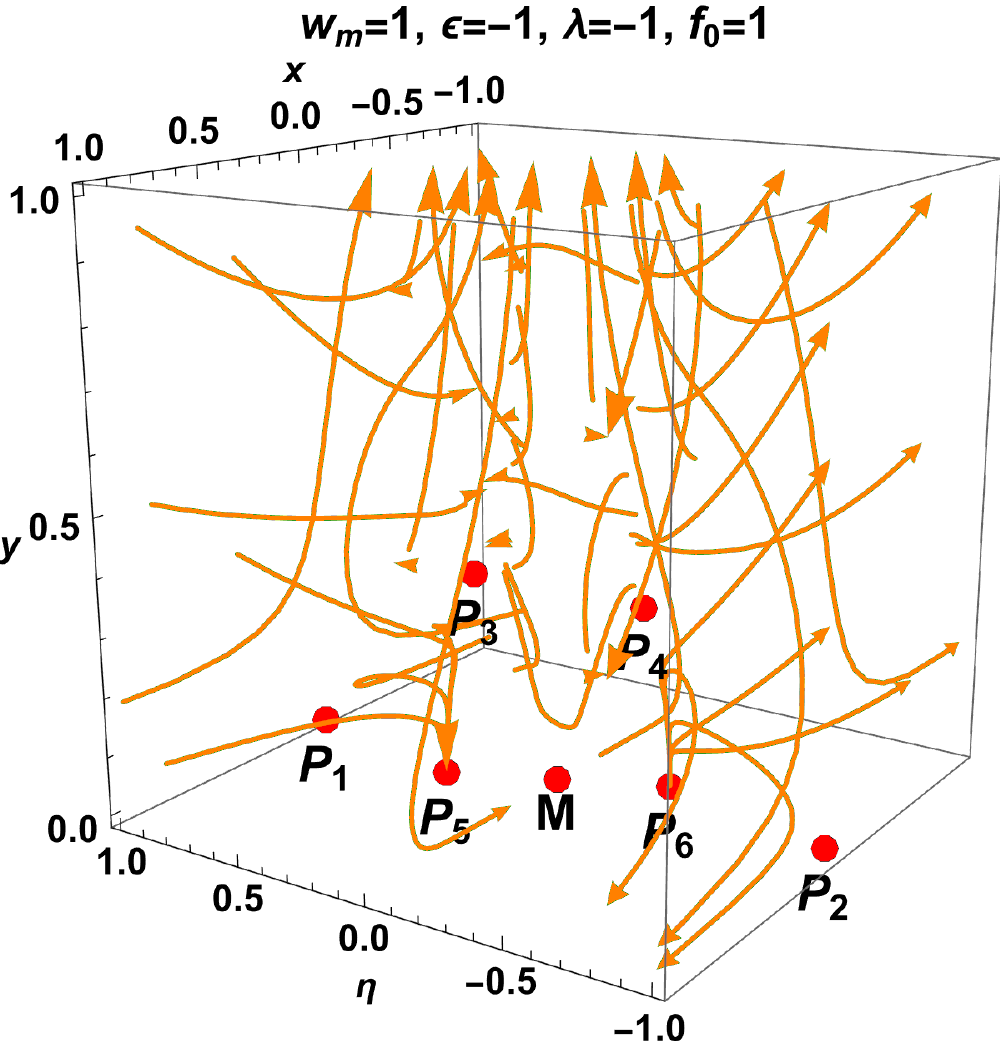}
    \caption{Phase-space analysis for system \eqref{newsyst-1}- \eqref{newsyst-3} for $\epsilon=-1$ and different values of the parameters $\lambda, f_0.$ Here we consider $Y>0$ and the three cases $w_m=0, \frac{1}{3}, 1.$}
    \label{fig:6}
\end{figure}

The equilibrium points for system \eqref{newsyst-1}-\eqref{newsyst-3} are the same as in section \ref{III-B} plus some additional points, the complete list of equilibrium points in the coordinates $(x,y,\eta)$ is the following. 
\begin{enumerate}
    \item $M=(0,0,0).$ The stability analysis and physical interpretation are the same as in section \ref{III-B}.
    \item $P_{1,2}=(0,0,\pm 1).$ The stability analysis and physical interpretation are the same as in section \ref{III-B}.
    \item $P_3=(0,\sqrt{\frac{\lambda }{\lambda -8 f_0}},\sqrt{\frac{\lambda }{\lambda -8 f_0}}).$ The existence conditions and physical interpretation are the same as in the section 
    \ref{III-B}; however, the second and third eigenvalues slightly change to  \newline $\left\{-\frac{3 \sqrt{\lambda } (w_m +1)}{\sqrt{\lambda -8 \text{f0}}},-\frac{\sqrt{\lambda } \left(3 \sqrt{3
   \lambda ^2-2}+\sqrt{51 \lambda ^2-18}\right)}{2 \sqrt{3 \lambda ^2-2} \sqrt{\lambda -8
   \text{f0}}},\frac{\sqrt{\lambda } \left(\sqrt{51 \lambda ^2-18}-3 \sqrt{3 \lambda ^2-2}\right)}{2 \sqrt{3
   \lambda ^2-2} \sqrt{\lambda -8 \text{f0}}}\right\},$ therefore the stability changes to 
   \begin{enumerate}
       \item a sink for 
       \begin{enumerate}
           \item $f_0<0$, $0<\lambda<\sqrt{\frac{2}{3}}$ or
           \item $f_0>0$, $-\sqrt{\frac{2}{3}}<\lambda<0,$
       \end{enumerate}
       \item a saddle for
       \begin{enumerate}
           \item $f_0<0$, $\lambda >\sqrt{\frac{2}{3}}$ or
           \item $f_0>0$, $ \lambda <-\sqrt{\frac{2}{3}}.$
       \end{enumerate}
   \end{enumerate}
    \item $P_4=(0,\sqrt{\frac{\lambda }{\lambda -8 f_0}},-\sqrt{\frac{\lambda }{\lambda -8 f_0}}).$ The existence conditions and physical interpretation are the same as in the section 
    \ref{III-B} however, the second and third eigenvalues slightly change to  \newline $\left\{\frac{3 \sqrt{\lambda } (w_m +1)}{\sqrt{\lambda -8 \text{f0}}},\frac{\sqrt{\lambda } \left(3 \sqrt{3
   \lambda ^2-2}-\sqrt{51 \lambda ^2-18}\right)}{2 \sqrt{3 \lambda ^2-2} \sqrt{\lambda -8
   \text{f0}}},\frac{\sqrt{\lambda } \left(3 \sqrt{3 \lambda ^2-2}+\sqrt{51 \lambda ^2-18}\right)}{2 \sqrt{3
   \lambda ^2-2} \sqrt{\lambda -8 \text{f0}}}\right\},$ therefore the stability changes to 
   \begin{enumerate}
       \item a source for 
       \begin{enumerate}
           \item $f_0<0$, $0<\lambda<\sqrt{\frac{2}{3}}$ or
           \item $f_0>0$, $-\sqrt{\frac{2}{3}}<\lambda<0,$
       \end{enumerate}
       \item a saddle for
       \begin{enumerate}
           \item $f_0<0$, $\lambda >\sqrt{\frac{2}{3}}$ or
           \item $f_0>0$, $ \lambda <-\sqrt{\frac{2}{3}}.$
       \end{enumerate}
   \end{enumerate}
    \item $P_5=\left(\sqrt{\frac{1}{20 \sqrt{\frac{10}{3}} f_0+5}},0,\sqrt{3} \sqrt{\frac{1}{4 \sqrt{30}
   f_0+3}}\right),$ with eigenvalues \newline $\left\{-\frac{3 (w_m+1)}{\sqrt{4 \sqrt{\frac{10}{3}} f_0+1}},-\frac{9 \left(4 \sqrt{10}
   f_0+\sqrt{3}\right)}{\left(4 \sqrt{30} f_0+3\right)^{3/2}},\frac{3 \lambda }{\sqrt{40 \sqrt{30}
   f_0+30}}\right\}.$ This point exists for $f_0 \geq 0,$ the asymptotic solution it describes is that of a de Sitter universe; also, we verify that $\omega_{\phi}=-1$ and $q=-1.$ The stability conditions are
   \begin{enumerate}
       \item a sink for $ f_0\geq0$, $ \lambda <0,$
       \item a saddle for $ f_0\geq0$, $ \lambda >0,$
       \item non-hyperbolic for $f_0\geq 0,$ $\lambda=0.$
   \end{enumerate}
    \item $P_6=\left(-\sqrt{\frac{1}{20 \sqrt{\frac{10}{3}} f_0+5}},0,-\sqrt{3} \sqrt{\frac{1}{4 \sqrt{30}
   f_0+3}}\right),$ with eigenvalues \newline $\left\{\frac{9 \left(4 \sqrt{10} f_0+\sqrt{3}\right)}{\left(4 \sqrt{30} f_0+3\right)^{3/2}},\frac{3
   (w_m+1)}{\sqrt{4 \sqrt{\frac{10}{3}} f_0+1}},-\frac{3 \lambda }{\sqrt{40 \sqrt{30}
   f_0+30}}\right\}.$ This point exists for $f_0 \geq 0,$ the asymptotic solution it describes is that of a de Sitter universe; also, we verify that $\omega_{\phi}=-1$ and $q=-1.$ The stability conditions are
   \begin{enumerate}
       \item a source for  $ f_0\geq0$, $ \lambda <0,$
       \item a saddle for $ f_0\geq0$, $ \lambda >0,$
       \item non-hyperbolic for  $ f_0\geq 0$, $ \lambda =0.$
   \end{enumerate}
    \item $P_7=\left(\sqrt{\frac{1}{5-20 \sqrt{\frac{10}{3}} f_0}},0,-\sqrt{3} \sqrt{\frac{1}{3-4 \sqrt{30}
   f_0}}\right),$ with eigenvalues $\lambda_1=\frac{3 \lambda }{\sqrt{30-40 \sqrt{30}
   f_0}}$ and $\lambda_{2,3}$ given by\newline $\left\{\frac{9 \left(\sqrt{\left(\sqrt{3}-4 \sqrt{10} f_0\right)^2} w_m-4 \sqrt{10} f_0
   (w_m+2)+\sqrt{3} (w_m+2)\right)}{2 \left(3-4 \sqrt{30} f_0\right)^{3/2}},-\frac{9
   \sqrt{\left(\sqrt{3}-4 \sqrt{10} f_0\right)^2} w_m+36 \sqrt{10} f_0 (w_m+2)-9 \sqrt{3}
   (w_m+2)}{2 \left(3-4 \sqrt{30} f_0\right)^{3/2}}\right\}.$   This point exists for $f_0 \leq 0,$ the asymptotic solution it describes is that of a de Sitter universe; also, we verify that $\omega_{\phi}=-1$ and $q=-1.$ The stability conditions are
   \begin{enumerate}
       \item a source for  $ f_0\leq0$, $ \lambda >0,$
       \item a saddle for $ f_0\leq0$, $ \lambda <0,$
       \item non-hyperbolic for  $ f_0\leq 0$, $ \lambda =0.$
   \end{enumerate}
    \item $P_8=\left(-\sqrt{\frac{1}{5-20 \sqrt{\frac{10}{3}} f_0}},0,\sqrt{3} \sqrt{\frac{1}{3-4 \sqrt{30}
   f_0}}\right),$ with eigenvalues $\lambda_1=-\frac{3 \lambda }{\sqrt{30-40 \sqrt{30} f_0}}$ and $\lambda_{2,3}$ given by \newline$\left\{\frac{9 \sqrt{\left(\sqrt{3}-4 \sqrt{10} f_0\right)^2} w_m+36 \sqrt{10} f_0 (w_m+2)-9
   \sqrt{3} (w_m+2)}{2 \left(3-4 \sqrt{30} f_0\right)^{3/2}},-\frac{9 \left(\sqrt{\left(\sqrt{3}-4
   \sqrt{10} f_0\right)^2} w_m-4 \sqrt{10} f_0 (w_m+2)+\sqrt{3} (w_m+2)\right)}{2
   \left(3-4 \sqrt{30} f_0\right)^{3/2}}\right\}.$ This point exists for $f_0 \leq 0,$ the asymptotic solution it describes is that of a de Sitter universe; also, we verify that $\omega_{\phi}=-1$ and $q=-1.$ The stability conditions are
   \begin{enumerate}
       \item a sink for  $ f_0\leq0$, $ \lambda >0,$
       \item a saddle for $ f_0\leq0$, $ \lambda <0,$
       \item non-hyperbolic for  $ f_0\leq 0$, $ \lambda =0.$
   \end{enumerate}
\end{enumerate}
In Figure \ref{fig:6} we present the stability analysis for system \eqref{newsyst-1}- \eqref{newsyst-3} with $\epsilon=-1$ and different values of the parameters $\lambda$ and $f_0.$ We consider $y>0$; however, the system is unbounded, suggesting non-trivial dynamics at infinity. We also considered the three cases $w_m=0$ (dust), $\frac{1}{3}$ (radiation) and $1$ (stiff matter). A summary of the results of this section is presented in Table \ref{tab:5}.

\begin{figure}[h]
    \centering
    \includegraphics[scale=0.45]{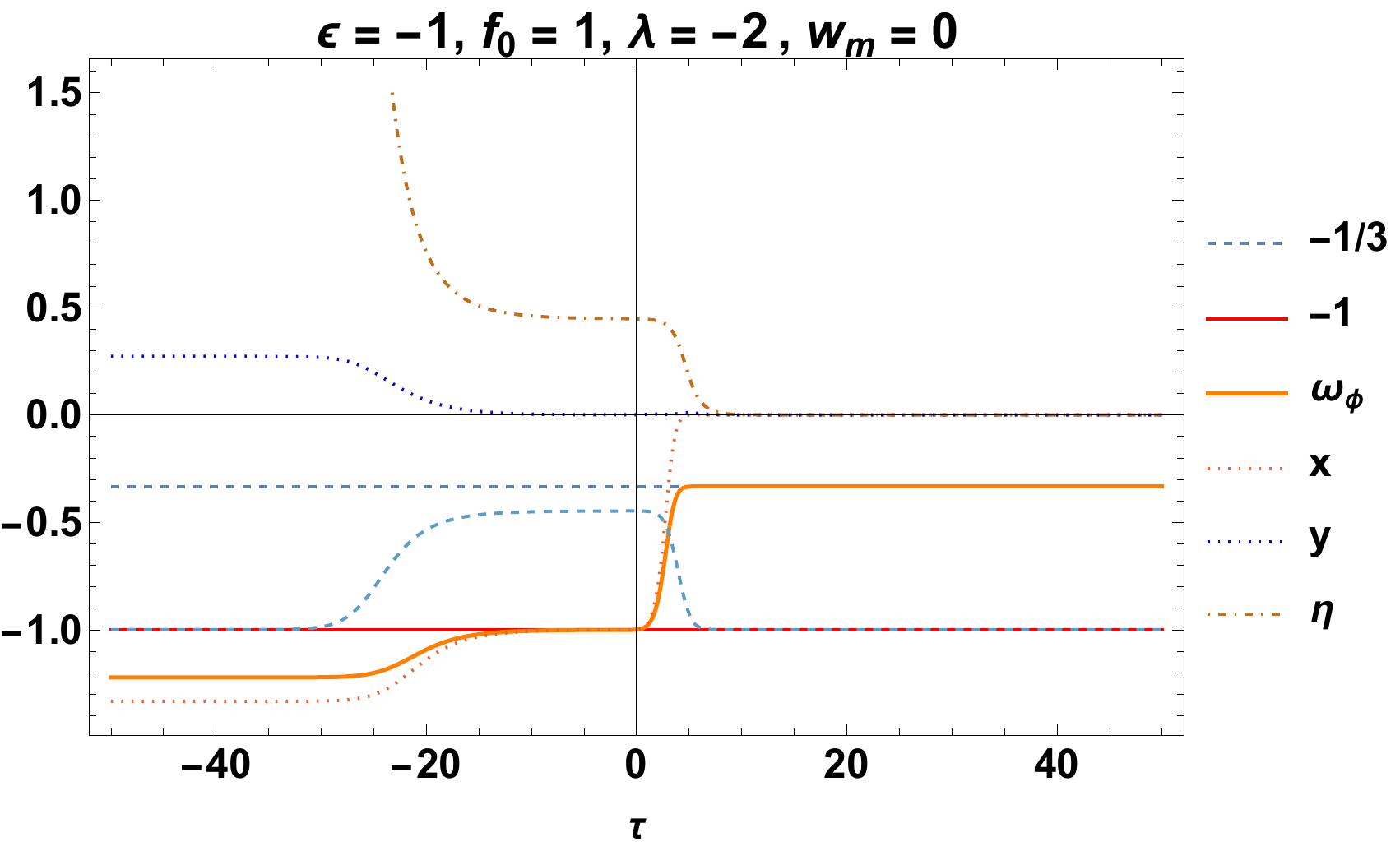}
    \includegraphics[scale=0.45]{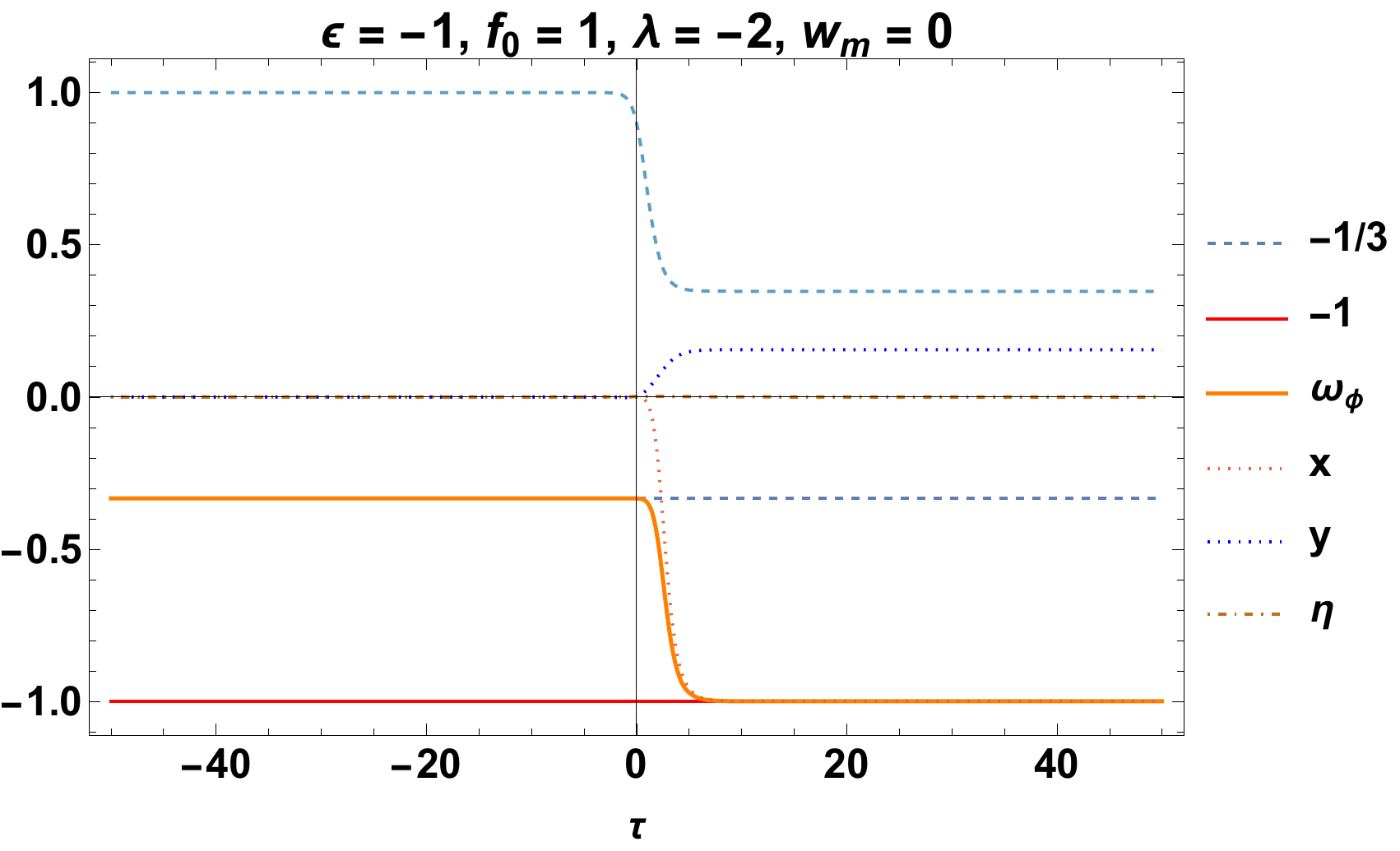}
    \caption{$\omega_\phi(\tau)$, $x(\tau)$, $y(\tau)$, and $\eta(\tau)$ evaluated at the solution of system \eqref{newsyst-1}- \eqref{newsyst-3} for $\epsilon=-1.$ The initial conditions for the left plot are $x(0)=0.001, \quad y(0)=\sqrt{\frac{\lambda }{\lambda -8 f_0}},  \quad \eta (0)=-\sqrt{\frac{\lambda }{\lambda -8 f_0}}$ (i.e., near~the saddle point $P_3$). The~solution is past asymptotic to $\omega_{\phi} =-1$ ($q=-1$), then remains near the de Sitter point $P_3$, then tending asymptotically to $\omega_{\phi} =-\frac{1}{3}$ (the Gauss-Bonnet point $P_2$) from~below. The initial conditions for the plot on the right are $x(0)=0.001,\quad y(0)=0.001,  \quad \eta (0)=0.9$ (i.e., near~the source point $P_1$). The solution is past asymptotic to $\omega_{\phi} =-\frac{1}{3}$ (zero acceleration), then it tends asymptotically to a de Sitter phase $\omega_{\phi}=-1, \quad q=-1$ describing a late-time acceleration.}
    \label{fig:weff2}
\end{figure}

In Figure \ref{fig:weff2} we present the expressions $\omega_\phi(\tau)$, $x(\tau)$, $y(\tau)$, and $\eta(\tau)$ evaluated at the solution of system \eqref{newsyst-1}- \eqref{newsyst-3} for $\epsilon=-1.$ The initial conditions for the left plot are $x(0)=0.001, \quad y(0)=\sqrt{\frac{\lambda }{\lambda -8 f_0}},  \quad \eta (0)=-\sqrt{\frac{\lambda }{\lambda -8 f_0}}$ (i.e., near~the saddle point $P_3$). The~solution is past asymptotic to $\omega_{\phi} =-1$ ($q=-1$), then remains near the de Sitter point $P_3$, then tending asymptotically to $\omega_{\phi} =-\frac{1}{3}$ (the Gauss-Bonnet point $P_2$) from~below. The initial conditions for the plot on the right are $x(0)=0.001,\quad y(0)=0.001,  \quad \eta (0)=0.9$ (i.e., near~the source point $P_1$). The solution is past asymptotic to $\omega_{\phi} =-\frac{1}{3}$ (zero acceleration), then it tends asymptotically to a de Sitter phase $\omega_{\phi}=-1, \quad q=-1$ describing a late-time acceleration.

\begin{table}[ht!]
    \caption{Equilibrium points of system \eqref{newsyst-1}-\eqref{newsyst-3} for $\epsilon=-1$ with their stability conditions. Also includes the value of $\omega_{\phi}$ and $q.$}
    \label{tab:5}
\newcolumntype{C}{>{\centering\arraybackslash}X}
\centering
\begin{tabularx}{\textwidth}{ccccccc}
\toprule 
  \text{Label}  & \; $x$& $y$& $\eta$ & \text{Stability}& $\omega_{\phi}$&$q$\\
  \midrule  
          $M$ & $0$ & $0$& $0$  & non-hyperbolic & indeterminate & indeterminate\\  \midrule 
          $P_1$ & $0$ & $0$ & $1$ & source for $0\leq w_m<1/3$ &&\\  
          &&&& saddle for $1/3<w_m\leq 1$ &&\\
          &&&& non-hyperbolic for $w_m=1/3$  & $-\frac{1}{3}$ & $0$\\  \midrule
          $P_2$ & $0$ & $0$ & $-1$ & sink for $0\leq w_m<1/3$ &&\\ 
          &&&& saddle for $1/3<w_m\leq 1$ &&\\ 
          &&&& non-hyperbolic for $w_m=1/3$  & $-\frac{1}{3}$ & $0$\\  \midrule
          $P_3$ & $0$ & $\sqrt{\frac{\lambda }{\lambda -8 f_0}}$ & $\sqrt{\frac{\lambda }{\lambda -8 f_0}}$ & saddle& $-1$ &$-1$\\  \midrule          
          $P_4$ & $0$ & $\sqrt{\frac{\lambda }{\lambda -8 f_0}}$ & $-\sqrt{\frac{\lambda }{\lambda -8 f_0}}$ & saddle  & $-1$ & $-1$\\ \midrule
          $P_5$ & $\sqrt{\frac{1}{20 \sqrt{\frac{10}{3}} f_0+5}}$ & $0$ & $\sqrt{3} \sqrt{\frac{1}{4 \sqrt{30}f_0+3}}$ & sink for $f_0\geq 0, \lambda<0$&&\\
          &&&& saddle for $f_0\geq 0, \lambda<0$  &&\\
          &&&& non-hyperbolic for $f_0\geq 0, \lambda=0$ & $-1$ & $-1$\\ \midrule
          $P_6$ & $\sqrt{\frac{1}{20 \sqrt{\frac{10}{3}} f_0+5}}$ & $0$ & $\sqrt{3} \sqrt{\frac{1}{4 \sqrt{30}f_0+3}}$ & source for $f_0\geq 0, \lambda<0$&&\\
          &&&& saddle for $f_0\geq 0, \lambda<0$  &&\\
          &&&& non-hyperbolic for $f_0\geq 0, \lambda=0$ & $-1$ & $-1$\\ \midrule     
          $P_7$ & $\sqrt{\frac{1}{5-20 \sqrt{\frac{10}{3}} f_0}}$ & $0$ & $-\sqrt{3} \sqrt{\frac{1}{3-4 \sqrt{30}  f_0}}$ &  source for $f_0\leq 0, \lambda>0$&&\\
          &&&& saddle for $f_0\leq 0, \lambda<0$  &&\\
          &&&& non-hyperbolic for $f_0\leq 0, \lambda=0$ & $-1$ & $-1$\\\midrule
          $P_8$ & $-\sqrt{\frac{1}{5-20 \sqrt{\frac{10}{3}} f_0}}$ & $0$ & $\sqrt{3} \sqrt{\frac{1}{3-4 \sqrt{30}f_0}}$ &  sink for $f_0\leq 0, \lambda>0$ &&\\
          &&&& saddle for $f_0\leq 0, \lambda<0$  &&\\
          &&&& non-hyperbolic for $f_0\leq 0, \lambda=0$& $-1$ & $-1$\\
        \bottomrule
    \end{tabularx}
\end{table}
\FloatBarrier
\FloatBarrier

\section{Dynamics at infinity}
\label{IV}
As suggested before, we are interested in the behaviour of the dynamical system at infinity; for that purpose, we define the new compact variables 
\begin{equation}
    X= \frac{x}{\sqrt{1+x^2+y^2}}, \quad   Y= \frac{y}{\sqrt{1+x^2+y^2}},
\end{equation}
and the new derivative 
\begin{equation}
    \frac{d f}{d s}= Z \frac{d f}{d\tau}, \quad Z= \sqrt{1-X^2-Y^2}.
\end{equation}
With these definitions, we obtain the following dynamical system 
\begin{align}
\label{cilindro-1}
 & \frac{d X}{d s}= -\frac{1}{L} \Big[\left(\eta ^2-1\right) \left(48 f_0 \lambda  X \eta +\sqrt{6} Z \left((\lambda -12 f_0 (w_m+1)) \eta ^2-\lambda
   \right)\right) Y^4 \nonumber \\
   & +\left(48 f_0 \epsilon  \lambda  \eta  \left(\eta ^2-1\right) X^3+\sqrt{6} Z \left(\left(96 \lambda  f_0^2-12
   (w_m-5) \epsilon  f_0+\epsilon  \lambda \right) \eta ^4-2 \epsilon  (\lambda -6 f_0 (w_m-5)) \eta ^2+\epsilon  \lambda \right) X^2 \right.
   \nonumber \\
   & \left.  +3 Z^2 \eta  \left(\left(192 w_m f_0^2+8 \lambda  f_0+(w_m+3) \epsilon \right) \eta ^4-2 ((w_m+3) \epsilon -4 f_0
   \lambda ) \eta ^2+(w_m+3) \epsilon -16 f_0 \lambda \right) X  \right.  \nonumber \\
   & \left. +\sqrt{6} Z^3 \left(\eta ^2-1\right) \left(4 (3 w_m
   f_0+f_0) \eta ^4+(\lambda -12 f_0 (w_m+1)) \eta ^2-\lambda \right)\right) Y^2 \nonumber \\
   & -Z^2 \eta  \left(-3 (w_m-1) \epsilon ^2
   \left(\eta ^2-1\right)^2 X^3+4 \sqrt{6} f_0 \epsilon  Z \eta  \left(\eta ^2-1\right) \left((6 w_m-2) \eta ^2+3 (w_m-5)\right)
   X^2 \right. \nonumber \\
   & \left. +3 Z^2 \left(\left(64 f_0^2+w_m \epsilon +\epsilon \right) \eta ^6-2 \left(96 w_m f_0^2+(w_m+2) \epsilon \right) \eta
   ^4+(w_m+5) \epsilon  \eta ^2-2 \epsilon \right) X \right. \nonumber \\
   & \left. -4 \sqrt{6} f_0 (3 w_m+1) Z^3 \eta ^3 \left(\eta ^2-1\right)\right)\Big],\\
 \label{cilindro-2} &  \frac{d Y}{d s}= \frac{Y}{L}  \Big[48 f_0 \epsilon  \lambda  \eta  \left(\eta ^2-1\right) X^4 \nonumber \\
  & +\sqrt{6} Z \left(\left(96 \lambda 
   f_0^2-12 (w_m-5) \epsilon  f_0+\epsilon  \lambda \right) \eta ^4-2 \epsilon  (\lambda -6 f_0 (w_m-5)) \eta ^2+\epsilon  \lambda \right) X^3 \nonumber \\
   & +3 \eta\left(16 f_0 \lambda  \left(\eta ^2-1\right) Y^2 \right. \nonumber \\ & \left. +Z^2 \left(\left(192 w_m f_0^2+\epsilon  (-w_m \epsilon +\epsilon +2)\right)
   \eta ^4  +2 \epsilon  ((w_m-1) \epsilon +8 f_0 \lambda -2) \eta ^2+\epsilon  (-w_m \epsilon +\epsilon -16 f_0 \lambda +2)\right)\right) X^2 \nonumber \\
   & +\sqrt{6}Z \left(\left(\eta ^2-1\right) \left((\lambda -12 f_0 (w_m+1)) \eta ^2-\lambda \right) Y^2 \right. \nonumber \\ & \left. +Z^2 \left(4 f_0 (-2
   \epsilon +w_m (6 \epsilon +3)+1) \eta ^6+\left(96 \lambda  f_0^2-4 (6 \epsilon  w_m+3 w_m-2 \epsilon +1) f_0+\epsilon  \lambda \right) \eta ^4-2
   \epsilon  \lambda  \eta ^2+\epsilon  \lambda \right)\right) X \nonumber \\
   & -3 Y^2 Z^2 \eta  \left(\eta ^2-1\right) \left((w_m \epsilon +\epsilon +8
   f_0 \lambda ) \eta ^2-(w_m+1) \epsilon \right) \nonumber \\
   & +3 Z^4 \eta ^3 \left(\left(64 f_0^2+w_m \epsilon +\epsilon \right) \eta ^4-2
   (w_m+1) \epsilon  \eta ^2+(w_m+1) \epsilon \right)\Big],\\
 \label{cilindro-3}&   \frac{d \eta}{d s}= \frac{\left(\eta ^2-1\right)}{L} \Big[8 \sqrt{6} f_0 (3 w_m-1)
   \epsilon  X Z \left(\eta ^2-1\right) \eta ^3  \nonumber \\
   & +3 Z^2 \left(\left(64 f_0^2+w_m \epsilon +\epsilon \right) \eta ^4-2
   (w_m+1) \epsilon  \eta ^2+(w_m+1) \epsilon \right) \eta ^2  \nonumber \\
   & -3 (w_m-1) \epsilon ^2 X^2 \left(\eta ^2-1\right)^2-3 Y^2 \left(\eta ^2-1\right)
   \left((w_m \epsilon +\epsilon +8 f_0 \lambda ) \eta ^2-(w_m+1) \epsilon \right)\Big],  
\end{align}
where 
\begin{equation}
    L= 2 \left(8 \sqrt{6} f_0 \epsilon  X \eta  \left(\eta
   ^2-1\right)+Z \left(\left(96 f_0^2+\epsilon \right) \eta ^4-2 \epsilon  \eta ^2+\epsilon \right)\right).
\end{equation}
To obtain the equilibrium points at infinity, we define the cylindrical coordinates $(\rho, \theta, \eta)$
\begin{align}
   X= \rho 
   \cos (\theta) , \quad Y= \rho \sin (\theta), \quad \eta=\eta,
\end{align}
such that $X^2+Y^2\rightarrow 1$ corresponds to  $\rho \rightarrow 1$. 

Then, as $\rho \rightarrow 1$ we have the leading terms
\begin{align}
& \frac{d \rho}{d s}=-\frac{1}{8 f_0 \epsilon }\sqrt{\frac{3}{2}} (1-\rho)\cos (\theta ) \Big[\eta^2 \left(8 f_0 \lambda  \tan ^2(\theta )+\epsilon  \left((w_m+1) \sec ^2(\theta )+(w_m-1) \epsilon
   -w_m-1\right)\right) \nonumber\\
   & \;\;\;\;\;\;\;\;\;\;\;\;\;\;\;\;\;\;\;\;\;     
   -16 f_0 \lambda  (\epsilon -1) \sin ^2(\theta )+\epsilon  \left(-\left((w_m+1) \sec ^2(\theta )\right)-w_m \epsilon +w_m+\epsilon
   +1\right)\Big], \label{eq30}\\
& \frac{d \theta}{d s}=\frac{\sqrt{\frac{3}{2}} \lambda  \sin (\theta ) ((\epsilon -1) \cos (2 \theta )+\epsilon +1)}{2 \epsilon }, \label{eq31}\\
& \frac{d \eta}{d s}=-\frac{\sqrt{\frac{3}{2}} \left(\eta
  ^2-1\right)}{16 f_0 \epsilon  \eta}  \Big[\eta^2 \left(\sin (\theta ) \tan (\theta ) (8 f_0 \lambda +w_m \epsilon +\epsilon )+(w_m-1) \epsilon ^2 \cos (\theta )\right) \nonumber \\
 & \;\;\;\;\;\;\;\;\;\;\;\;\;\;\;\;\;\;\;\;\;\;\;\;\;\;\;\;\;   -\epsilon 
   ((w_m+1) \sin (\theta ) \tan (\theta )+(w_m-1) \epsilon  \cos (\theta ))\Big]. \label{eq32}
\end{align}
\subsection{Analysis at infinity for $\epsilon=1$}
\label{IV-A}

Recall that the equilibrium points for this case are the same ones as the finite regime in section \ref{III-B} plus the following additional points described in this section. Since $Y>0$, we set $\theta\in [0,\pi]$. 
Therefore, the points at infinity for $\epsilon=+1$ in the coordinates $(\rho,\theta,\eta)$ are 
\begin{enumerate}
    \item $Q_{1,2}=(1,0,\pm1),$ with eigenvalues $\left\{0,0,\sqrt{\frac{3}{2}} \lambda \right\}.$ These points satisfy $\omega_{\phi}=-\frac{1}{3},$ $q=0,$ this means that the points described a universe dominated by the Gauss-Bonnet term.
    \item $Q_{3,4}=(1,\pi,\pm1),$ with eigenvalues $\left\{0,0,-\sqrt{\frac{3}{2}} \lambda \right\}.$ These points satisfy $\omega_{\phi}=-\frac{1}{3},$ $q=0,$ the physical interpretation is the same as the previous points.
\end{enumerate}
The third eigenvalue corresponds to the $\theta$-axis. 
To analyze the non-hyperbolic nature of the critical points in the plane $(\rho, \eta)$, we consider the variable $\tau$ as the independent variable. 
This is equivalent to divide the system \eqref{eq30}-\eqref{eq32} by $\sqrt{1-\rho^2}$. 

\begin{figure}[h]
    \centering
    \includegraphics[scale=0.5]{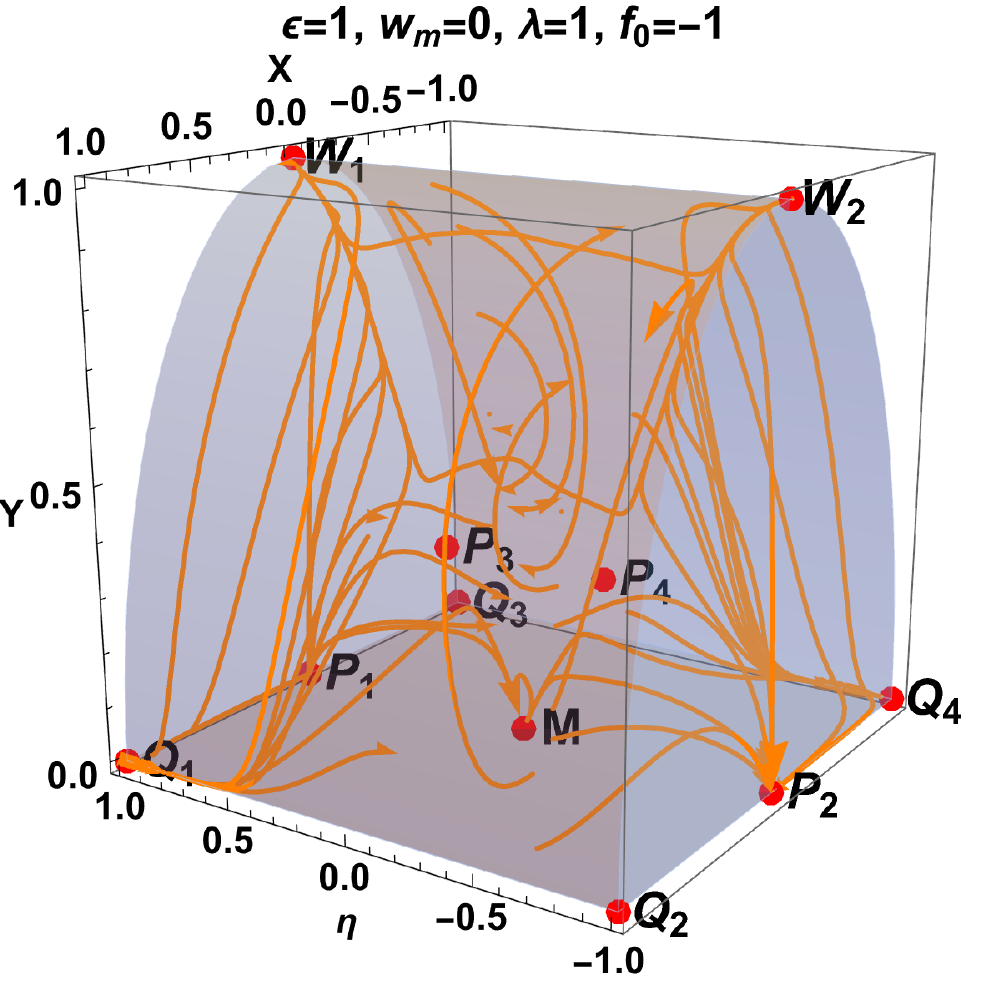}
    \includegraphics[scale=0.5]{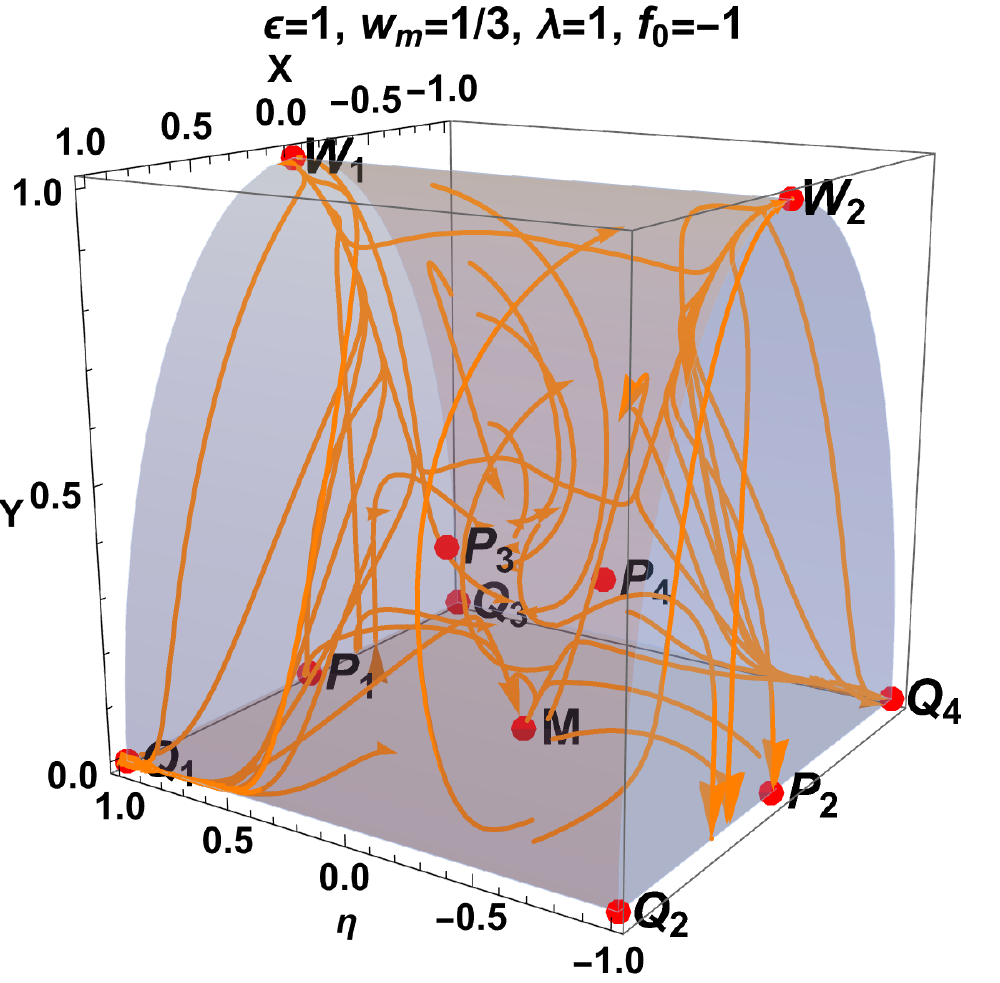}
    \includegraphics[scale=0.5]{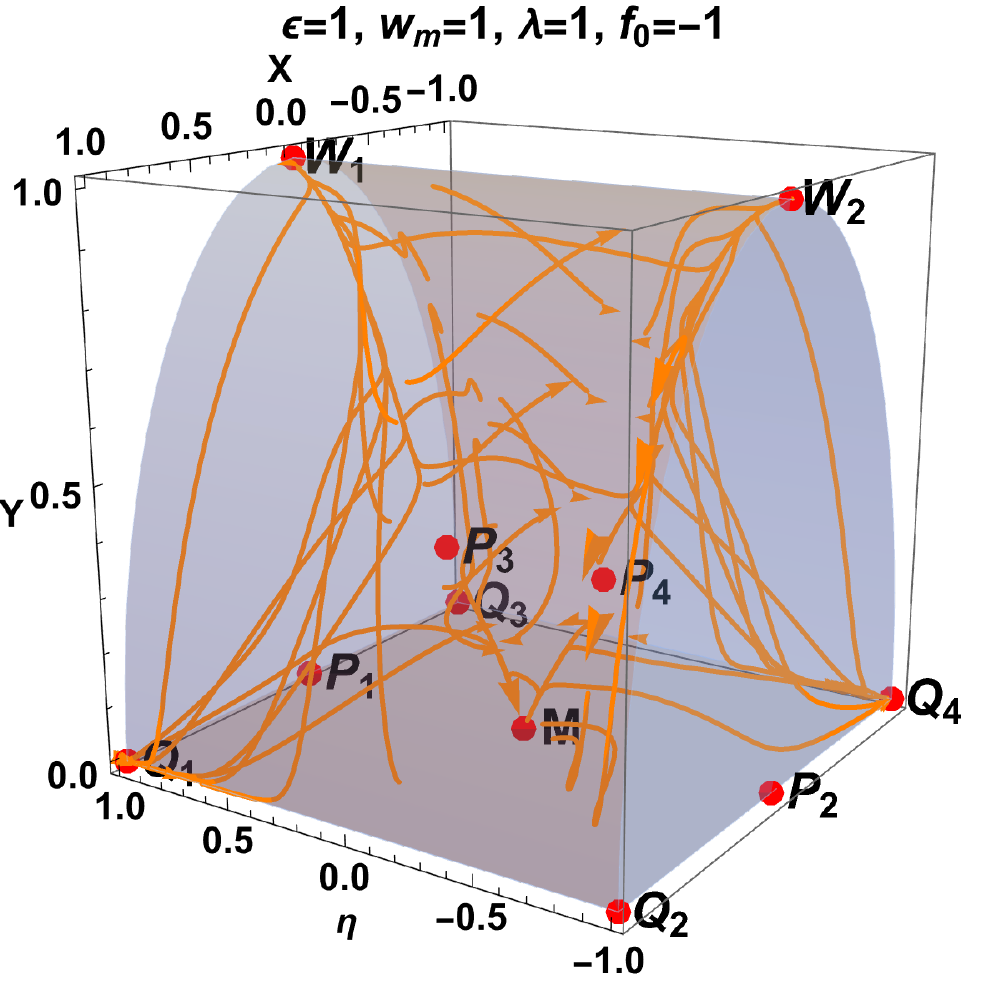}
    \includegraphics[scale=0.5]{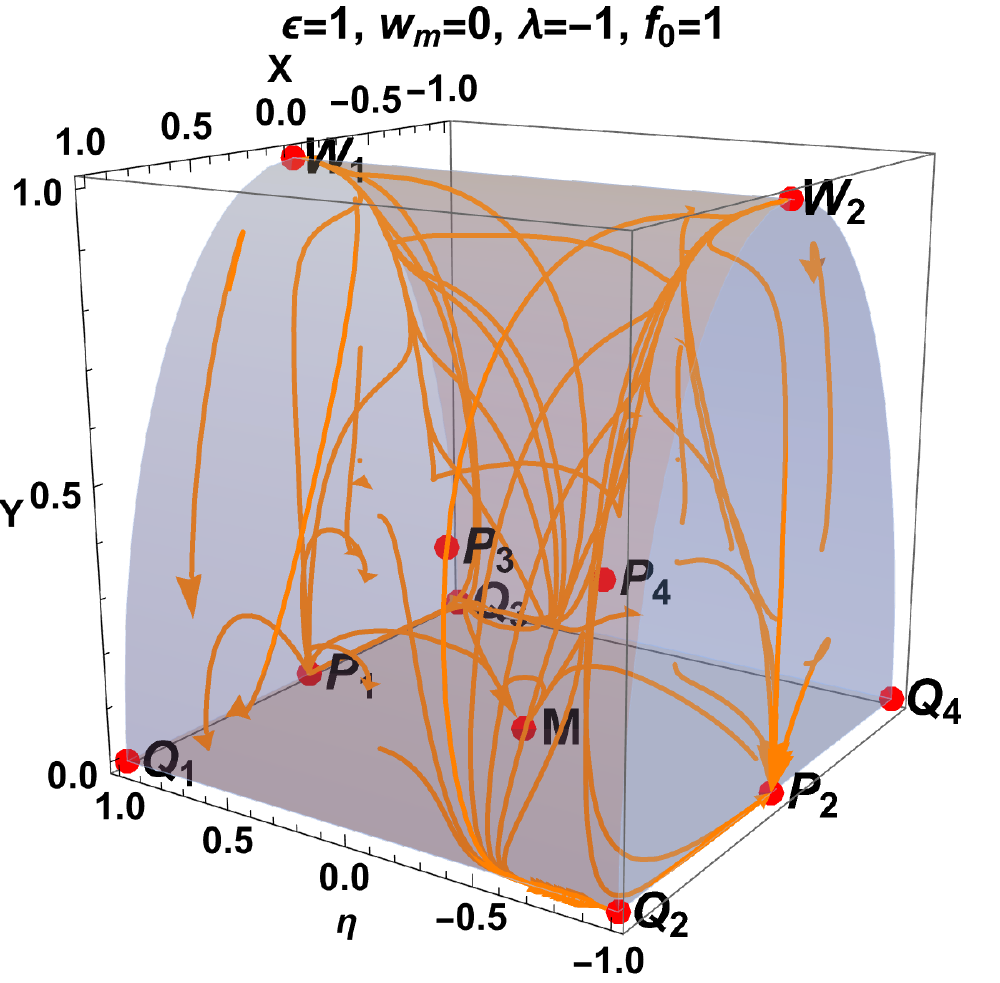}
    \includegraphics[scale=0.5]{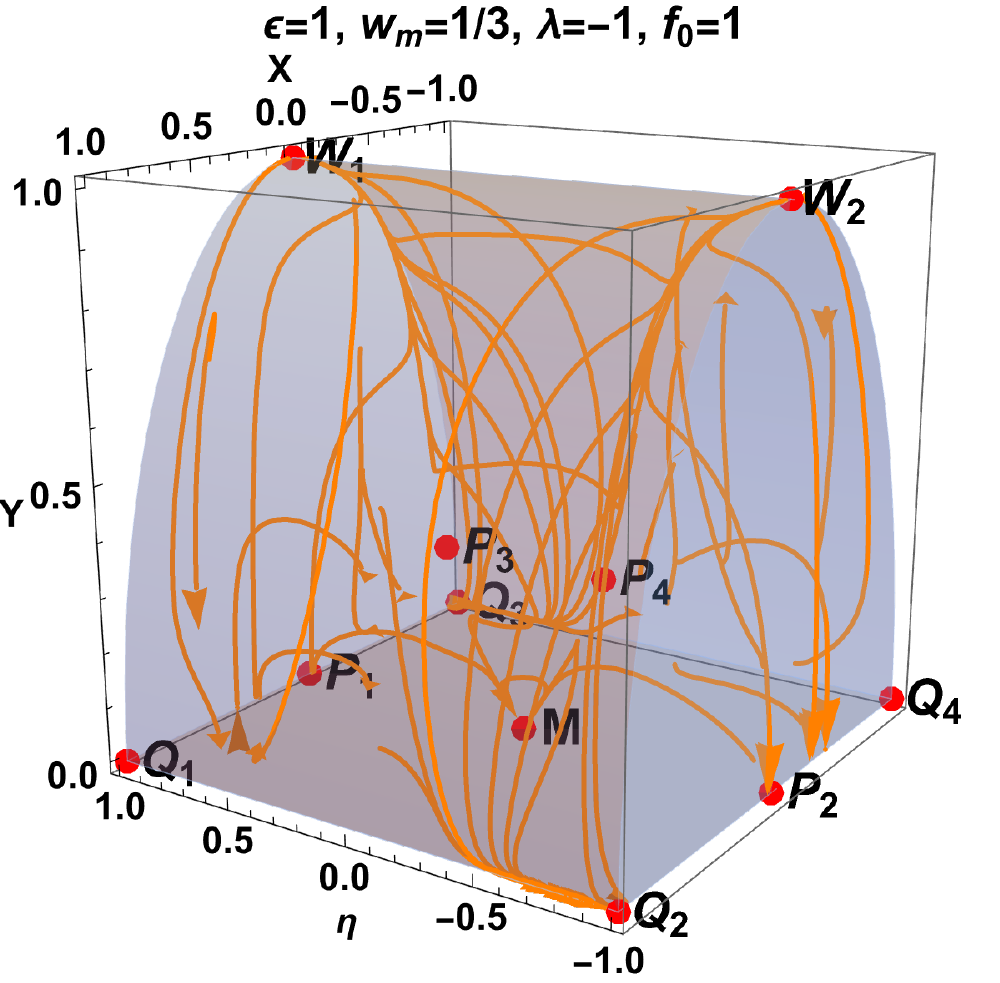}
    \includegraphics[scale=0.5]{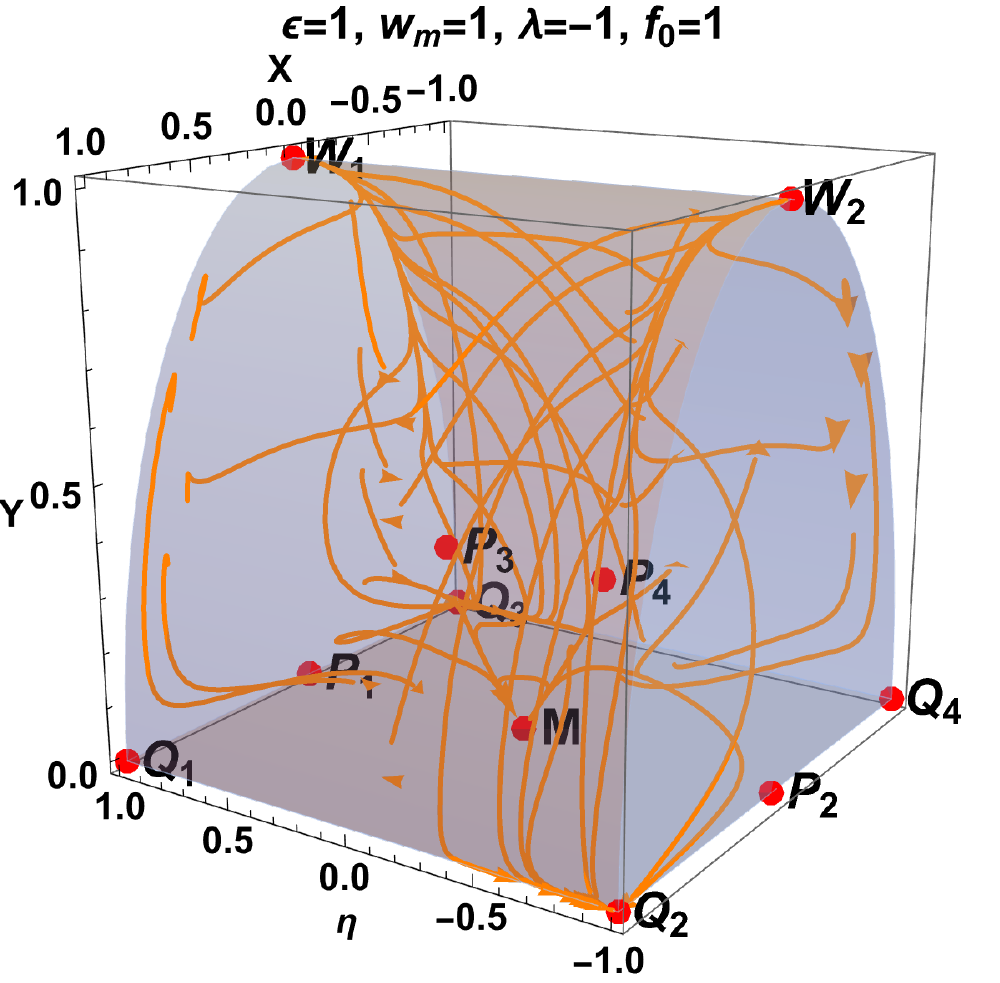}
    \caption{Phase-space analysis for system \eqref{cilindro-1}-\eqref{cilindro-3} for $\epsilon=1$ and different values of the parameters $\lambda, f_0.$ Here we consider $Y>0$ and the three cases $w_m=0, \frac{1}{3}, 1.$ The points $W_{1,2}$ are singularities where both the numerator and denominator of the equations of the system vanish.}
    \label{fig:cilindroepsilon=1}
\end{figure}

The points $Q_{1,2}$ in the re-scaled system have the eigenspace given by 
\begin{equation}
    \left(
\begin{array}{ccc}
 \pm 2 & \pm 2 (3 w_m -1) & \text{sign}(\lambda) \infty \\
 \{0,0,1\} & \{1,0,0\} & \{0,1,0\} \\
\end{array}
\right).
\end{equation}
For the points  $Q_{3,4}$ in the rescaled system has the eigenspace given by 
\begin{equation}
    \left(
\begin{array}{ccc}
 \pm 2 & \pm 2 (3 w_m -1) & -\text{sign}(\lambda) \infty \\
 \{0,0,1\} & \{1,0,0\} & \{0,1,0\} \\
\end{array}
\right).
\end{equation}
The last column with infinity entries is an artefact of the division by $\sqrt{1-\rho^2}$ as $\rho\rightarrow 1$. 
Therefore, by combining the two approaches, we obtain that,
\begin{enumerate}
    \item $Q_{1,2}$ with eigenvalues $\left\{\pm 2,\pm 2(3w_m-1), \sqrt{\frac{3}{2}}\lambda\right\}$ satisfy the following
    \begin{enumerate}
        \item $Q_{1}$ is a source ($Q_{2}$ is a sink) for $\lambda>0,$ $\frac{1}{3}<w_m\leq 1.$
        \item They are saddles for 
        \begin{enumerate}
            \item $\lambda\neq 0,$ $0\leq w_m<\frac{1}{3},$
            \item $\lambda<0,$ $\frac{1}{3}<w_m\leq 1,$
        \end{enumerate}
        \item non-hyperbolic for $\lambda=0$ or  $w_m=\frac{1}{3}.$
    \end{enumerate}
    \item $Q_{3,4}$ with eigenvalues $\left\{\pm 2, \pm 2(3w_m-1), -\sqrt{\frac{3}{2}}\lambda\right\}$ satisfy the following. 
    \begin{enumerate}
        \item $Q_{3}$ is a source ($Q_4$ is a sink) for $\lambda<0,$ $\frac{1}{3}<w_m\leq 1.$
        \item They are saddles for 
        \begin{enumerate}
            \item $\lambda\neq 0,$ $0\leq w_m<\frac{1}{3},$
            \item $\lambda>0,$ $\frac{1}{3}<w_m\leq 1,$
        \end{enumerate}
        \item They are non-hyperbolic for $\lambda=0$ or  $w_m=\frac{1}{3}.$
    \end{enumerate}

\end{enumerate}
In Figure \ref{fig:cilindroepsilon=1} we present the phase-space analysis for system \eqref{cilindro-1}-\eqref{cilindro-3} for $\epsilon=1$ and different values of the parameters $\lambda, f_0.$ Where we defined the region $0\leq X^2+Y^2\leq 1,$ $Y>0$ and $-1\leq \eta\leq 1$ defining half a cylinder. We also considered the three cases $w_m=0$ (dust), $\frac{1}{3}$ (radiation), and $ 1$ (stiff matter).  

\FloatBarrier
\subsection{Analysis at infinity for $\epsilon=-1$}
\label{IV-B}
\begin{figure}[h]
    \centering
    \includegraphics[scale=0.6]{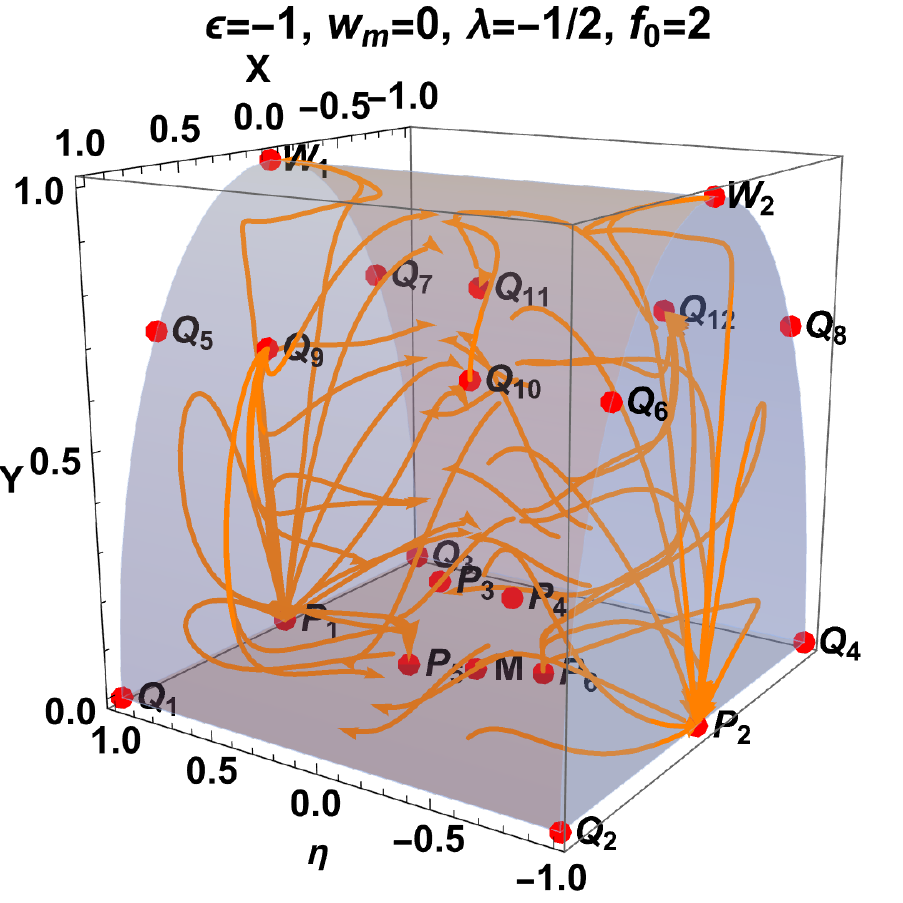}
   \includegraphics[scale=0.6]{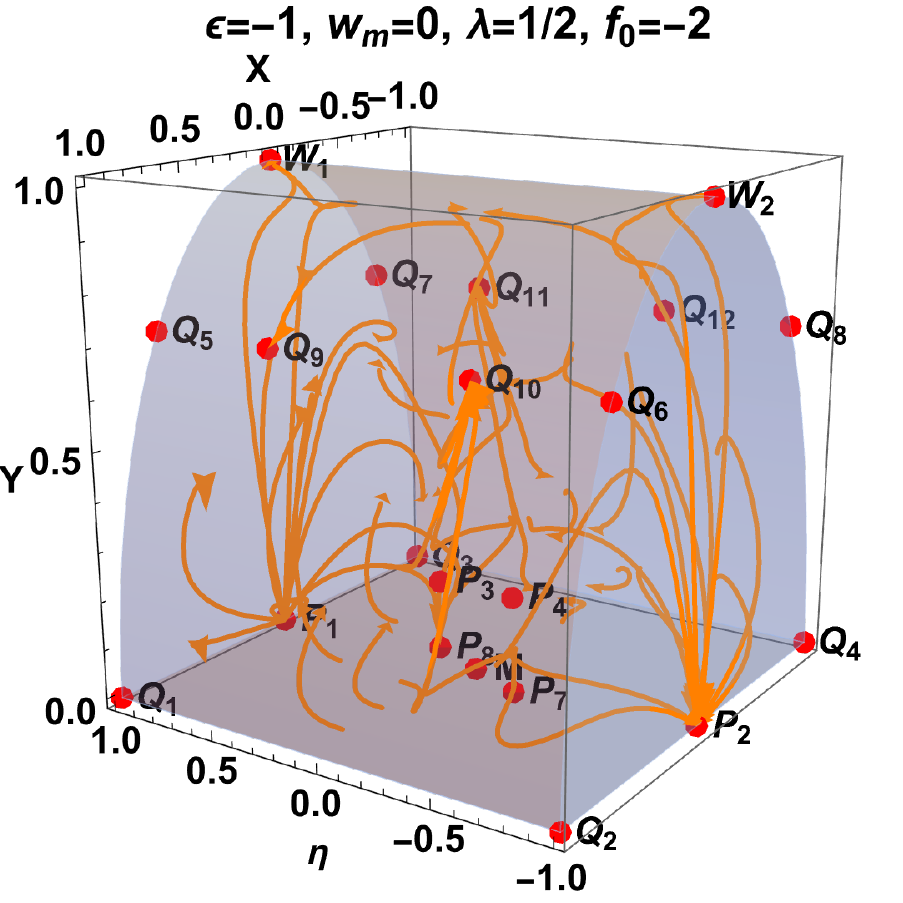}\\
    \includegraphics[scale=0.6]{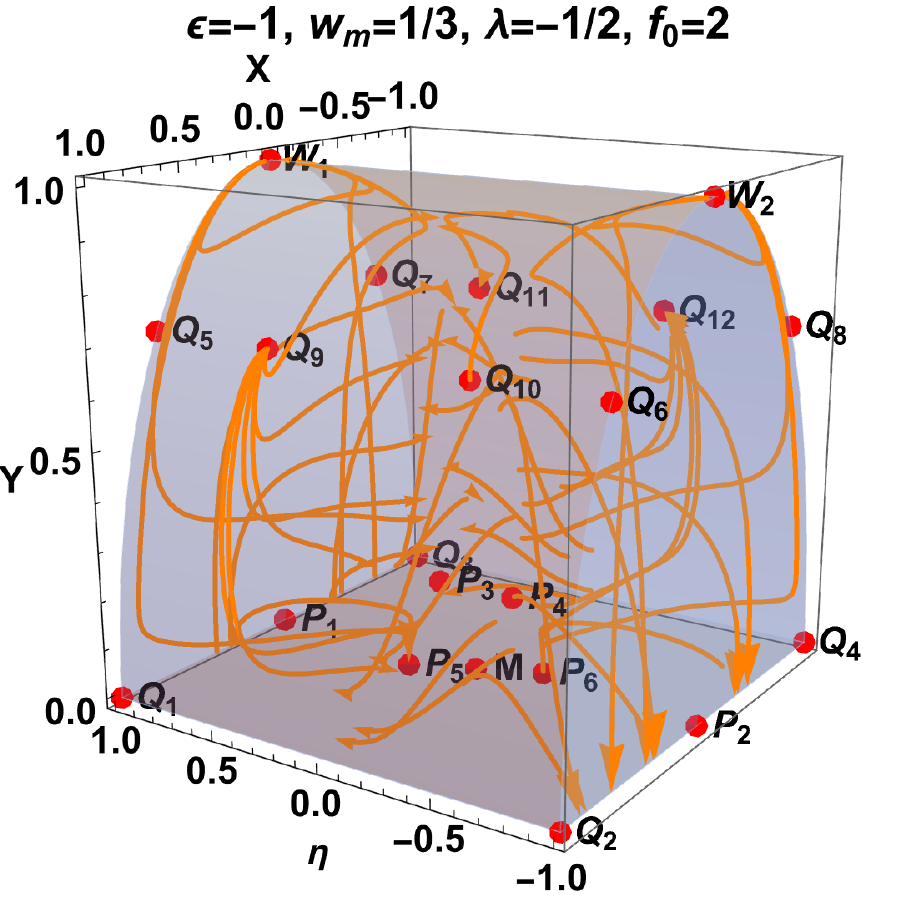}
    \includegraphics[scale=0.6]{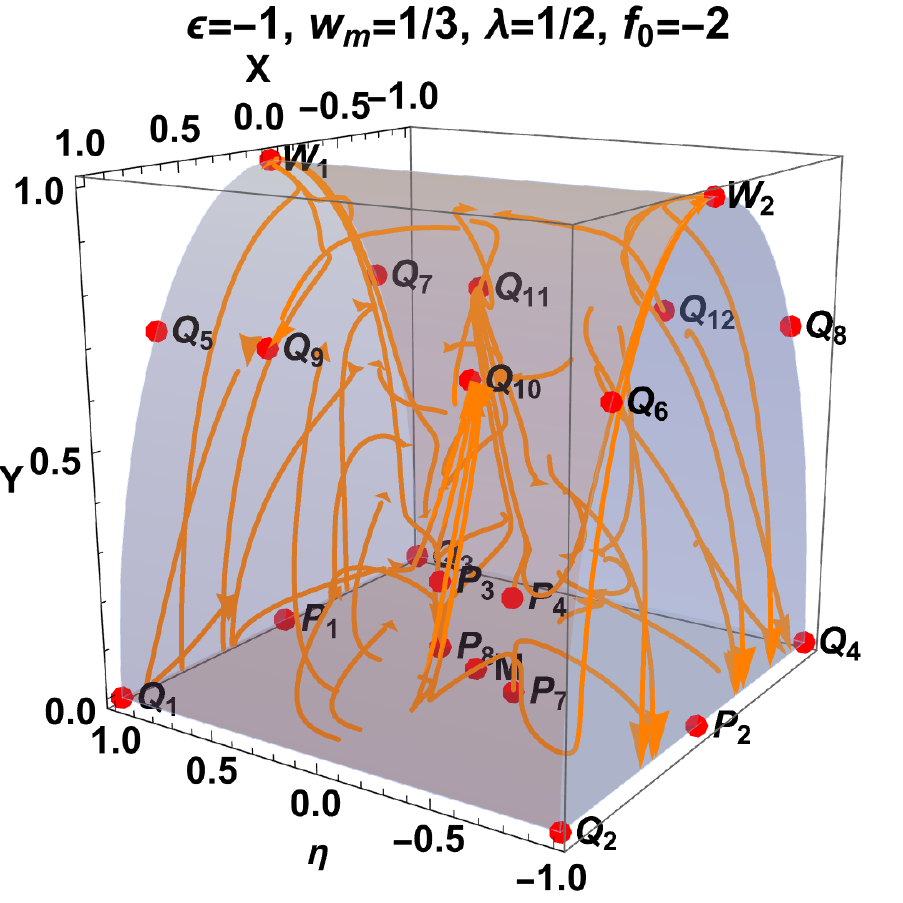}
     \caption{Phase-space analysis for system \eqref{cilindro-1}-\eqref{cilindro-3} for $\epsilon=-1$ and different values of the parameters $\lambda, f_0.$ Here we consider $Y>0$ and the two cases $w_m=0, \frac{1}{3}.$ The points $W_{1,2}$ are singularities where both the numerator and denominator of the equations of the system vanish.}
    \label{fig:7}
\end{figure}

As in the previous section, we consider $\theta \in [0,\pi].$ The equilibrium points for this case are the same ones from section \ref{III-C} plus the following the points at infinity for $\epsilon=-1$ in the coordinates $(\rho,\theta,\eta)$, say,
 \begin{enumerate}
    \item $Q_{1,2}=(1,0,\pm 1),$ with eigenvalues $\{0,0,\sqrt{\frac{2}{3}}\lambda\}.$ These points verify that $\omega_{\phi}=-\frac{1}{3}$ and $q=0.$ These points describe a universe dominated by the Gauss-Bonnet term, and the analysis is the same as in section \ref{IV-A}. 
    \item $Q_{3,4}=(1,\pi,\pm 1),$ with eigenvalues $\{0,0,\sqrt{\frac{2}{3}}\lambda\}.$ These points verify that $\omega_{\phi}=-\frac{1}{3}$ and $q=0.$ These points describe a universe dominated by the Gauss-Bonnet term, and the analysis is the same as in section \ref{IV-A}. 
    \item $Q_{5,6}=(1,\frac{\pi}{4},\pm 1),$ with eigenvalues $\left\{0,-\frac{\sqrt{3} \lambda }{2},-\frac{\sqrt{3} \lambda }{4}\right\}.$ These points verify that $\omega_{\phi}=-\frac{1}{3}$ and $q=0.$ These points describe a universe dominated by the Gauss-Bonnet term. Using a strategy similar to that shown in section \ref{IV-A} that is, re-scaling the system dividing by $\sqrt{1-\rho^2}$, we obtain that the stability of points $Q_{5,6}$ is given by the eigenvalues $\{\pm 2, - \frac{\sqrt{3} \lambda }{2},- \frac{\sqrt{3} \lambda }{4}\}.$ 
    \begin{enumerate}
        \item $Q_5$ is a source ($Q_6$ is a saddle) for $\lambda<0,$
        \item $Q_5$ is a saddle ($Q_6$ is a sink) for $\lambda>0,$
        \item they are non-hyperbolic for $\lambda=0.$
    \end{enumerate}
    \item $Q_{7,8}=(1,\frac{3\pi}{4},\pm 1),$ with eigenvalues $\left\{0,\frac{\sqrt{3} \lambda }{4},\frac{\sqrt{3} \lambda }{2}\right\}.$ These points verify that $\omega_{\phi}=-\frac{1}{3}$ and $q=0.$ These points describe a universe dominated by the Gauss-Bonnet term. Using a strategy similar to that shown in section \ref{IV-A} that is, re-scaling the system dividing by $\sqrt{1-\rho^2}$, we obtain that the stability of the points $Q_{5,6}$ is given by the eigenvalues $\{\pm 2,  \frac{\sqrt{3} \lambda }{2}, \frac{\sqrt{3} \lambda }{4}\}.$     
    \begin{enumerate}
        \item $Q_7$ is a source ($Q_8$ is a saddle) for $\lambda>0,$
        \item $Q_7$ is a saddle ($Q_8$ is a sink) for $\lambda<0,$
        \item they are non-hyperbolic for $\lambda=0.$
    \end{enumerate}
    \item $Q_{9,10}=(1,\frac{\pi}{4},\pm \frac{1}{\sqrt{1-4f_0 \lambda}}),$ with eigenvalues $\left\{-\frac{\sqrt{3} \lambda }{2},-\sqrt{3} \lambda ,-\sqrt{3} \lambda \right\}.$ These points verify that $\omega_{\phi}=w_m-\frac{4}{3}$ and $q=\frac{3 (w_m-1)}{2},$ in Figure \ref{fig:weff-and-q} we see the plot of these observables and note for instance that: for $w_m=0$ the values are $\omega_{\phi}=-\frac{4}{3},$ $q=-\frac{3}{2}$; for $w_m=\frac{1}{3}$ the values are $\omega_{\phi}=-1,$ $q=-1$ that is, they are de Sitter points; for $w_m=1$ the values are $\omega_{\phi}=-\frac{1}{3},$ $q=$ that is, they are Gauss-Bonnet points. These points are 
    \begin{enumerate}
        \item sources for $\lambda<0,$
        \item sinks for $\lambda>0,$
        \item nonhyperbolic for $\lambda=0.$
    \end{enumerate}
    \begin{figure}[h]
        \centering
        \includegraphics[scale=0.7]{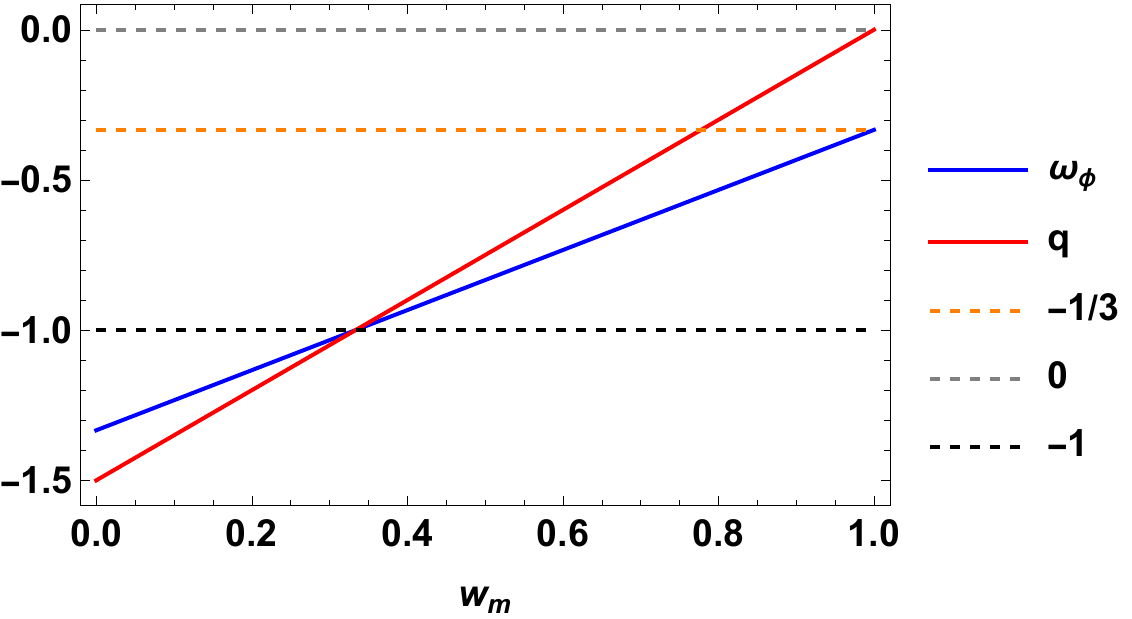}
        \caption{Plot of $\omega_{\phi}$ and $q$ where we set $w_m\in [0,1].$}
        \label{fig:weff-and-q}
    \end{figure}
    \FloatBarrier
    \item $Q_{11,12}=(1,\frac{3\pi}{4},\pm \frac{1}{\sqrt{1-4f_0 \lambda}}),$ with eigenvalues $\left\{\frac{\sqrt{3} \lambda }{2},\sqrt{3} \lambda ,\sqrt{3} \lambda \right\}.$ These points also verify that $\omega_{\phi}=w_m-\frac{4}{3}$ and $q=\frac{3 (w_m-1)}{2},$ again in Figure \ref{fig:weff-and-q} we see the plot of these observables this means that the interpretation is the same as in the previous points. These points are 
    \begin{enumerate}
        \item sources for $\lambda>0,$
        \item sinks for $\lambda<0,$
        \item nonhyperbolic for $\lambda=0.$
    \end{enumerate}
 \end{enumerate}
Figure \ref{fig:7} shows a phase-space analysis for system \eqref{cilindro-1}-\eqref{cilindro-3} for $\epsilon=-1$ and different values of the parameters $\lambda, f_0.$ Here we consider $Y>0$ and the two cases $w_m=0$ (dust) and $\frac{1}{3}$ (radiation).   The points $W_{1,2}$ are singularities where both the numerator and denominator of the equations of the system vanish.
    
\FloatBarrier
 
   \section{Dynamical system analysis for: $y=0, \lambda=0$}\label{V}

   In this section, we study the case where the model has no scalar field potential, equivalent to setting $\lambda=y=0$ in system \eqref{newsyst-1}-\eqref{newsyst-3}. With these assumptions, we have a reduced 2-dimensional system given by
   \begin{align}
       &\frac{dx}{d\tau}=\frac{1}{K}\Bigg[\eta  \left(3 x \left(\eta ^6 \left(64
   f_0^2+w_m \epsilon +\epsilon
   \right)-2 \eta ^4 \left(96 f_0^2
   w_m+(w_m+2) \epsilon
   \right)+(w_m+5) \epsilon  \eta ^2-2
   \epsilon \right)\right)\nonumber \\  &+\eta\left(-4 \sqrt{6} f_0 (3
   w_m+1) \eta ^3 \left(\eta
   ^2-1\right)+4 \sqrt{6} f_0 \epsilon 
   \eta  \left(\eta ^2-1\right) x^2
   \left((6 w_m-2) \eta ^2+3
   (w_m-5)\right) \right. \nonumber \\ & \left. -3 (w_m-1)  \left(\eta ^2-1\right)^2 x^3\right)\Bigg], \label{no-potential-1}\\
       &\frac{d\eta}{d\tau}=\frac{1}{K}\Bigg[\left(\eta ^2-1\right) \left(192 f_0^2
   \eta ^6+8 \sqrt{6} f_0 (3
   w_m-1) \epsilon  \left(\eta
   ^2-1\right) \eta ^3 x \right. \nonumber \\ & \left. +3 \epsilon 
   \left(\eta ^2-1\right)^2
   \left((w_m+1) \eta ^2+x^2
   (\epsilon -w_m \epsilon )\right)\right)\Bigg]. \label{no-potential-2}
   \end{align}
   
   \subsection{Dynamical analysis for $\epsilon=1$}
   \label{V-A}
   The equilibrium points for system \eqref{no-potential-1}-\eqref{no-potential-2} in the coordinates $(x,\eta)$ are the following.

\begin{figure}[h!]
       \centering
       \includegraphics[scale=0.355]{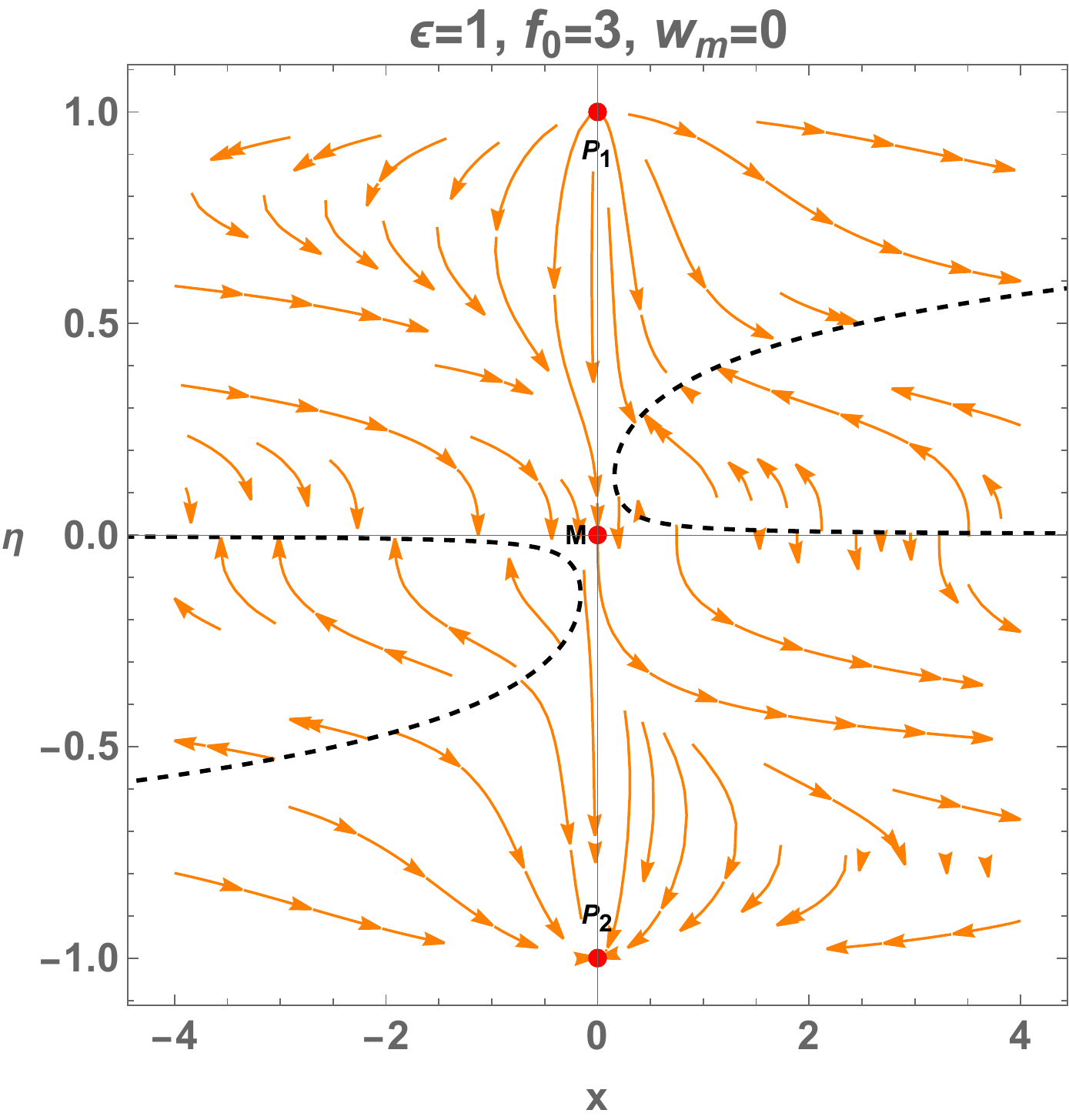}
       \includegraphics[scale=0.355]{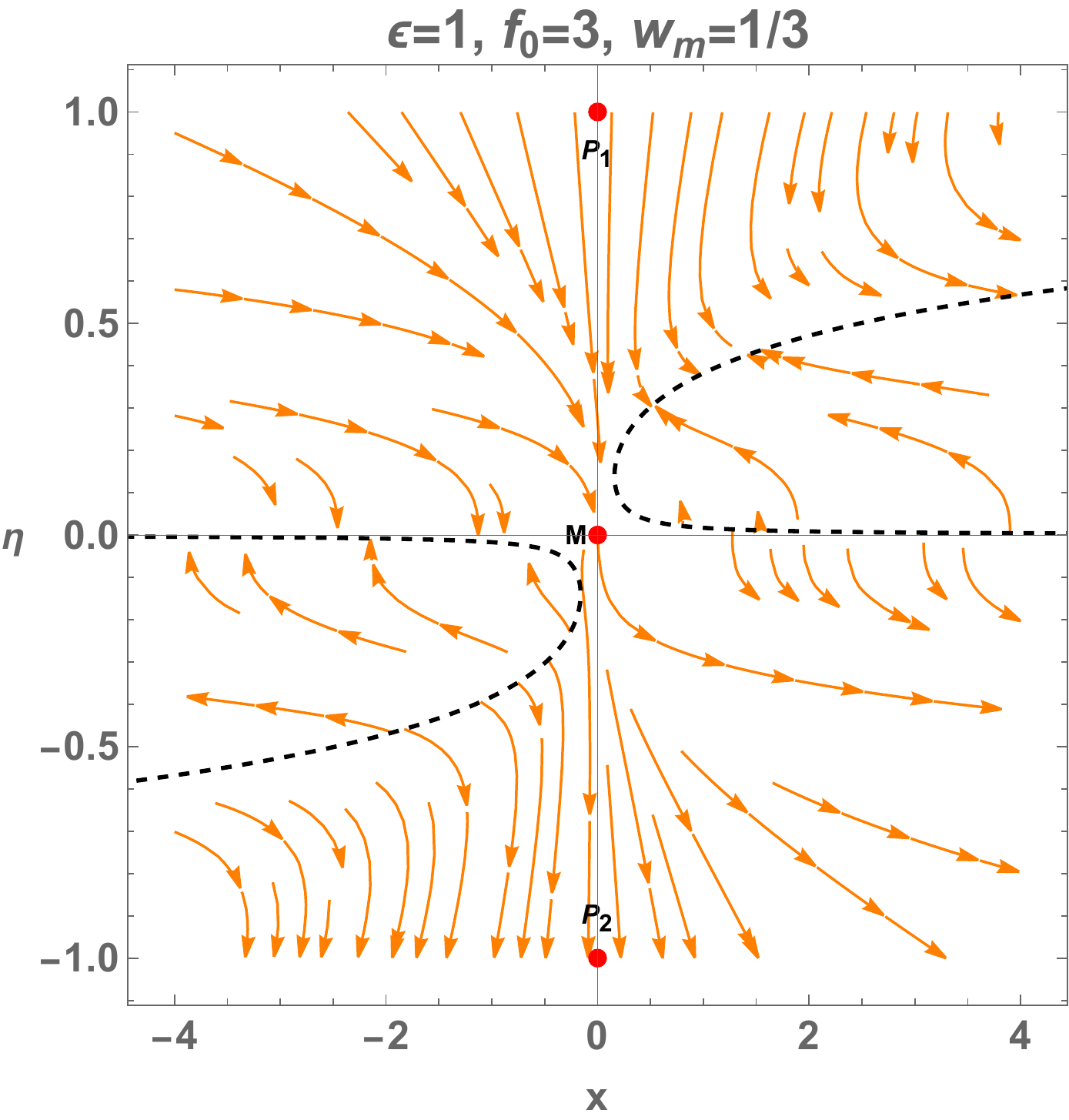}
       \includegraphics[scale=0.355]{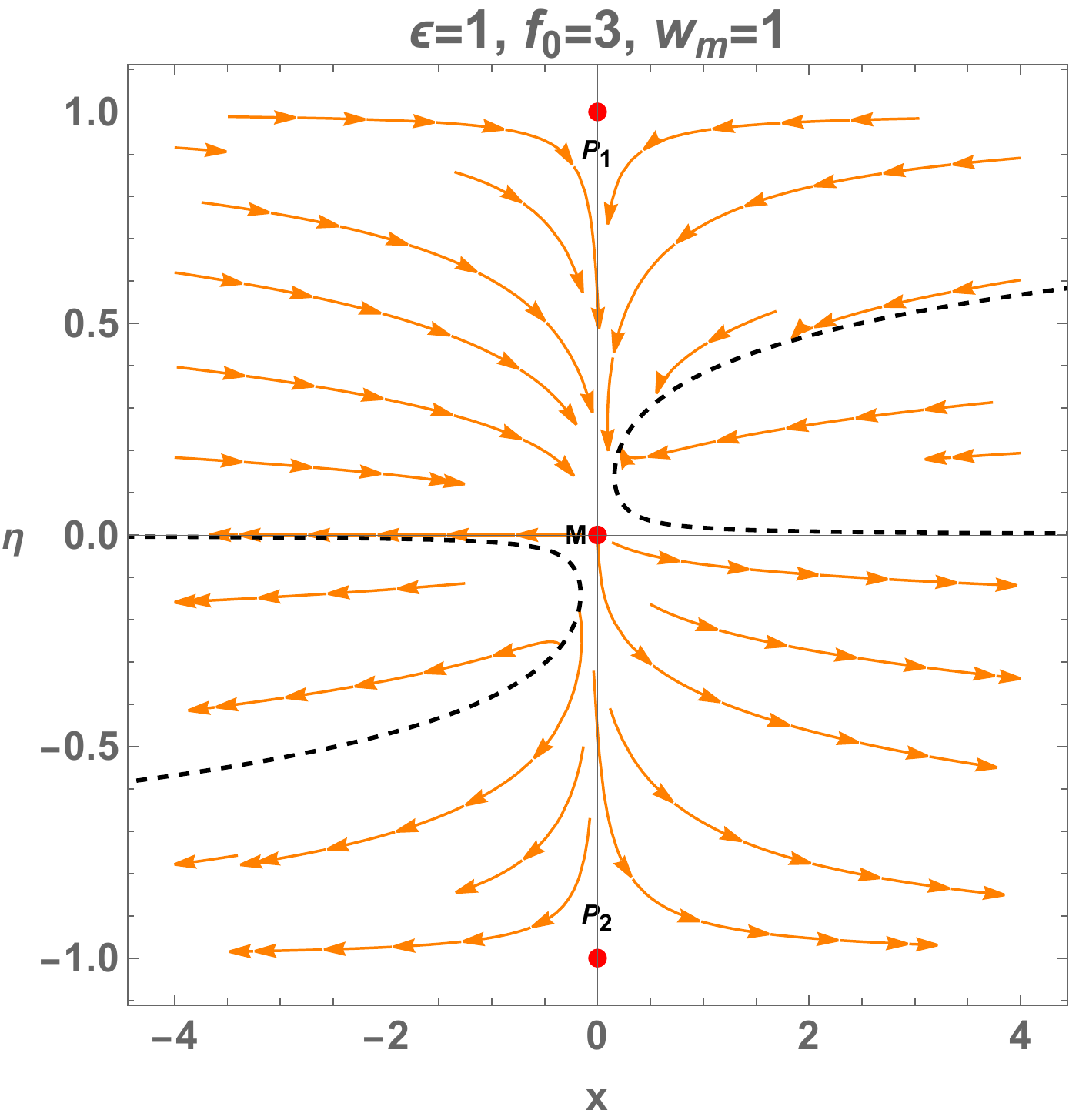}
       \caption{Phase portrait for \eqref{no-potential-1}-\eqref{no-potential-2} for $\epsilon=1,$ for $w_m=0, \frac{1}{3}, 1.$ The black-dashed lines correspond to the values where $K=0,$ which corresponds to singular curves where the flow direction and the stability change. }
       \label{fig:1}
   \end{figure}

   \begin{enumerate}
       \item $M=(0,0),$ with eigenvalues $\{0,0\}.$ The asymptotic solution is that of the Minkowski spacetime. 
       \item $P_{1,2}=(0,\pm 1),$ with eigenvalues $\{\pm 2,\pm (1-3 w_m)\}.$ The asymptotic solution described by $P_{1,2}$ is a universe dominated by the Gauss-Bonnet term. We also verify that $\omega_{\phi}=-\frac{1}{3},$ and $q=0.$  
       These points are 
       \begin{enumerate}
           \item $P_1$ is a source ($P_2$ is a sink) for $0\leq w_m<\frac{1}{3}$,
           \item saddles for $\frac{1}{3}< w_m\leq 1$,
           \item non-hyperbolic for $w_m=\frac{1}{3}.$
       \end{enumerate}
   \end{enumerate}
   In Fig. \ref{fig:1} we present a phase portrait of system \eqref{no-potential-1}-\eqref{no-potential-2} for $\epsilon=1,$ for the three values $w_m=0$ (dust), $\frac{1}{3}$ (radiation) and $1$ (stiff matter). A summary of the analysis performed in this section is given in Table \ref{tab:1}.
   \begin{table}[ht!]
    \caption{Equilibrium points of system \eqref{no-potential-1}, \eqref{no-potential-2}  for $\epsilon=+1$ with their stability conditions. Also includes the value of $\omega_{\phi}$ and $q.$}
    \label{tab:1}
\newcolumntype{C}{>{\centering\arraybackslash}X}
\centering
    \setlength{\tabcolsep}{4.6mm}
\begin{tabularx}{\textwidth}{cccccc}
\toprule 
  \text{Label}  & \; $x$& $\eta$ & \text{Stability}& $\omega_{\phi}$&$q$\\
  \midrule  
  $M$ & $0$ & $0$  & non-hyperbolic & \text{indeterminate} & \text{indeterminate}\\  \midrule 
  $P_{1}$ & $0$ & $1$ &        source for $0\leq w_m<1/3$ && \\
          &&& saddle for $1/3<w_m\leq 1$ && \\
          &&& non-hyperbolic for $w_m=1/3$ & $-\frac{1}{3}$& $0$\\  \midrule 
  $P_{2}$ & $0$ & $-1$ & sink for $0\leq w_m<1/3$ && \\
          &&& saddle for $1/3<w_m\leq 1$ && \\
          &&& non-hyperbolic for $w_m=1/3$ & $-\frac{1}{3}$& $0$\\  
\bottomrule
    \end{tabularx}
\end{table}
\FloatBarrier
   \subsection{Dynamical analysis for $\epsilon=-1$}
   \label{V-B}
The equilibrium points of system \eqref{no-potential-1}-\eqref{no-potential-2} for $\epsilon=-1$ in the coordinates $(x,y)$ are the same as in the previous section plus some additional points, the complete list is the following

\begin{enumerate}
    \item $M=(0,0),$ with eigenvalues$\{0,0\}.$ This is a Minkowski point, and the analysis is the same as in section \ref{V-A}.
    \item $P_1=(0,\pm 1),$ with eigenvalues  $\{\pm 2,\pm(1-3 w_m)\}.$ These are Gauss-Bonnet points; the analysis is the same as in section \ref{V-A}.
    \item $P_3=\Big(\frac{1}{\sqrt{20 \sqrt{\frac{10}{3}} f_0+5}},\frac{\sqrt{3}}{\sqrt{4 \sqrt{30} f_0+3}}\Big),$ with eigenvalues $\left\{-\frac{3 (w_m+1)}{\sqrt{4 \sqrt{\frac{10}{3}} f_0+1}},-\frac{9 \left(4 \sqrt{10} f_0+\sqrt{3}\right)}{\left(4 \sqrt{30} f_0+3\right)^{3/2}}\right\}.$ This point exists for $f_0\geq 0$ and it describes a de Sitter universe since $\omega_{\phi}=-1$ and $q=-1.$
   This point is a sink for all  $f_0\geq 0$ and $w_m\in [0,1].$
    \item $P_4=(-\frac{1}{\sqrt{20 \sqrt{\frac{10}{3}} f_0+5}},-\frac{\sqrt{3}}{\sqrt{4 \sqrt{30} f_0+3}}\Big),$ with eigenvalues $\left\{\frac{3
   (w_m+1)}{\sqrt{4 \sqrt{\frac{10}{3}} f_0+1}},\frac{9 \left(4 \sqrt{10} f_0+\sqrt{3}\right)}{\left(4 \sqrt{30} f_0+3\right)^{3/2}}\right\}.$ This point exists for $f_0\geq 0$ and it describes a de Sitter universe since $\omega_{\phi}=-1$ and $q=-1.$
   This point is a source for all  $f_0\geq 0$ and $w_m\in [0,1].$
    \item $P_5=\Big(\frac{1}{\sqrt{5-20 \sqrt{\frac{10}{3}} f_0}},-\frac{\sqrt{3}}{\sqrt{3-4 \sqrt{30}
   f_0}}\Big),$ with eigenvalues \newline $\left\{\frac{9 \left(\sqrt{\left(\sqrt{3}-4 \sqrt{10} f_0\right)^2} w_m-4 \sqrt{10} f_0
   (w_m+2)+\sqrt{3} (w_m+2)\right)}{2 \left(3-4 \sqrt{30} f_0\right)^{3/2}},-\frac{9
   \sqrt{\left(\sqrt{3}-4 \sqrt{10} f_0\right)^2} w_m+36 \sqrt{10} f_0 (w_m+2)-9 \sqrt{3}
   (w_m+2)}{2 \left(3-4 \sqrt{30} f_0\right)^{3/2}}\right\}.$ This point exists for $f_0\leq 0$ and it describes a de Sitter universe since $\omega_{\phi}=-1$ and $q=-1.$
   This point is a source for all  $f_0\leq 0$ and $w_m\in [0,1].$
    \item $P_6=\Big(-\frac{1}{\sqrt{5-20 \sqrt{\frac{10}{3}} f_0}},\frac{\sqrt{3}}{\sqrt{3-4 \sqrt{30}
   f_0}}\Big),$ with eigenvalues \newline $\left\{\frac{9 \sqrt{\left(\sqrt{3}-4 \sqrt{10} f_0\right)^2} w_m+36 \sqrt{10} f_0
   (w_m+2)-9 \sqrt{3} (w_m+2)}{2 \left(3-4 \sqrt{30} f_0\right)^{3/2}},-\frac{9
   \left(\sqrt{\left(\sqrt{3}-4 \sqrt{10} f_0\right)^2} w_m-4 \sqrt{10} f_0
   (w_m+2)+\sqrt{3} (w_m+2)\right)}{2 \left(3-4 \sqrt{30} f_0\right)^{3/2}}\right\}.$ This point exists for $f_0\leq 0$ and it describes a de Sitter universe since $\omega_{\phi}=-1$ and $q=-1.$
   This point is a sink for all  $f_0\leq 0$ and $w_m\in [0,1].$
\end{enumerate}

\begin{figure}[h!]
    \centering
    \includegraphics[scale=0.35]{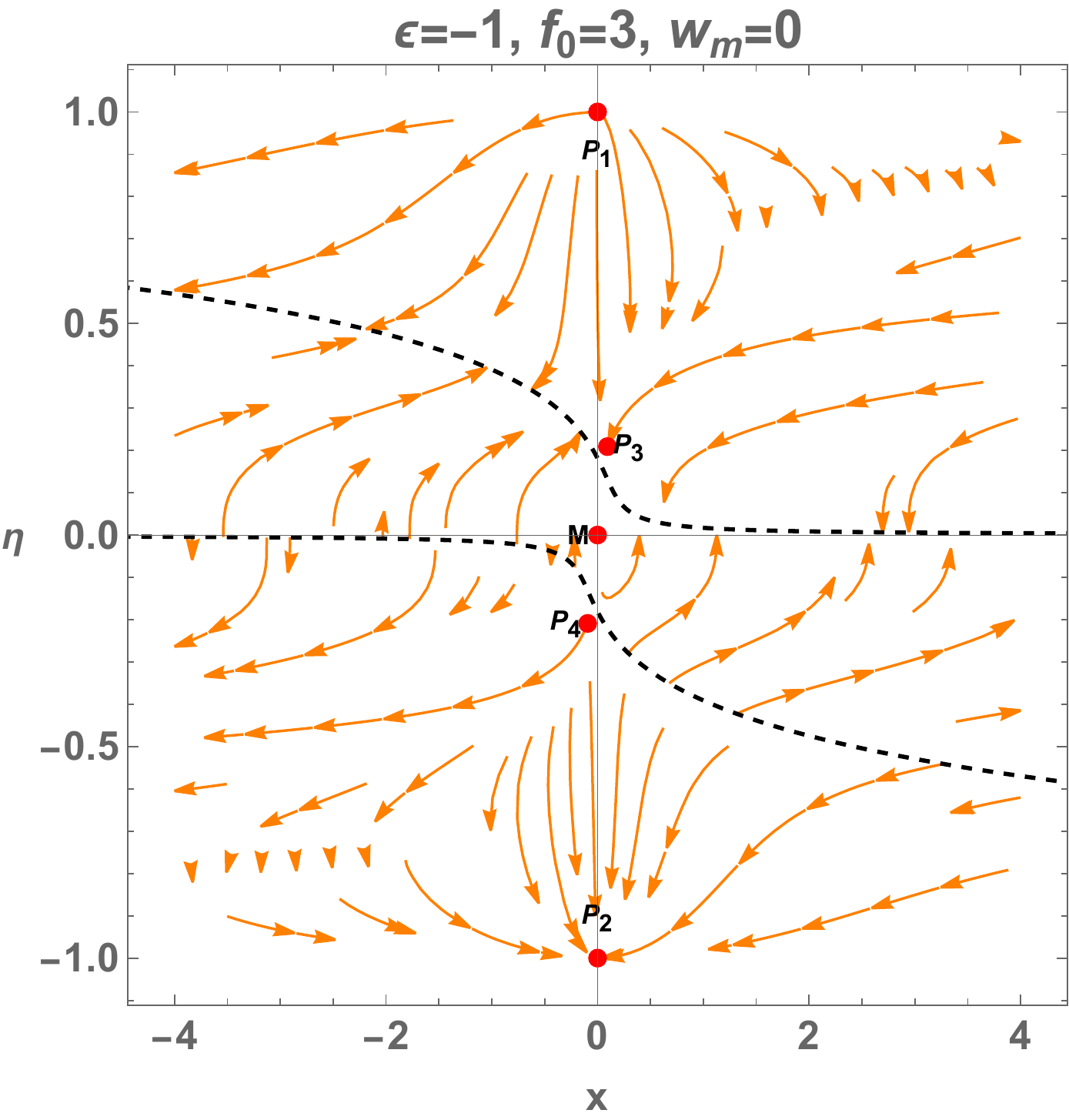}
    \includegraphics[scale=0.35]{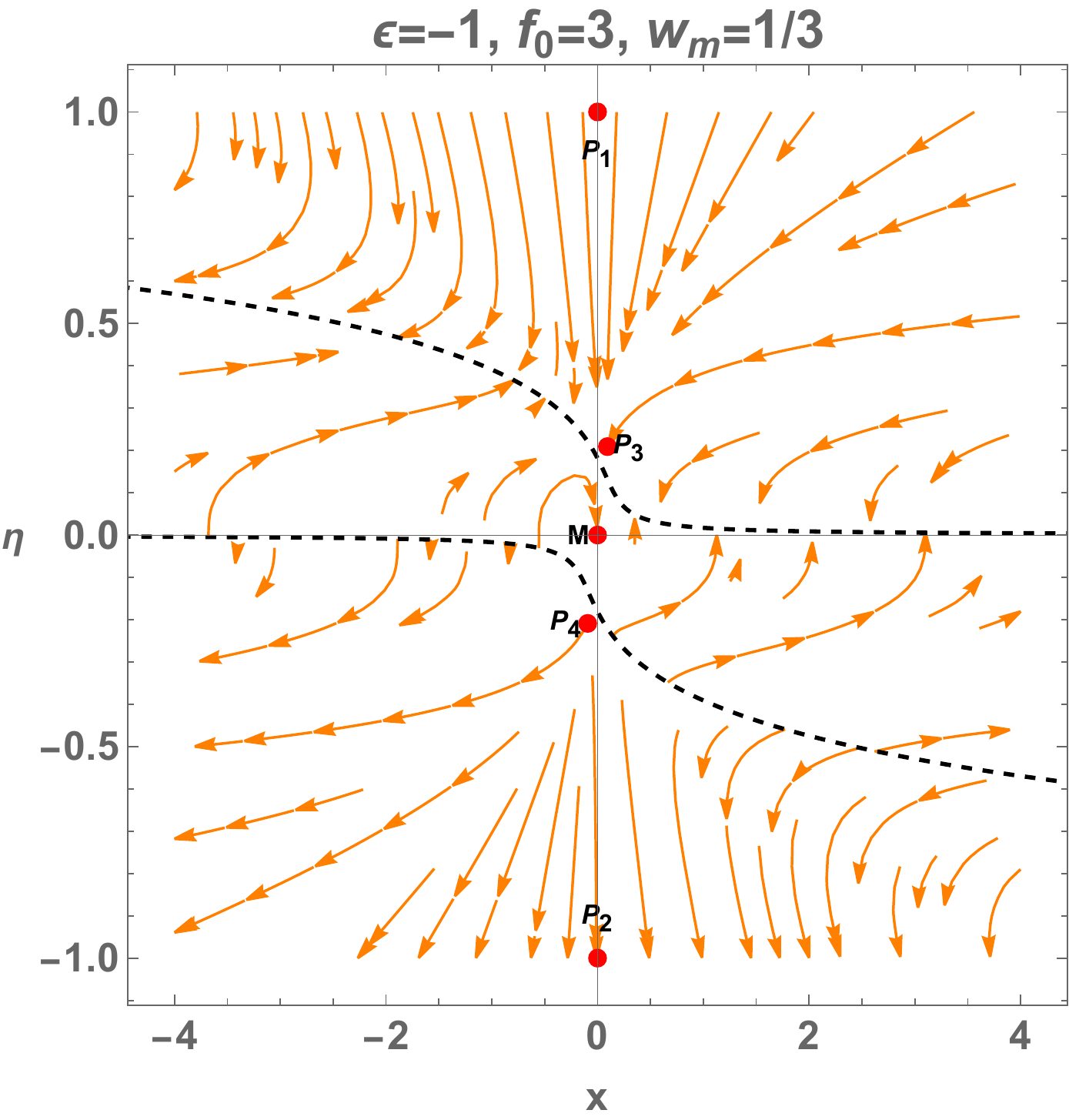}
    \includegraphics[scale=0.35]{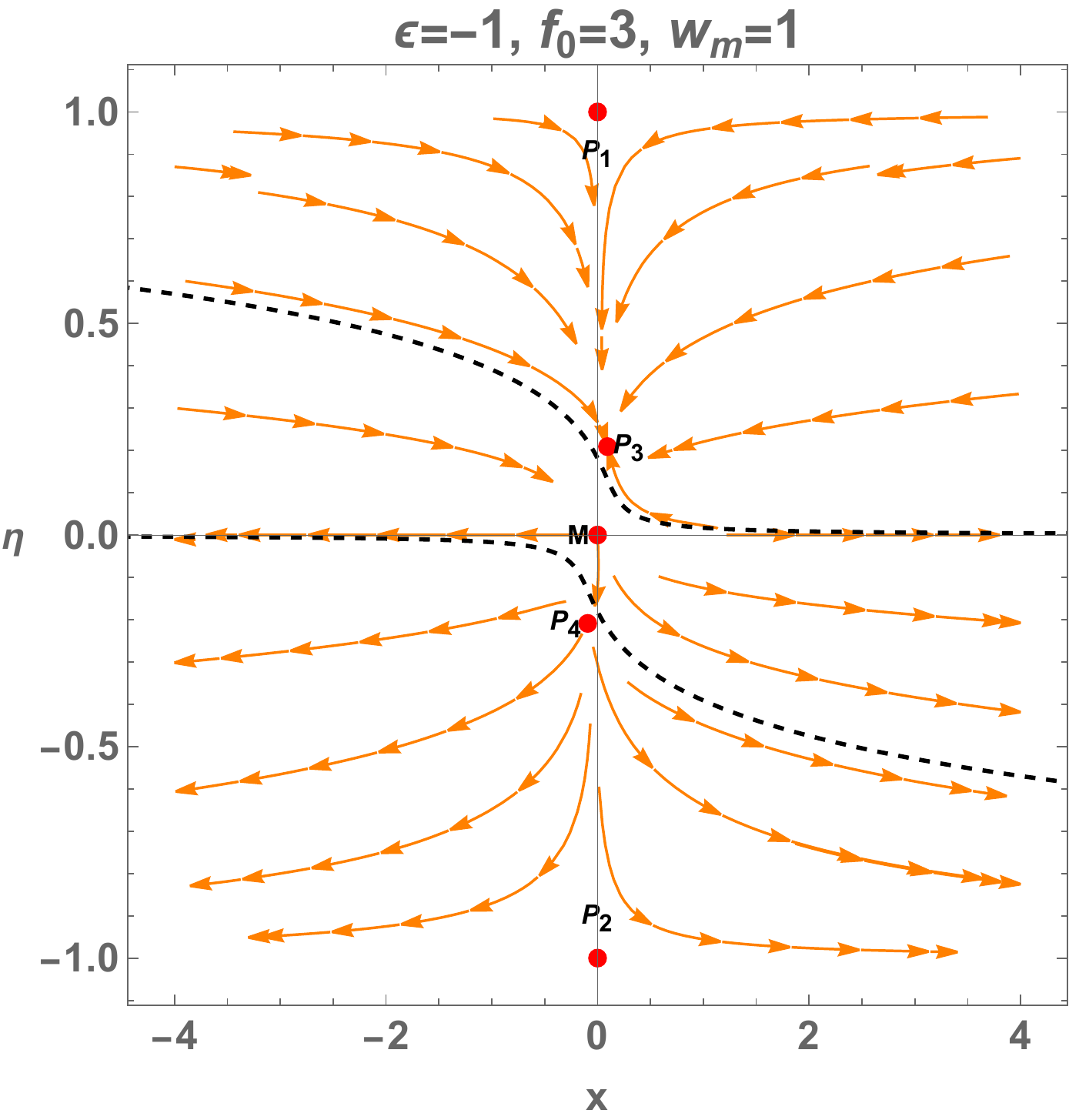}
    \includegraphics[scale=0.35]{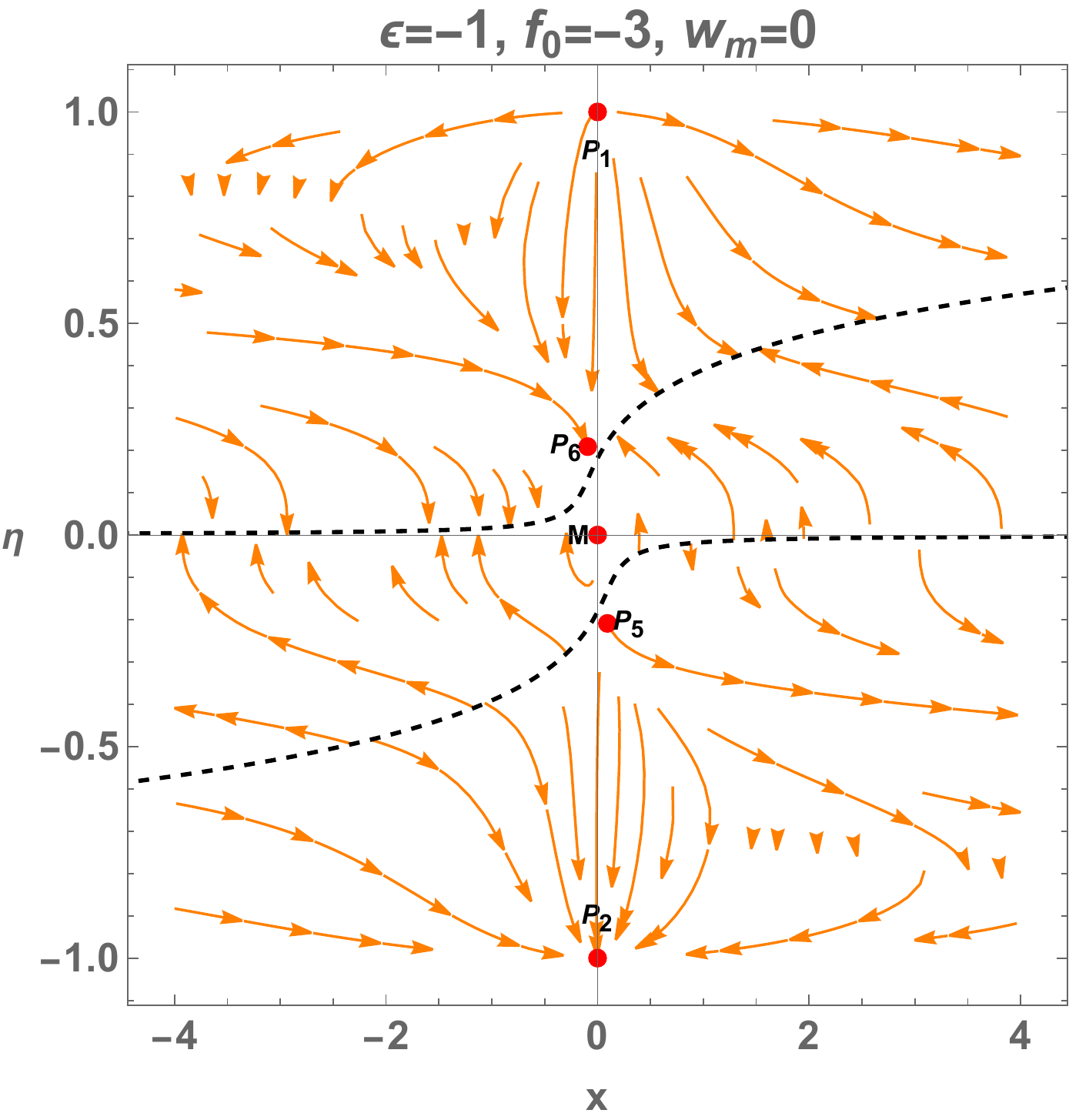}
    \includegraphics[scale=0.35]{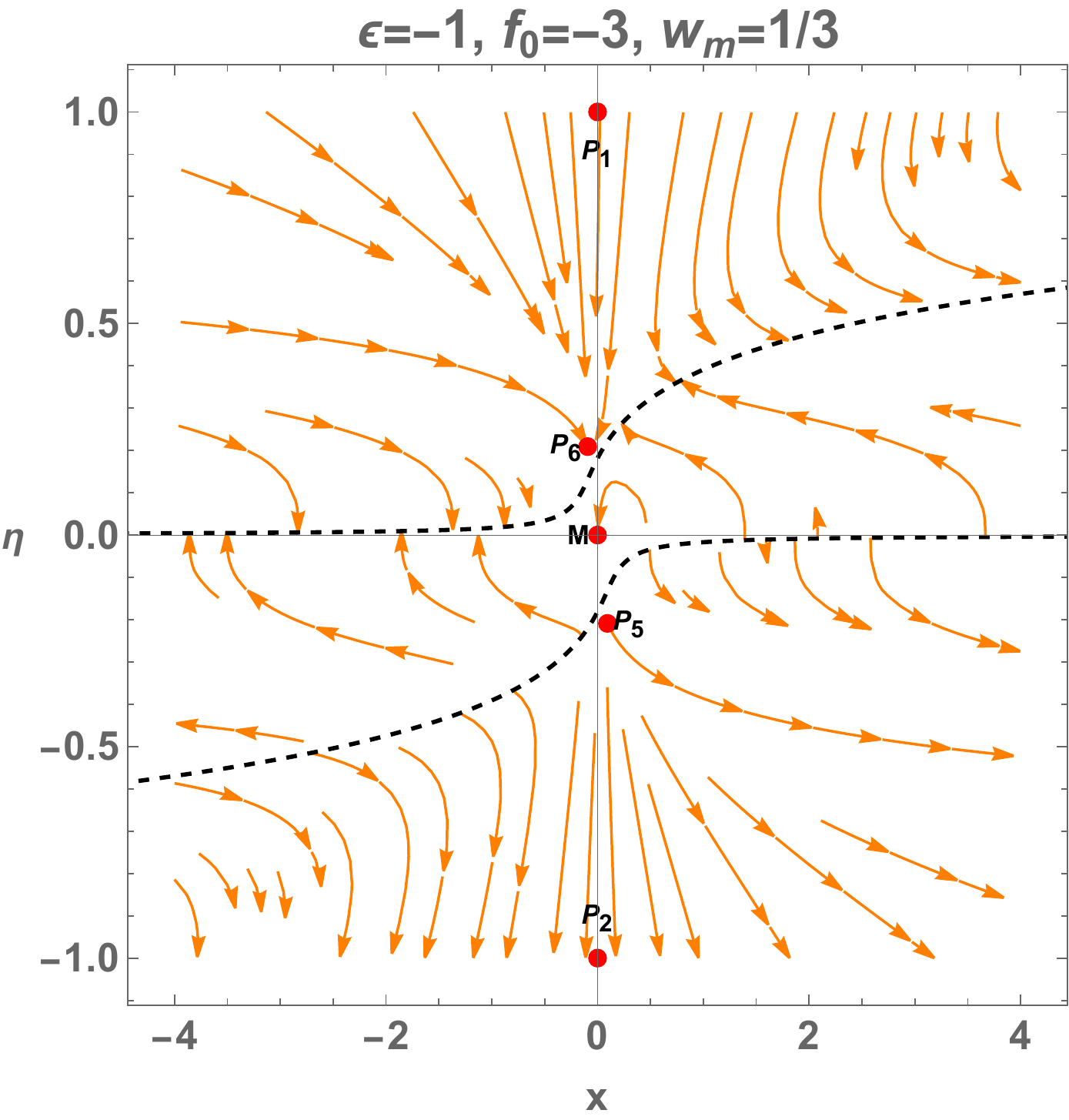}
    \includegraphics[scale=0.35]{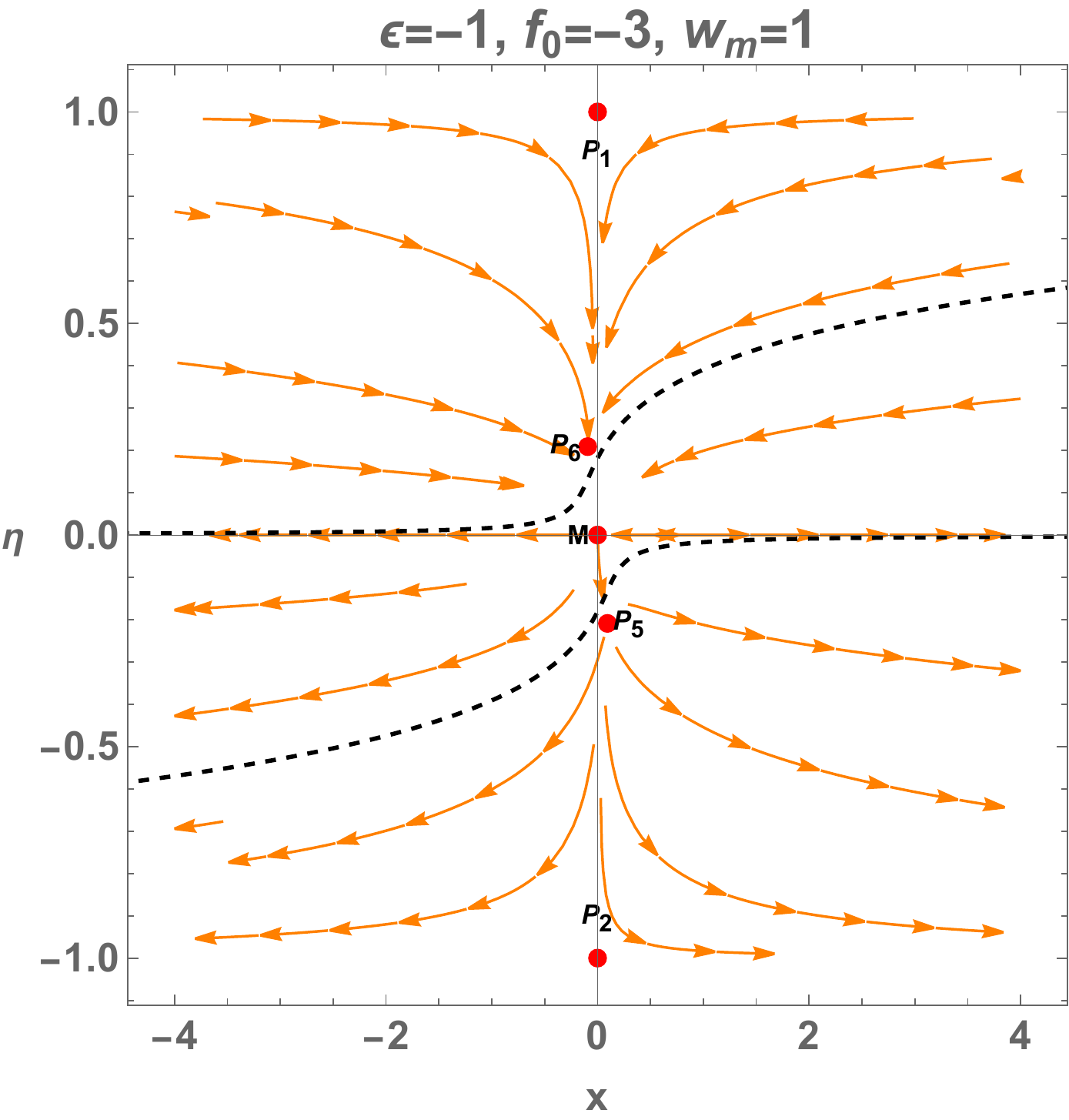}
    \caption{Phase portrait for \eqref{no-potential-1}-\eqref{no-potential-2} for $\epsilon=-1,$ for $w_m=0, \frac{1}{3}, 1$ and $f_0=\pm 3.$ The black-dashed lines correspond to the values where $K=0,$ which corresponds to singular curves where the flow direction and the stability change.}
    \label{fig:2}
\end{figure}

In Fig. \ref{fig:2}, we present different phase portraits for system \eqref{no-potential-1}-\eqref{no-potential-2} for $\epsilon=-1$ with two configurations: in the top row, we fix $f_0=3$ to show points $P_{3,4};$ we also show $P_{5,6}$ by setting $f_0=-3$ in the bottom row. The values for the EoS parameter used for the plots are $w_m=0$ (dust), $\frac{1}{3}$ (radiation) and $1$ (stiff matter). A summary of the analysis of this section is presented in Table \ref{tab:2}.

\begin{table}[ht!]
    \caption{Equilibrium points of system \eqref{no-potential-1}, \eqref{no-potential-2}  for $\epsilon=-1$ with their stability conditions. Also includes the value of $\omega_{\phi}$ and $q.$}
    \label{tab:2}
\newcolumntype{C}{>{\centering\arraybackslash}X}
\centering
    \setlength{\tabcolsep}{2.6mm}
\begin{tabularx}{\textwidth}{cccccc}
\toprule 
  \text{Label}  & \; $x$& $\eta$ & \text{Stability}& $\omega_{\phi}$&$q$\\
  \midrule  
  $M$ & $0$ & $0$  & non-hyperbolic & \text{indeterminate} & \text{indeterminate}\\  \midrule 
  $P_{1}$ & $0$ & $1$ & source for $0\leq w_m<1/3$ && \\
          &&& saddle for $1/3<w_m\leq 1$ && \\
          &&& non-hyperbolic for $w_m=1/3$ & $-\frac{1}{3}$& $0$\\  \midrule 
  $P_{2}$ & $0$ & $-1$ & sink for $0\leq w_m<1/3$ && \\
          &&& saddle for $1/3<w_m\leq 1$ && \\
          &&& non-hyperbolic for $w_m=1/3$ & $-\frac{1}{3}$& $0$\\  \midrule 
  $P_{3}$ & $\frac{1}{\sqrt{20 \sqrt{\frac{10}{3}} f_0+5}}$& $\frac{\sqrt{3}}{\sqrt{4 \sqrt{30} f_0+3}}$ & $\text{sink}$& $-1$ & $-1$\\  \midrule 
  $P_{4}$ & $-\frac{1}{\sqrt{20 \sqrt{\frac{10}{3}} f_0+5}}$& $-\frac{\sqrt{3}}{\sqrt{4 \sqrt{30} f_0+3}}$ & $\text{source}$& $-1$ & $-1$\\  \midrule 
  $P_{5}$ & $\frac{1}{\sqrt{5-20 \sqrt{\frac{10}{3}} f_0}}$ & $-\frac{\sqrt{3}}{\sqrt{3-4 \sqrt{30}
   f_0}}$  & $\text{source}$& $-1$& $-1$\\  \midrule 
  $P_{6}$ & $-\frac{1}{\sqrt{5-20 \sqrt{\frac{10}{3}} f_0}}$ & $\frac{\sqrt{3}}{\sqrt{3-4 \sqrt{30}
   f_0}}$  & $\text{sink}$& $-1$& $-1$\\
\bottomrule
    \end{tabularx}
\end{table}

   \FloatBarrier
   \subsection{ Infinity analysis for system \eqref{no-potential-1}-\eqref{no-potential-2} for $\epsilon=1$}\label{V-C}

The numerical results in sections \ref{V-A}-\ref{V-B} suggest non-trivial dynamics when $x \rightarrow \pm \infty$. For that reason, we introduce the compacted variable 
\begin{equation}\label{ucompact}
    u= \frac{x}{\sqrt{1 + x^2}}, 
\end{equation}
with inverse 
\begin{equation}
    x= \frac{u}{\sqrt{1 -u^2}}, 
\end{equation}
and the new time variable 
\begin{equation}
  {f}^{\prime} =  \sqrt{1 - u^2} \frac{df}{d \tau}.
\end{equation}

Note that for $u \rightarrow \pm 1$ we have dynamics for $x\rightarrow \infty.$
 
Using the compacted variable \eqref{ucompact} together with system \eqref{no-potential-1}-\eqref{no-potential-2} we obtain the following compacted dynamical system 
\begin{align}
    &u'=\frac{\eta  \left(1-u^2\right)}{K_1} \Bigg\{3 \left(1-u^2\right) u \left(64 f_0^2 \eta ^4 \left(\eta ^2-3
   w_m\right)+\left(\eta ^2-1\right)^2 \epsilon  \left(\eta ^2 (w_m+1)-2\right)\right) \nonumber \\
   & -4 \sqrt{6} f_0 \eta ^3 \left(\eta ^2-1\right) \left(1-u^2\right)^{3/2} (3 w_m+1) \nonumber \\
   & +4 f_0 \eta  \left(\eta ^2-1\right) \sqrt{6-6 u^2} u^2 \epsilon  \left(\eta ^2 (6 w_m-2)+3 (w_m-5)\right)  -3 \left(\eta ^2-1\right)^2 u^3 (w_m-1)\Big\}, \label{compact-1}\\
    &\eta'=\frac{\left(\eta ^2-1\right)}{K_1}  \Bigg\{3 \eta ^2 \left(1-u^2\right)
   \left(64 f_0^2 \eta ^4+\left(\eta ^2-1\right)^2 (w_m+1) \epsilon \right) \nonumber \\
   & +8 f_0 \left(\eta
   ^2-1\right) \eta ^3 u \sqrt{6-6 u^2} (3 w_m-1) \epsilon -3 \left(\eta ^2-1\right)^2 u^2
   (w_m-1)\Bigg\}, \label{compact-2}
\end{align}

where $K_1:=K_1(u,\eta,\epsilon,f_0)=2 \left(\sqrt{1-u^2} \left(96 f_0^2 \eta ^4+\left(\eta ^2-1\right)^2 \epsilon \right)+8 \sqrt{6} f_0 \eta
    \left(\eta ^2-1\right) u \epsilon \right)$.

   Setting $\epsilon=1$ in system \eqref{compact-1}-\eqref{compact-2} and re-scaling the system dividing by $\sqrt{1-u^2},$ the equilibrium points are the same ones as in section \ref{V-A} plus new points at infinity that satisfy $u=\pm 1.$ We present first the new points followed by the points from the finite regime. 
\begin{enumerate}
    \item $Q_{1,2}=(1,\pm 1),$ with eigenvalues $\{\pm 2,\pm 2(3w_m-1)\}.$ These points describe a universe dominated by the Gauss-Bonnet term; we also verify that $\omega_{\phi}=-\frac{1}{3}$ and  $q=0.$ These points are  
    \begin{enumerate}
          \item $Q_1$ is a source ($Q_2$ is a sink) for $\frac{1}{3}< w_m\leq 1,$
           \item a saddle for $0\leq w_m<\frac{1}{3},$
           \item non-hyperbolic for $w_m=\frac{1}{3}.$
       \end{enumerate}
    \item $Q_{3,4}=(-1,\pm 1),$ with eigenvalues $\{\pm 2,\pm 2(3w_m-1)\}.$ These are Gauss-Bonnet points, and the analysis is the same as $Q_1$ and $Q_2$ respectively.
    \item $M=(0,0),$ see section \ref{V-A}.
    \item $P_1=(0,1),$ see section \ref{V-A}.
    \item $P_2=(0,-1),$ see section \ref{V-A}.
\end{enumerate}

\begin{figure}[h!]
    \centering
    \includegraphics[scale=0.3]{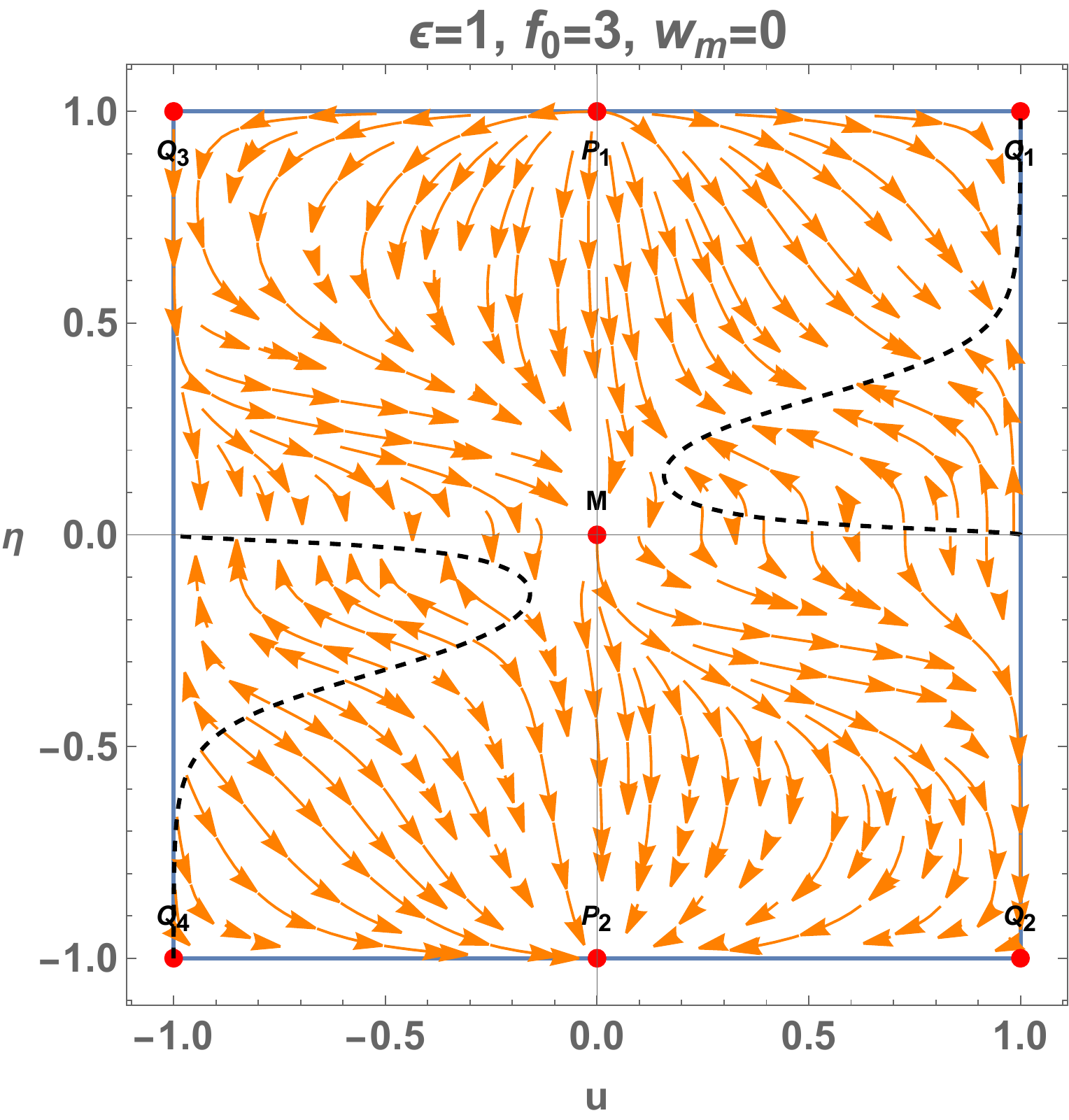}
    \includegraphics[scale=0.3]{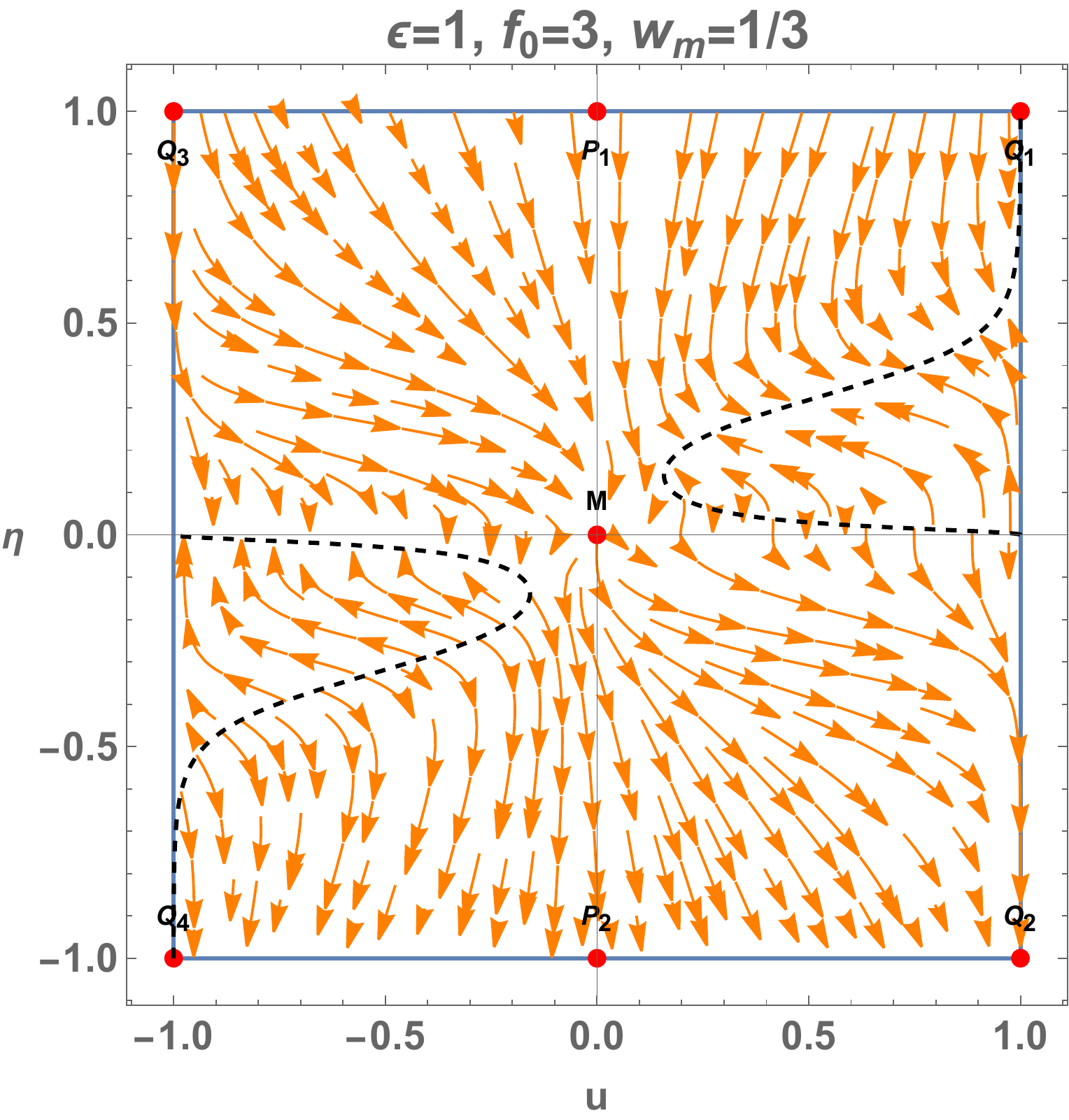}
    \includegraphics[scale=0.3]{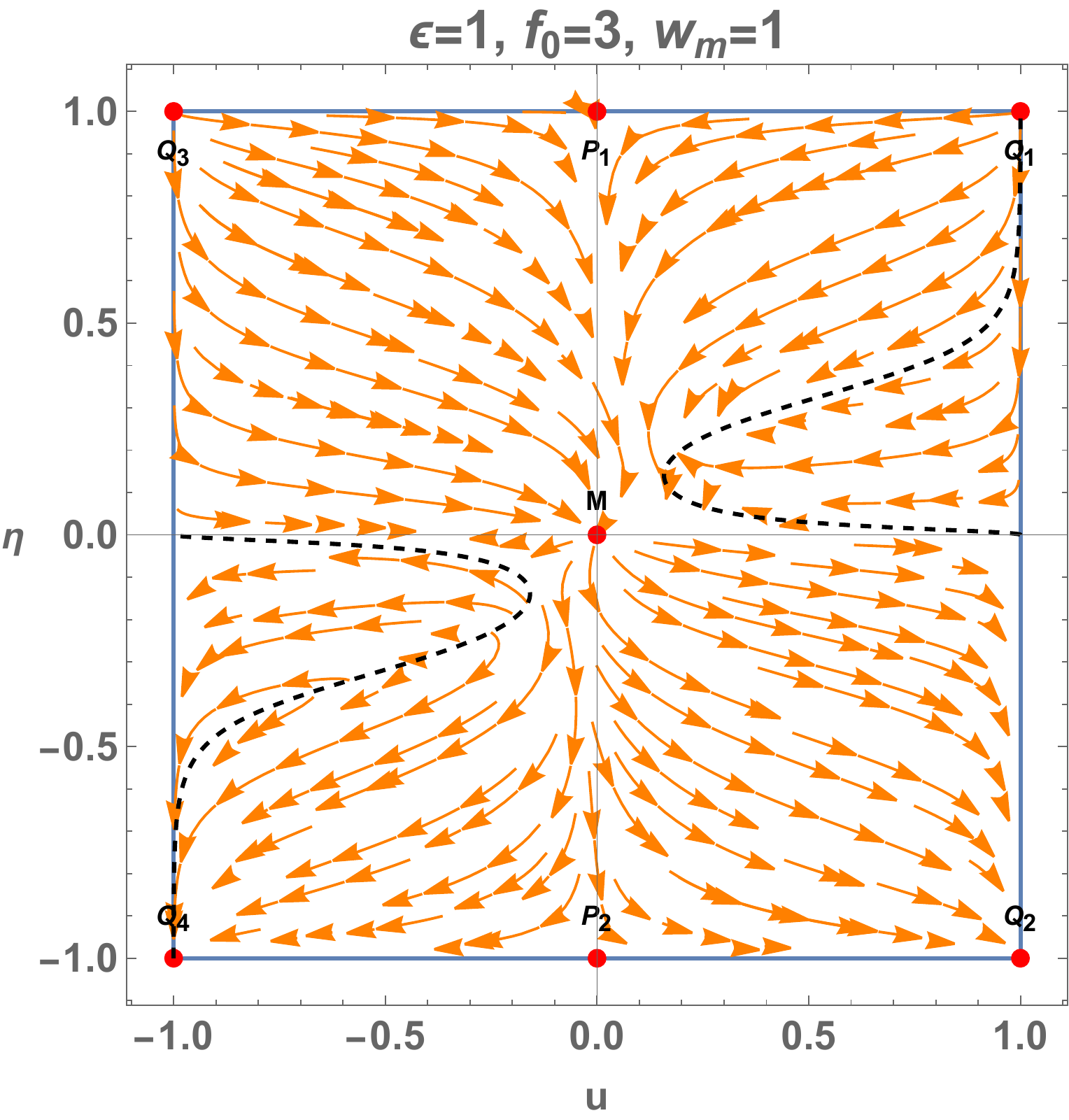}
    \caption{Phase-plot analysis for system \eqref{compact-1}, \eqref{compact-2} for $\epsilon=1$ and $f_0=3.$ We also consider the cases $w_m=0, \frac{1}{3}, 1.$ The dashed black lines correspond to singularities where the flow changes direction.}
    \label{fig:3}
\end{figure}

In Fig. \ref{fig:3}, we present various phase portraits for system \eqref{compact-1}-\eqref{compact-2} for $\epsilon=1$ and different values of the EoS parameter $w_m=0$ (dust), $\frac{1}{3}$ (radiation) and $1$ (stiff matter). These plots contain the finite regime points $M$ and $P_i$, and the infinite regime points $Q_i$. In Table \ref{tab:3}, we present a summary of the stability analysis only for the points on the infinite regime; this table can be complemented with the information from Table \ref{tab:1}.

\begin{table}[ht!]
    \caption{Equilibrium points of system \eqref{compact-1}-\eqref{compact-2} for $\epsilon=\pm 1$ with their stability conditions. Also includes the value of $\omega_{\phi}$ and $q.$}
    \label{tab:3}
\newcolumntype{C}{>{\centering\arraybackslash}X}
\centering
    \setlength{\tabcolsep}{7.6mm}
\begin{tabularx}{\textwidth}{cccccc}
\toprule 
  \text{Label}  & \; $x$& $\eta$ & \text{Stability}& $\omega_{\phi}$&$q$\\
  \midrule  
  $Q_1$ & $1$ & $1$  & saddle for $0\leq w_m<1/3$ && \\
          &&& source for $1/3<w_m\leq 1$ && \\
          &&& non-hyperbolic for $w_m=1/3$  & $-\frac{1}{3}$ & $0$\\  \midrule 
 $Q_2$ & $1$ & $-1$  & saddle for $0\leq w_m<1/3$ && \\
          &&& sink for $1/3<w_m\leq 1$ && \\
          &&& non-hyperbolic for $w_m=1/3$  & $-\frac{1}{3}$ & $0$\\  \midrule 
  $Q_3$ & $-1$ & $1$  & saddle for $0\leq w_m<1/3$ && \\
          &&& source for $1/3<w_m\leq 1$ && \\
          &&& non-hyperbolic for $w_m=1/3$  & $-\frac{1}{3}$ & $0$\\   \midrule
           $Q_4$ & $-1$ & $-1$  & saddle for $0\leq w_m<1/3$ && \\
          &&& sink for $1/3<w_m\leq 1$ && \\
          &&& non-hyperbolic for $w_m=1/3$  & $-\frac{1}{3}$ & $0$\\ 
\bottomrule
    \end{tabularx}
\end{table}

   \FloatBarrier
      \subsection{Infinity analysis for system \eqref{no-potential-1}-\eqref{no-potential-2} for $\epsilon=-1$}\label{V-D}

  Setting $\epsilon=-1$ in system \eqref{compact-1}-\eqref{compact-2}, the equilibrium points are the same ones as in section \ref{V-B} plus new points at infinity that satisfy $u=\pm 1.$ As before, we present first the new points followed by the points from the finite regime.
  
\begin{figure}[h!]
    \centering
    \includegraphics[scale=0.3]{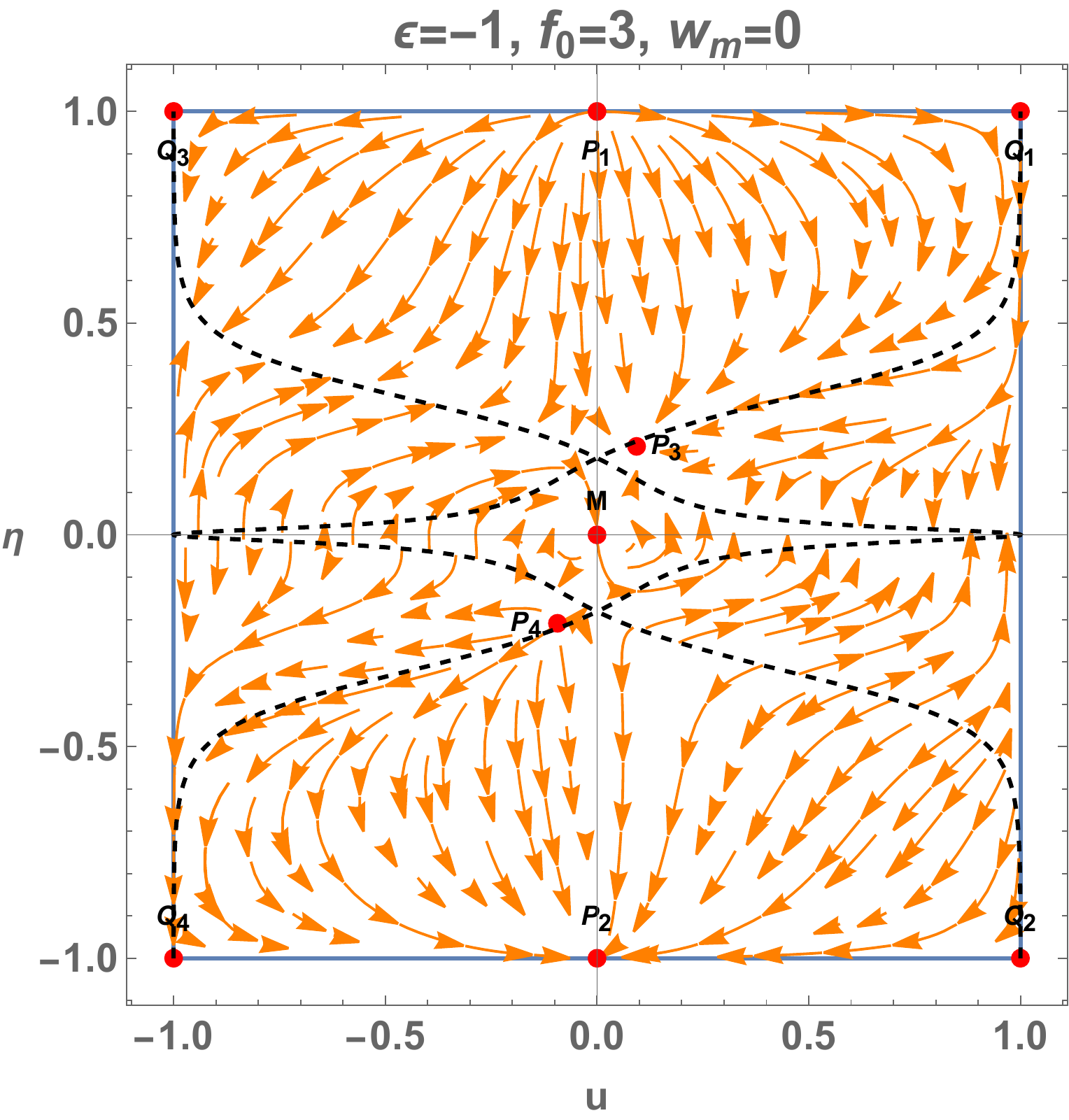}
    \includegraphics[scale=0.3]{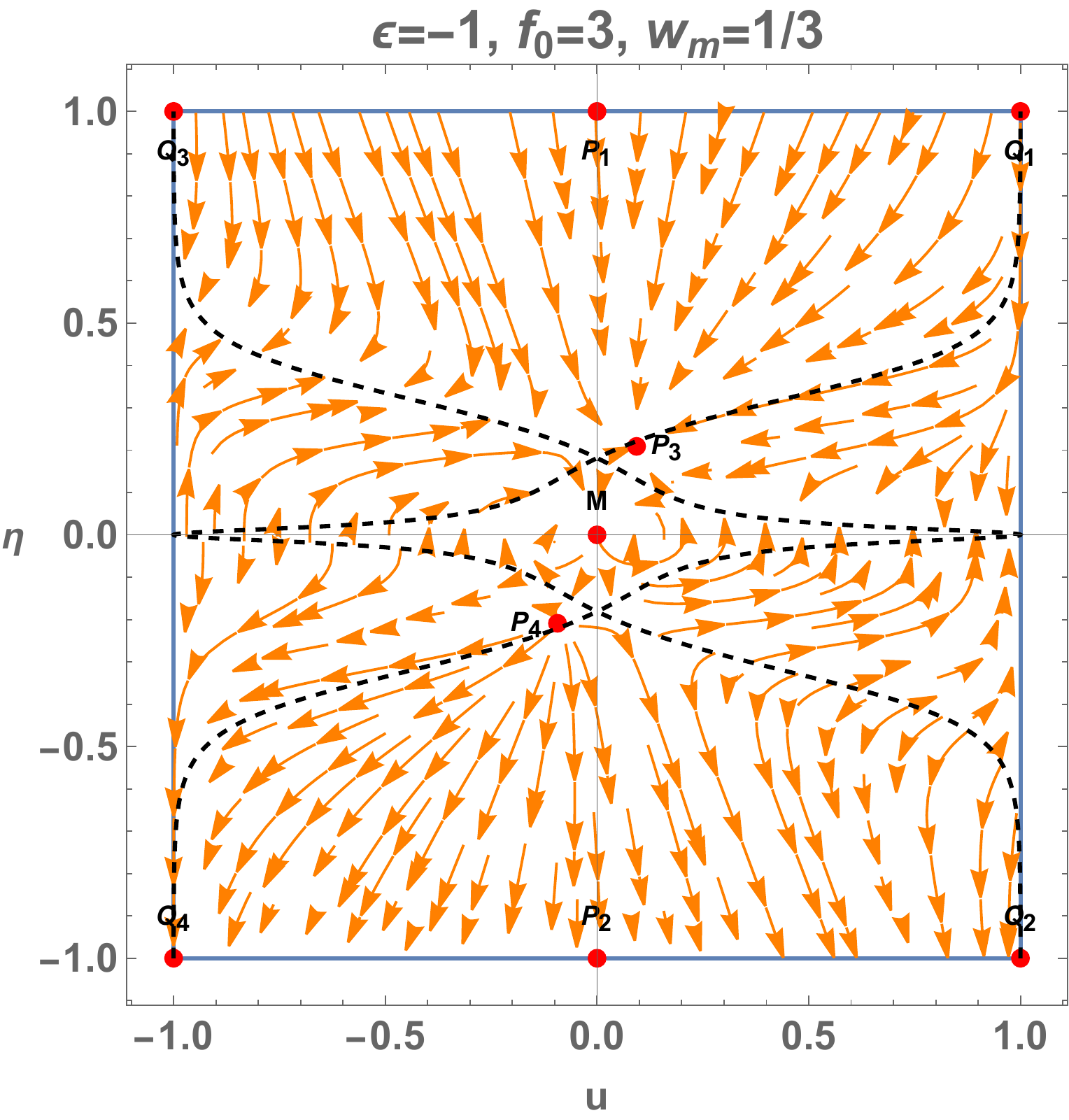}
    \includegraphics[scale=0.3]{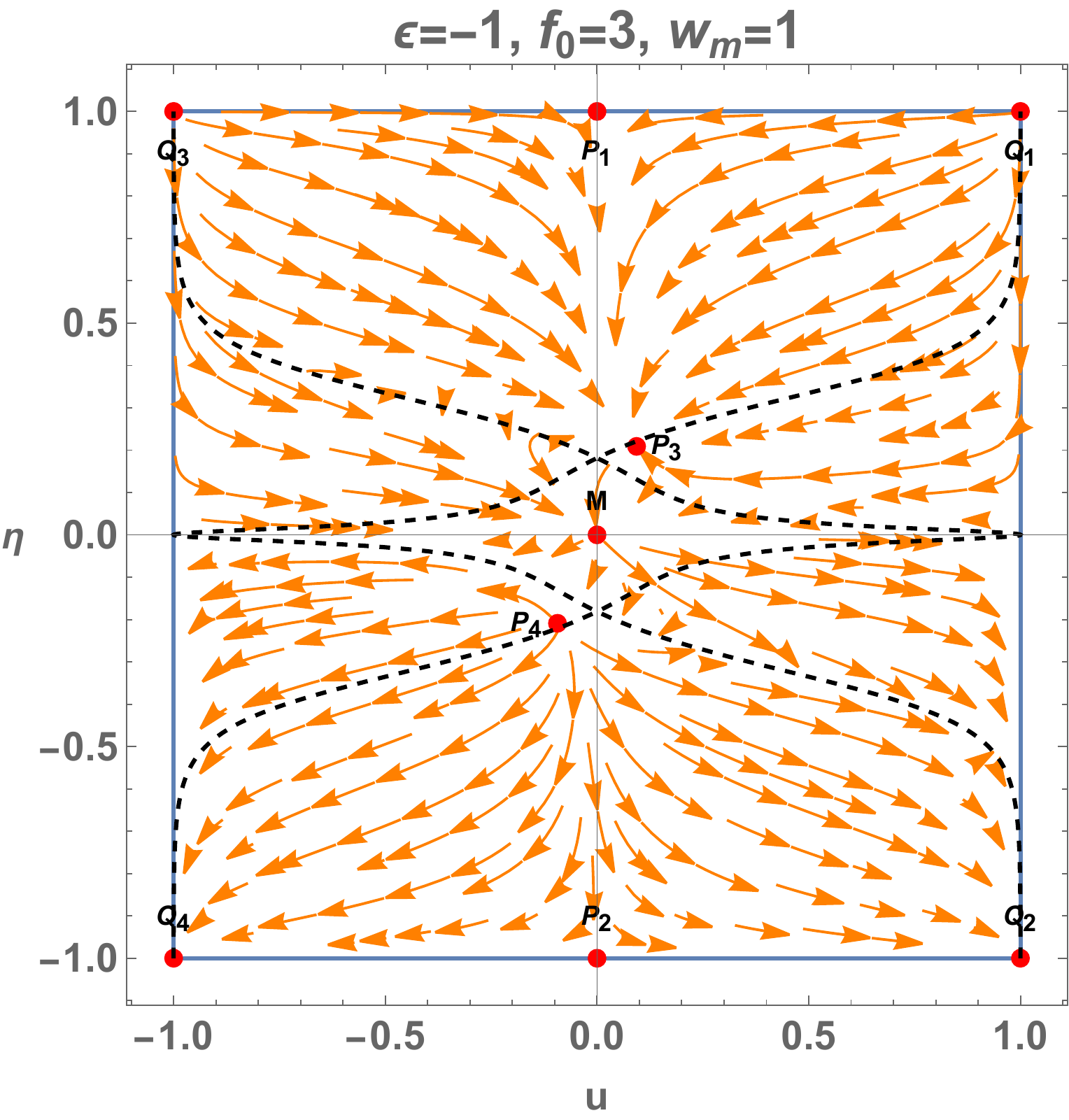}
    \includegraphics[scale=0.3]{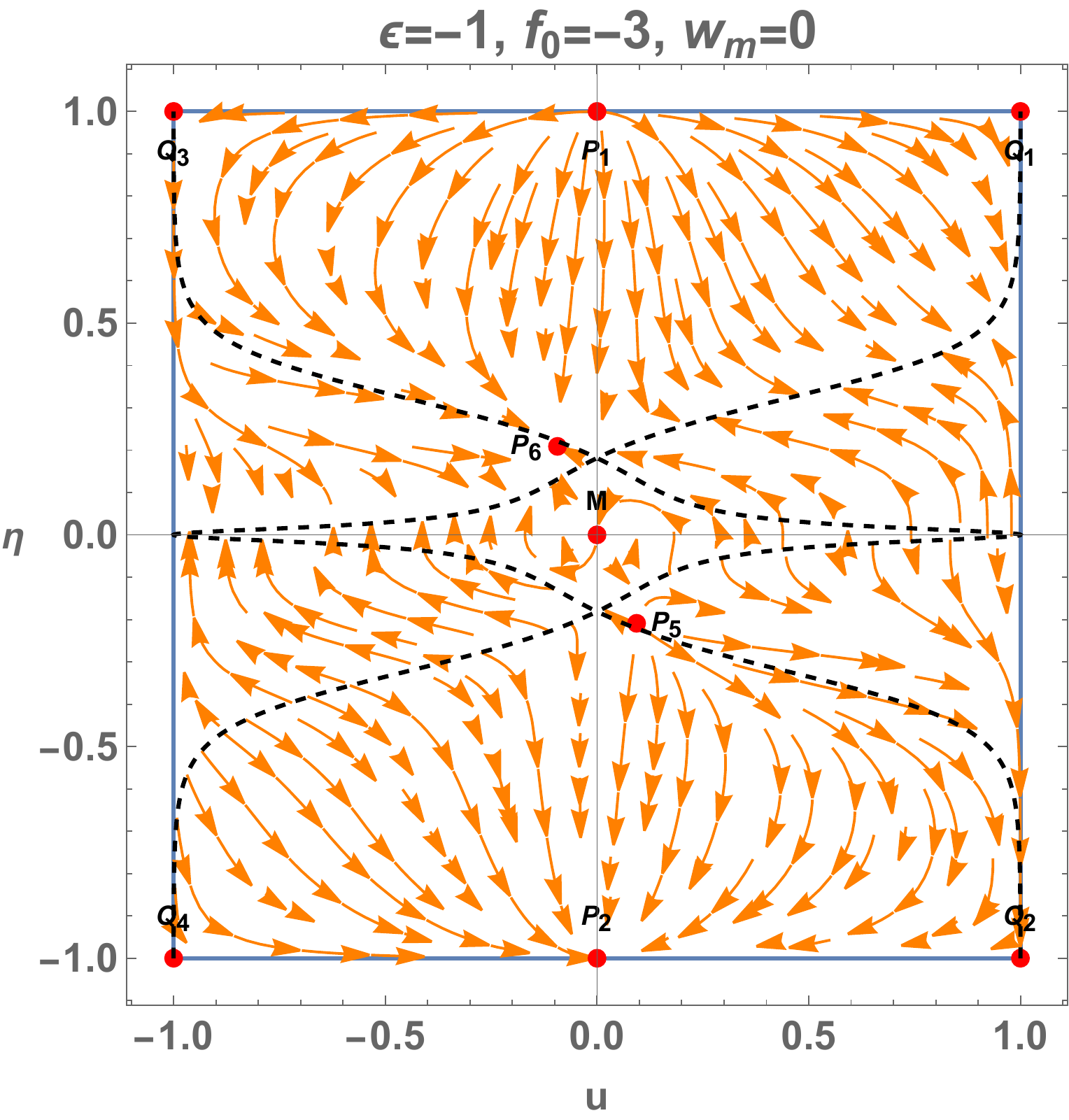}
    \includegraphics[scale=0.3]{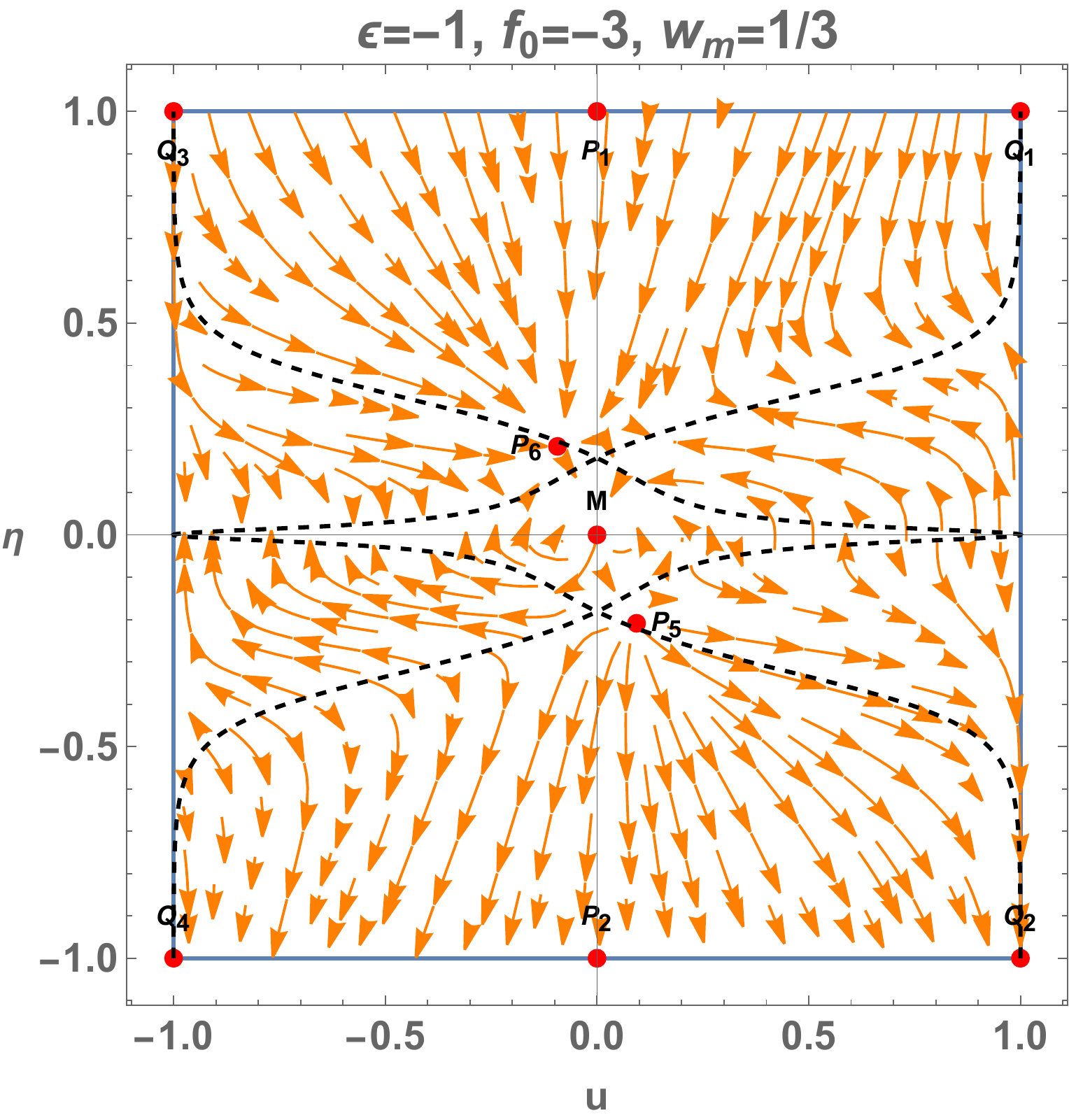}
    \includegraphics[scale=0.3]{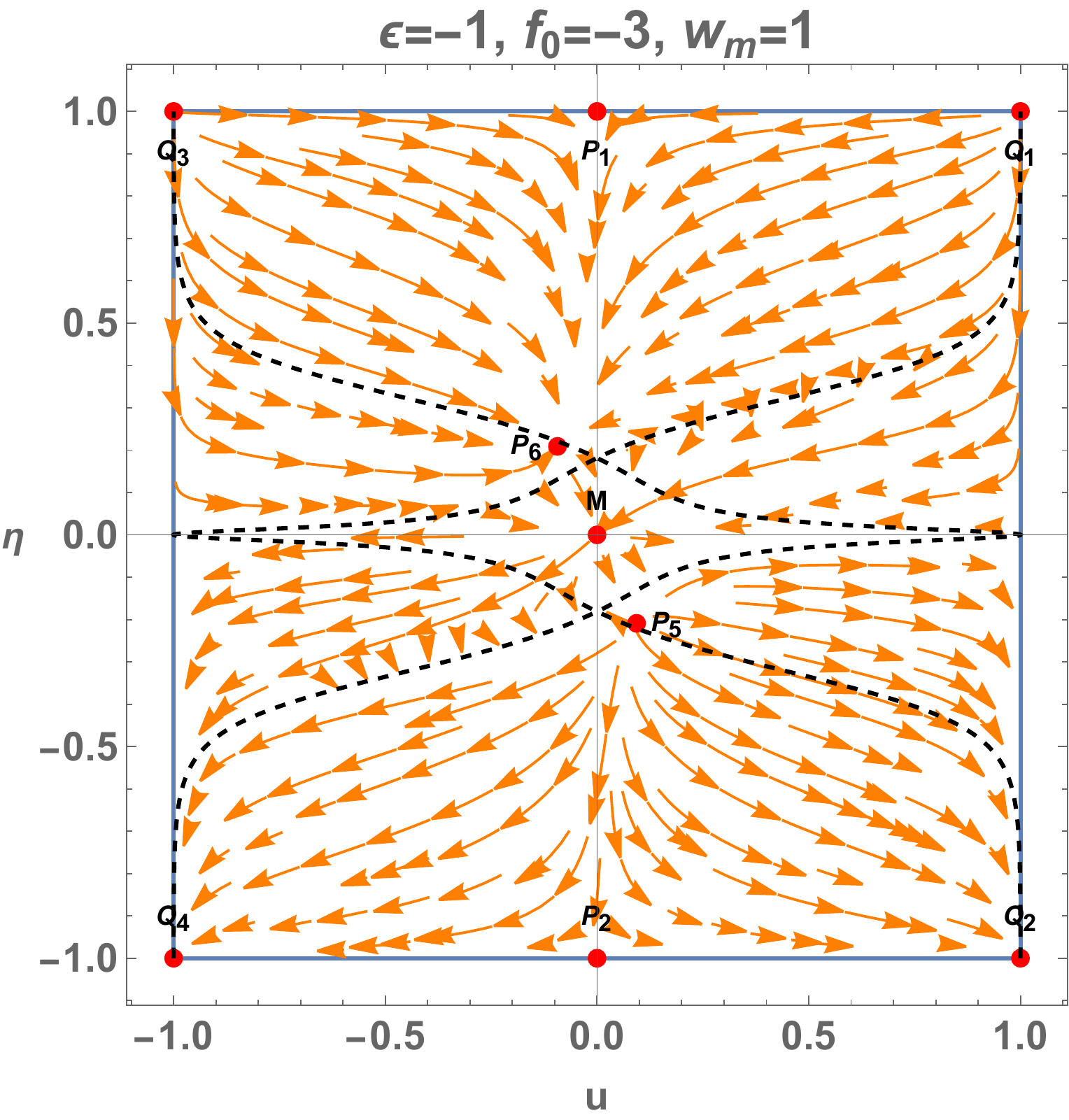}
    \caption{Phase-plot analysis for system \eqref{compact-1}, \eqref{compact-2} for $\epsilon=-1$ and $f_0=3.$ We also consider the cases $w_m=0, \frac{1}{3}, 1.$ The dashed black lines correspond to singularities where the flow changes direction.}
    \label{fig:4}
\end{figure}

\begin{enumerate}
    \item $Q_{1,2}=(1,\pm 1),$ with eigenvalues $\{\pm 2,\pm 2(3w_m-1)\}.$ These points are Gauss-Bonnet points, and the analysis is the same as in section \ref{V-C}.
    \item $Q_{3,4}=(-1,\pm 1),$ with eigenvalues $\{\pm 2,\pm 2(3w_m-1)\}.$ These points are Gauss-Bonnet points, and the analysis is the same as in section \ref{V-C}.
    \item $M=(0,0),$ see section \ref{V-A}.
    \item $P_1=(0,1),$ see section \ref{V-A}.
    \item $P_2=(0,-1),$ see section \ref{V-A}.
    \item $P_3=\Big(\frac{1}{\sqrt{20 \sqrt{\frac{10}{3}} f_0+5}},\frac{\sqrt{3}}{\sqrt{4 \sqrt{30} f_0+3}}\Big),$ see section \ref{V-B}.
    \item $P_4=\Big(-\frac{1}{\sqrt{20 \sqrt{\frac{10}{3}} f_0+5}},-\frac{\sqrt{3}}{\sqrt{4 \sqrt{30} f_0+3}}\Big),$ see section \ref{V-B}.
    \item $P_5=(0,-1),$ see section \ref{V-B}.
    \item $P_6=(0,-1),$ see section \ref{V-B}.
\end{enumerate}
      In Fig. \ref{fig:4}, we present various phase portraits for system \eqref{compact-1}-\eqref{compact-2} for $\epsilon=-1$ different values of the EoS parameter $w_m=0$ (dust), $\frac{1}{3}$ (radiation) and $1$ (stiff matter). As before, we set two values for $f_0$ to show the points $P_{3,4}$ and $P_{5,6}.$ These plots contain the finite regime points $M$ and $P_i$ as well as the infinite regime points $Q_i$. Note that the infinite regime points are the same for both values of $\epsilon,$; therefore, we present the summary of the stability analysis in Table \ref{tab:3} once again, but in this case, the information can be complemented with Table \ref{tab:2}.

   \FloatBarrier
    
\section{Conclusions}
\label{con}
In this paper, we have extended our previous study \cite{dn13} by introducing an ideal gas which can describe the radiation, dark matter, or dark energy, depending on the equation of state, in the Einstein-Gauss-Bonnet scalar field model in a four-dimensional cosmology. In addition, we performed a detailed analysis of the phase space and reconstructed the asymptotic behaviour of the physical parameters.

New dimensionless variables different from that of the $H$-normalization have been introduced. We wrote the field equations in the equivalent form of a four-dimensional algebraic-differential system of first-order equations. Because of the algebraic constraint, the dimension of the latter system is reduced to three. Moreover, for $\rho_m=0$, we recover the two-dimensional system investigated in \cite{dn13}.

We determined the equilibrium points for the field equations in the finite and infinite regimes. For the latter, we define a set of compact variables. Then, we calculated the asymptotic behaviour of the physical parameters for each equilibrium point. For the linear coupling between the scalar field and the Gauss-Bonnet component, asymptotic solutions exist that describe the de Sitter spacetime or a universe dominated by the Gauss-Bonnet scalar. 

We have shown that the stability properties of the equilibrium points depend on the nature of the ideal gas (its equation of state parameter), the scalar field and the scalar $f_0$ of the coupling function for the Gauss-Bonnet term. 

For the general case described in section \ref{III-B} with $\epsilon=1$ we obtained the following results: the Gauss-Bonnet point $P_1$ is a source for $0\leq w_m<\frac{1}{3}$; the Gauss-Bonnet point $P_2$ is a sink for $0\leq w_m<\frac{1}{3}.$
For the general case described in section \ref{III-C} with $\epsilon=-1$ we obtained the following results: the Gauss-Bonnet point $P_1$ is a source for $0\leq w_m<\frac{1}{3}$; the Gauss-Bonnet point $P_2$ is a sink for $0\leq w_m<\frac{1}{3}$; 
the de Sitter point $P_3$ is a sink for $f_0<0$ and $0<\lambda <\sqrt{\frac{2}{3}}$; the de Sitter point  $P_4$ is a source for $f_0<0$ and $0<\lambda <\sqrt{\frac{2}{3}}$; the de Sitter point $P_5$ is a sink for $f_0\geq 0$ and $\lambda<0$; the de Sitter point $P_6$ is a source for $f_0\geq 0$ and $\lambda<0$; the de Sitter point $P_7$ is a source for $f_0\leq 0$ and $\lambda>0$; the de Sitter point $P_8$ is a sink for $f_0\leq 0$ and $\lambda>0$;

The numerical results suggested that there must be non-trivial dynamics at infinity; given this, in section \ref{IV-A}, we investigated the infinity behaviour for $\epsilon=1$ and obtained the following results:
the Gauss-Bonnet point $Q_1$ is a source for $\lambda>0$ and $\frac{1}{3}<w_m\leq 1$; the Gauss-Bonnet point $Q_2$ is a sink for $\lambda>0$ and $\frac{1}{3}<w_m\leq 1$; the Gauss-Bonnet point $Q_3$ is a source for $\lambda<0$ and $\frac{1}{3}<w_m\leq 1$;  the Gauss-Bonnet point $Q_4$ is a sink for $\lambda<0$ and $\frac{1}{3}<w_m\leq 1$.

Similarly we obtained the following results for $\epsilon=-1$ in  section \ref{IV-B} for the behaviour at infinity:
the Gauss-Bonnet point $Q_1$ is a source for $\lambda>0$ and $\frac{1}{3}<w_m\leq 1$; the Gauss-Bonnet point $Q_2$ is a sink for $\lambda>0$ and $\frac{1}{3}<w_m\leq 1$; the Gauss-Bonnet point $Q_3$ is a source for $\lambda<0$ and $\frac{1}{3}<w_m\leq 1$;  the Gauss-Bonnet point $Q_4$ is a sink for $\lambda<0$ and $\frac{1}{3}<w_m\leq 1$;  the Gauss-Bonnet point $Q_5$ is a source for $\lambda<0$; the Gauss-Bonnet point $Q_6$ is a sink for $\lambda>0$; the Gauss-Bonnet point $Q_7$ is a source for $\lambda>0$; the Gauss-Bonnet point $Q_8$ is a sink for $\lambda<0$; the Gauss-Bonnet points $Q_{9,10}$ are sources for $\lambda<0$ and sinks for $\lambda>0$; the Gauss-Bonnet points $Q_{11,12}$ are sources for $\lambda>0$ and sinks for $\lambda<0$.

In section \ref{V} we study the case where $y=0$ and $\lambda=0.$ There we studied a two-dimensional system for the variables $x$ and $\eta.$ Setting $\epsilon=1$ in section \ref{V-A} we obtained the following results: the Gauss-Bonnet point $P_1$ is a source for $0\leq w_m <\frac{1}{3}$; the Gauss-Bonnet point $P_2$ is a sink for $0\leq w_m <\frac{1}{3}$; 
On the other hand, setting $\epsilon=-1$ in section \ref{V-B} we obtained the following results: the Gauss-Bonnet point $P_1$ is a source for $0\leq w_m <\frac{1}{3}$; the Gauss-Bonnet point $P_2$ is a sink for $0\leq w_m <\frac{1}{3}$; the de Sitter point $P_3$ is a sink for $f_0\geq 0$ and $0\leq w_m <1$; the de Sitter point $P_4$ is a source for $f_0\geq 0$ and $0\leq w_m <1$; the de Sitter point $P_5$ is a sink for $f_0\leq 0$ and $0\leq w_m <1$; the de Sitter point $P_6$ is a sink for $f_0\leq 0$ and $0\leq w_m <1$.

As in the general case, in sections \ref{V-C} and \ref{V-D} we investigated the behaviour at infinity for the reduced system setting $\epsilon=\pm 1$ and obtained the following results: the Gauss-Bonnet point $Q_1$ is a source for $\frac{1}{3}<w_m\leq 1$; the Gauss-Bonnet point $Q_2$ is a sink for $\frac{1}{3}<w_m\leq 1$; the Gauss-Bonnet point $Q_3$ is a source for $\frac{1}{3}<w_m\leq 1$;  the Gauss-Bonnet point $Q_4$ is a sink for $\frac{1}{3}<w_m\leq 1$.

Observe that when $x=y=0$, we acquire  $z= \eta^2$, which means $\Omega_{m}= \rho_m/(3H^2)=z/\eta^2=1$, and we have matter-dominated solutions. Accordingly, the gravitational models can admit a cosmological solution where the matter source dominates, $\Omega_{m}=1$ (see Figure \ref{fig:weff1}). 

For investigating the viability of the theory, it is desirable to have complete cosmological  dynamics~\cite{Avelino:2013wea}; it should describe an early radiation-dominated era, later entering into an epoch of mater domination and finally reproducing the present acceleration of the Universe. In the dynamical systems language, complete cosmological dynamics can be understood as an orbit connecting a past attractor, also called a source, with a late-time attractor, also called a sink, that passes through some saddle points, such that radiation precedes matter domination. Some solutions interpolating between critical points can provide information on the intermediate stages of the evolution, with interest in orbits corresponding to a specific cosmological history~\cite{dn1,dn2}.

 To present one possible evolution of the physical model, Figure \ref{fig:weff1} displays the expressions $\omega_\phi(\tau)$, $x(\tau)$, $y(\tau)$, and $\eta(\tau)$ evaluated at a solution of system \eqref{newsyst-1}- \eqref{newsyst-3} for $\epsilon=1$ for the initial conditions for the left plot are $x(0)=0.001, \quad y(0)=\sqrt{\frac{\lambda }{\lambda -8 f_0}},  \quad \eta (0)=-\sqrt{\frac{\lambda }{\lambda -8 f_0}}$. The~solution is past asymptotic to $\omega_{\phi} =-1$ ($q=-1$), then remains near the de Sitter point, then tending asymptotically to $\omega_{\phi} =-\frac{1}{3}$ (the Gauss-Bonnet point) from~below. The initial conditions for the plot on the right are $x(0)=0.001,\quad y(0)=0.001,  \quad \eta (0)=0.9$. The solution is past asymptotic to $\omega_{\phi} =-\frac{1}{3}, q=0$ (zero acceleration), then it grows to $\omega_{\phi},q>0$, finally, it tends asymptotically to $\omega_{\phi}=0, \quad q=\frac{1}{2}$ describing a matter-dominated solution.

In the same lines, Figure \ref{fig:weff2} presents the expressions $\omega_\phi(\tau)$, $x(\tau)$, $y(\tau)$, and $\eta(\tau)$ evaluated at the solution of system \eqref{newsyst-1}- \eqref{newsyst-3} for $\epsilon=-1.$ The initial conditions for the left plot are $x(0)=0.001, \quad y(0)=\sqrt{\frac{\lambda }{\lambda -8 f_0}},  \quad \eta (0)=-\sqrt{\frac{\lambda }{\lambda -8 f_0}}$. The~solution is past asymptotic to $\omega_{\phi} =-1$ ($q=-1$), then remains near the de Sitter point, then tending asymptotically to $\omega_{\phi} =-\frac{1}{3}$ (the Gauss-Bonnet point) from~below. The initial conditions for the plot on the right are $x(0)=0.001,\quad y(0)=0.001,  \quad \eta (0)=0.9$. The solution is past asymptotic to $\omega_{\phi} =-\frac{1}{3}$ (zero acceleration), then it tends asymptotically to a de Sitter phase $\omega_{\phi}=-1, \quad q=-1$ describing a late-time acceleration.
 
Finally, one topic to be considered in further studies is reconstructing the cosmological history using different coupling functions between the scalar field and the Gauss-Bonnet scalar. 

\section*{Acknowledgments}

Alfredo David  Millano was supported by Agencia Nacional de Investigación y Desarrollo (ANID) Subdirección de Capital Humano/Doctorado Nacional/año 2020 folio 21200837, Gastos operacionales Proyecto de tesis/2022 folio 242220121, and by Vicerrectoría de Investigación y Desarrollo Tecnológico (VRIDT) at Universidad Católica del Norte. GL was funded through Concurso De Pasantías De Investigación Año 2022, Resolución VRIDT No. 040/2022 and Resolución VRIDT No. 054/2022. He also thanks the support of Núcleo de Investigación Geometría Diferencial y Aplicaciones, Resolución VRIDT N°096/2022, and Andronikos Paliathanasis acknowledges VRIDT-UCN through Concurso de Estadías de Investigación, Resolución VRIDT N°098/2022.

\appendix

\end{document}